\documentclass[pra,onecolumn,floatfix,superscriptaddress,longbibliography,notitlepage, nofootinbib]{revtex4-1}
\pdfoutput=1

\usepackage[utf8]{inputenc}
\usepackage{braket}
\usepackage{amsmath}
\DeclareMathOperator{\Tr}{Tr}
\usepackage{dcolumn}   
\usepackage{bm}        
\usepackage{amssymb}
\usepackage{mathtools}
\usepackage{tikz}
\usepackage{qcircuit}
\usepackage{appendix}
\usepackage{hyperref}
\usepackage{subcaption}
\usepackage{amsmath,amsfonts,amssymb}
\usepackage{mathtools}
\usepackage{thmtools,thm-restate}
\usepackage{algorithm}
\usepackage{mathrsfs}
\usepackage{tikz}
\usepackage{subcaption}
\usepackage{pgfplots}
\usetikzlibrary{3d}
\usepackage{tkz-euclide}
\usepackage{float}
\usepackage{algpseudocode}
\pgfplotsset{compat=1.17}
\usetikzlibrary{3d, arrows.meta, decorations.pathmorphing}

\newcommand{\xbf}{\textbf{x}}

\newcommand{\Ibb}{\mathbb{I}}

\newcommand{\Rbb}{\mathbb{R}}
\newcommand{\Wbf}{\textbf{W}}
\newcommand{\bbf}{\textbf{b}}
\newcommand{\Bbf}{\textbf{B}}
\newcommand{\wbf}{\textbf{w}}
\newcommand{\Xbf}{\textbf{X}}

\newtheorem{definition}{Definition}
\newtheorem{theorem}{Theorem}
\newtheorem{lemma}{Lemma}

\newtheorem{method}{Algorithm}

\usepackage{listofitems} 
\usetikzlibrary{arrows.meta} 
\usepackage[outline]{contour} 
\contourlength{1.4pt}

\usepackage{xcolor}
\colorlet{myred}{red!80!black}
\colorlet{myblue}{blue!80!black}
\colorlet{mygreen}{green!60!black}
\colorlet{myorange}{orange!70!red!60!black}
\colorlet{mydarkred}{red!30!black}
\colorlet{mydarkblue}{blue!40!black}
\colorlet{mydarkgreen}{green!30!black}

\tikzset{
  >=latex, 
  node/.style={thick,circle,draw=myblue,minimum size=22,inner sep=0.5,outer sep=0.6},
  node in/.style={node,green!20!black,draw=mygreen!30!black,fill=mygreen!25},
  node hidden/.style={node,blue!20!black,draw=myblue!30!black,fill=myblue!20},
  node convol/.style={node,orange!20!black,draw=myorange!30!black,fill=myorange!20},
  node out/.style={node,red!20!black,draw=myred!30!black,fill=myred!20},
  connect/.style={thick,mydarkblue}, 
  connect arrow/.style={-{Latex[length=4,width=3.5]},thick,mydarkblue,shorten <=0.5,shorten >=1},
  node 1/.style={node in}, 
  node 2/.style={node hidden},
  node 3/.style={node out}
}

\usetikzlibrary{arrows.meta}

\def\BibTeX{{\rm B\kern-.05em{\sc i\kern-.025em b}\kern-.08em
    T\kern-.1667em\lower.7ex\hbox{E}\kern-.125emX}}

\begin{document}
\title{Quantum Algorithms Without Coherent Quantum Access}
\author{Nhat A. Nghiem}
\affiliation{Department of Physics and Astronomy, State University of New York at Stony Brook, Stony Brook, NY 11794-3800, USA}
\affiliation{C. N. Yang Institute for Theoretical Physics, State University of New York at Stony Brook, Stony Brook, NY 11794-3840, USA}

\begin{abstract}
    Demonstrating quantum advantage has been a pressing challenge in the field. Most claimed quantum speedups rely on a subroutine in which classical information can be accessed in a coherent quantum manner, which imposes a crucial constraint on the implementability of these quantum algorithms. It has even been shown that without such an access, the quantum computer cannot be stronger than the classical counterparts. Thus, whether a quantum computer can be useful for practical applications is still open. In this work, we develop several variants of quantum algorithms. Our key framework employs a classical preprocessing step and a quantum procedure to perform gradient descent. We then translate such algorithm into an algorithm for solving linear systems, performing least-square fitting, building a support vector machine, performing supervised cluster assignment, training neural network, and solving for ground-state/excited-state energy, performing principle component analysis, with end-to-end applications of quantum algorithms. The classical preprocessing and the quantum procedure of our framework are shown to have logarithmic complexity in the dimension of input data, and quantum coherent access to input data is not required. Thus, our framework suggests an alternatively efficient route for quantum computers to handle real-world problems.   
\end{abstract}
\maketitle

\section{Introduction}

The study of quantum algorithms has advanced rapidly, yielding many methods to address diverse computational challenges \cite{grover1996fast, shor1999polynomial, feynman2018simulating, lloyd1996universal, berry2007efficient, berry2012black, berry2014high, berry2015hamiltonian, low2017optimal, low2019hamiltonian, childs2022quantum, o2016scalable, cerezo2021variational, babbush2023exponential, babbush2018low, mitarai2023perturbation, robert2021resource, kitaev1995quantum, aharonov2006polynomial, childs2010relationship, childs2021high, childs2017quantum, childs2021quantum, lloyd2013quantum, lloyd2014quantum, lloyd2016quantum, lloyd2020quantum, hauke2020perspectives, liu2021efficient, liu2018quantum, liu2024towards, durr1996quantum, deutsch1992rapid, bauer2020quantum, brassard1997quantum, brassard2002quantum, jordan2012quantum, leyton2008quantum, nachman2021quantum, jordan2005fast, garnerone2012adiabatic, gilyen2022quantum, miessen2023quantum, haah2021quantum, arrazola2019quantum, kerenidis2019quantum, hallgren2007polynomial}. A central objective is to establish quantum advantage—showing that quantum computers solve problems more efficiently than classical ones. Several rigorous demonstrations already exist. Bravyi et al.~\cite{bravyi2018quantum, bravyi2020quantum} showed that constant-complexity noisy quantum circuits can outperform classical methods on linear algebraic tasks. Maslov \cite{maslov2021quantum} proved advantages in space-restricted computation, and Liu et al.~\cite{liu2021rigorous} identified a supervised learning setting where quantum classifiers surpass classical ones. These results exemplify quantum advantage and fuel the pursuit of quantum supremacy. Growing interest also surrounds machine learning and data science \cite{lloyd2013quantum, lloyd2014quantum, lloyd2016quantum, lloyd2020quantum, wiebe2012quantum, wiebe2014quantum, mitarai2018quantum, farhi2018classification, farhi2014quantum, zhou2020quantum}, where high-dimensional data often strains classical resources. By contrast, quantum systems naturally operate in exponentially large spaces using only logarithmic qubits, making them especially promising for these domains.

Quantum speedups—ranging from polynomial to exponential—are often based on algorithms requiring oracle or black-box access, which lets quantum computers process classical data coherently. This access underpins many claimed exponential advantages, such as solving linear systems in logarithmic time \cite{harrow2009quantum, childs2017quantum}, least-squares fitting over exponentially large datasets \cite{wiebe2014quantum}, and quantum support vector machines \cite{rebentrost2014quantum}. In practice, however, oracle access is difficult to realize. Quantum random access memory (QRAM) has been proposed \cite{giovannetti2008architectures, giovannetti2008quantum}, but scalable implementations remain out of reach, limiting applicability. Moreover, Tang \cite{tang2019quantum, tang2021quantum} showed that if classical computers gain certain sampling access—analogous to QRAM—many supposed quantum exponential advantages collapse \cite{gilyen2018quantum, tang2018quantum, chia2018quantum, ding2021quantum, chia2022sampling}. This is often called the input issue. A second challenge is the output issue: the results of algorithms like the quantum linear solver \cite{harrow2009quantum,childs2017quantum} or quantum data fitting \cite{wiebe2012quantum} are quantum states encoding solutions in amplitudes. Extracting useful information requires tomography, which can erase the advantage \cite{aaronson2015read}. Together, these challenges raise doubts about whether quantum algorithms can meaningfully surpass classical methods in practice.

In this work, driven by the pursuit of (practical) quantum advantage, we draw upon the insight that real-world applications often involve extremely high-dimensional data. The natural ability of quantum mechanical systems to store exponentially large amounts of information should, therefore, be more effectively leveraged. A specific problem that has motivated our investigation is gradient descent—a widely used optimization method in machine learning and data science due to its efficiency and simplicity. 
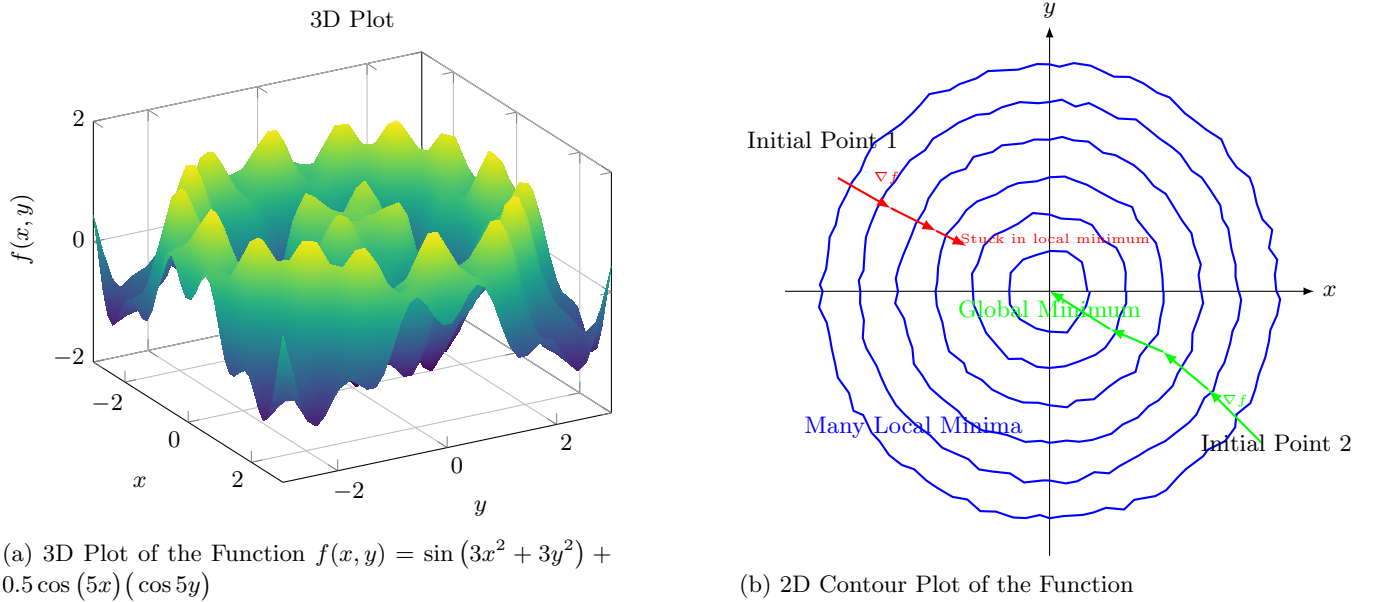
\begin{figure}[H]
    \centering
    \begin{subfigure}[b]{0.45\textwidth}
        \centering
        \begin{tikzpicture}
            \begin{axis}[
                view={60}{30},
                xlabel={$x$},
                ylabel={$y$},
                zlabel={$f(x, y)$},
                colormap/viridis,
                grid=major,
                domain=-3:3,
                y domain=-3:3,
                samples=40,
                zmin=-2, zmax=2,
                title={3D Plot}
            ]
                \addplot3[
                    surf,
                    shader=interp
                ]
                {sin(deg(3 * sqrt(x*x + y*y))) + 0.5 * cos(deg(5 * x)) * cos(deg(5 * y))};
            \end{axis}
        \end{tikzpicture}
        \caption{3D Plot of the Function $f(x,y) = \sin \big( 3x^2 + 3y^2\big) + 0.5 \cos \big( 5x\big) \big( \cos 5y \big)$}
    \end{subfigure}
    \hfill
    \begin{subfigure}[b]{0.45\textwidth}
        \centering
        \begin{tikzpicture}
            \draw[->] (-3.5, 0) -- (3.5, 0) node[right] {$x$};
    \draw[->] (0, -3.5) -- (0, 3.5) node[above] {$y$};

    \foreach \r in {0.5, 1, 1.5, 2, 2.5, 3} {
        \draw[blue, thick, smooth, decoration={random steps, segment length=2mm, amplitude=0.5mm}, decorate] (0, 0) circle (\r);
    }

    \draw[->, thick, red] (-2.8, 1.5) -- (-2.1, 1.1) node[midway, above right] {\tiny $\nabla f$};
    \draw[->, thick, red] (-2.1, 1.1) -- (-1.5, 0.8);
    \draw[->, thick, red] (-1.5, 0.8) -- (-1.1, 0.6) node[midway, right] {\tiny Stuck in local minimum};

    \draw[->, thick, green] (2.8, -2) -- (2.1, -1.3) node[midway, above] {\tiny $\nabla f$};
    \draw[->, thick, green] (2.1, -1.3) -- (1.5, -0.8);
    \draw[->, thick, green] (1.5, -0.8) -- (0.8, -0.5);
    \draw[->, thick, green] (0.8, -0.5) -- (0, 0) node[below] {Global Minimum};

    \node at (-3, 2) {Initial Point 1};
    \node at (3, -2) {Initial Point 2};
    \node[blue] at (-1.8, -1.8) {Many Local Minima};
        \end{tikzpicture}
        \caption{2D Contour Plot of the Function}
    \end{subfigure}
    \label{fig: 3d2d}
    \caption{Illustration of gradient descent method applied to a multi-variate function. The function $f(x,y)$ drawn on the left exhibits many extrema, as indicated by the contour on the right figure. The initialization is critical in the performance of gradient descent, as different initialization may result in different point of extrema. The extreme case is when the function is convex, then the local minima is also global minima, and gradient descent is guaranteed to converge to global minima.    }
\end{figure}
More formally, we are given an objective function $f: \Rbb^n \rightarrow \Rbb$ of $n$ variables, and the goal is to find a point $\xbf \equiv (x_1,x_2,...,x_n)$ where the value of the given function is (locally) minimized. The algorithmic procedure of gradient descent is as follows. We begin with some random guess $\xbf_0$, then iterate the following procedure:
\begin{align}
    \xbf_{t+1} = \xbf_t - \eta \bigtriangledown f(\xbf_t) 
    \label{eqn: 1}
\end{align}
where $\eta >0$ is a hyperparameter. 
Gradient descent relies on computing partial derivatives of a function, which can be approximated numerically from function evaluations. Iterations proceed until the update $\xbf_t$ stabilizes, i.e., $\nabla f(\xbf_t)!\to!0$, indicating convergence to a minimum. In practice, the method runs for a fixed number of steps $T$, with complexity determined by evaluating $f$, its gradient, and the iteration count. Despite its simplicity, gradient descent underpins many scientific and engineering tasks, including solving linear systems, least-squares fitting, training neural networks, and building support vector machines. Several efforts have sought quantum adaptations \cite{rebentrost2019quantum, nghiem2024simple}, but current methods apply only to restricted polynomial classes and still face the input–output issues discussed earlier. Thus, the practical value of quantum gradient descent remains an open question.

In this work, we generalize quantum gradient descent to a broader class of polynomials while overcoming prior limitations. With classical preprocessing, we design an explicit quantum circuit that achieves exponential speedup without relying on black-box or oracle access, thereby resolving the input issue. Since many tasks—linear systems, least-squares fitting, SVMs, neural networks, ground-state energy, and PCA—can be cast as optimization problems, our algorithm naturally applies to them. Moreover, we address the output issue \cite{aaronson2015read} by showing, through concrete examples, how the quantum solution state can be directly leveraged, e.g., predicting new inputs in least-squares fitting. While these applications have been explored separately before \cite{harrow2009quantum, childs2017quantum, rebentrost2014quantum, zlokapa2021quantum}, they relied on oracle access, limiting practicality. In contrast, our approach requires only the coefficients and degrees of the target function, ensuring a provable and realistic quantum advantage. This contributes to the growing body of results \cite{bravyi2018quantum, bravyi2020quantum, maslov2021quantum, liu2021rigorous} demonstrating quantum advantage in practical settings and further motivates the pursuit of quantum computing. The next section outlines our framework for quantum gradient descent, including objectives, key insights, techniques, algorithm sketch, and outcomes. We study three classes of $n$-variable functions $f(\xbf): \Rbb^n \to \Rbb$, each with distinct gradient structures. Full details of the algorithm appear in Appendix \ref{sec: mainalgorithm}, with supporting definitions and tools provided in Appendix \ref{sec: prelim}.

\section{Quantum Gradient Descent Without Coherent Quantum Access}

\subsection{Key Technique}
\label{sec: keytechnique}

In this section, we describe the key technique that underlies our subsequent algorithms. 

\noindent
\textbf{State preparation.} Let the coefficients $\{ a_1,a_2,...,a_n\}$ where each $a_i \in \mathbb{C}$ are known, there exists a variety of amplitude encoding techniques \cite{grover2000synthesis,grover2002creating,plesch2011quantum, schuld2018supervised, nakaji2022approximate,marin2023quantum,zoufal2019quantum,zhang2022quantum, mcardle2022quantum}, which allow us to efficiently prepare the state:
\begin{align}
    \ket{\Phi}= \frac{1}{||a||} \sum_{i=1}^n a_i \ket{i-1}
\end{align}
where $||a|| = \sqrt{  \sum_{i=1}^n|a_i|^2}$. For instance, Ref.~\cite{zhang2022quantum} constructed a universal protocol for amplitude encoding. By universal, it means that this method can work with all coefficients $\{a_i\}_{i=1}^n$, as long as they are classically known. Specifically, the method of \cite{zhang2022quantum} uses a circuit of complexity $\Theta(\log n)$, $\mathcal{O}(s_a)$ ancilla qubits ($s_a$ is the number of nonzero elements in $\{a_i\}_{i=1}^n$), and $\mathcal{O}(\log n)$ classical prep-processing time. At the same time, the method in \cite{mcardle2022quantum} can prepare the state $\ket{\Phi}$, achieving (asymptotically) the same circuit complexity, but using significantly less ancilla qubits and  classical preprocessing time. However, the method only works if the amplitudes $\{a_i\}_{i=1}^n$ obeys a certain functional output, e.g., $a_i = f(i)$ for some definite-parity function $f(.)$. Similarly, the other works \cite{grover2000synthesis,grover2002creating,plesch2011quantum, schuld2018supervised, nakaji2022approximate,marin2023quantum,zoufal2019quantum} provide a few class of states that can be efficiently prepared without using many ancillas and classical preprocessing time. For subsequent usage, we recapitulate the state preparation procedure above in the following lemma:
\begin{lemma}[Efficient State Preparation]
\label{lemma: statepreparation}
    Let $  \ket{\Phi}= \frac{1}{||a||} \sum_{i=1}^n a_i \ket{i-1}$ and $\{a_i\}_{i=1}^n$ are known. Then: 
    \begin{itemize}
        \item (General case 1) $\ket{\Phi}$ can be prepared with a quantum circuit of complexity $\mathcal{O}(\log n)$, $\mathcal{O}(s_a)$ ancilla qubits and $\mathcal{O}\left( \log n\right)$ classical preprocessing time \cite{zhang2022quantum}.
        \item (Structured case 2) If $a_i$ satisfies $a_i = f(i)$ for some definite-parity function $f(.)$, then it can be prepared with a circuit of complexity $\mathcal{O}\left( \log n\right)$, $\mathcal{O}(1)$ ancilla qubits and without classical pre-processing time \cite{mcardle2022quantum}.
        \item (Structured case 3 ) If $\ket{\Phi}$ has structure as any of the following works \cite{grover2000synthesis,grover2002creating,plesch2011quantum, schuld2018supervised, nakaji2022approximate,marin2023quantum,zoufal2019quantum,prakash2014quantum}, then $\ket{\Phi}$ can be prepared with a circuit of complexity $\mathcal{O}\left( \log n\right)$, $\mathcal{O}(1)$ ancilla qubit and no classical preprocessing.
    \end{itemize}
\end{lemma}

\noindent
\textbf{Block-encoding and quantum singular value transformation (QSVT).} Roughly speaking, a unitary $U$ is said be a block encoding of $A$ (with operator norm $|A| \leq 1$) if $U$ contains $A$ in the top left corner, i.e., $U = \begin{pmatrix}
        A & \cdot \\
        \cdot & \cdot 
\end{pmatrix}$, where $(\cdot)$ refers to possibly non-zero entries. Block encoding is central to the recently breakthrough quantum singular value transformation framework \cite{gilyen2019quantum}, which has been shown to have the potential to unify many quantum algorithms. We refer to Appendix \ref{sec: prelim} for a more concrete and formal definition of block encoding, as well as important related techniques involving block-encoded operator. Here we recapitulate a few simple but very useful properties for our subsequent construction. Suppose that $U_1$ is a block encoding of $A_1$, $U_2$ is a block encoding of $A_2$, then for some known $\alpha, \beta \leq 1$, we can construct another unitary a unitary block encoding of $\alpha_1 A_1, \alpha_2 A_2$ (Lemma \ref{lemma: scale}, trivial scaling) $ \alpha A_1 + \beta A_2$ (Lemma \ref{lemma: sumencoding}, Linear combination), of $ A_1 A_2$ (Lemma \ref{lemma: product}, Multiplication), and also of $A_1 \otimes A_2$ (Ref. \cite{camps2020approximate}, Tensor product). Additionally, for a factor $\gamma > 1$ and with a guarantee $\gamma A \leq \frac{1}{2}$, it is possible to construct the block encoding of $\gamma A$ (Lemma \ref{lemma: amp_amp}, Amplification). 

The last recipe we need is the following result:
\begin{lemma}[Theorem 2 in \cite{rattew2023non}, \cite{guo2024nonlinear}; Block Encoding into a Diagonal Matrix]
\label{lemma: diagonal}
     Given an $\log(n)$-qubit quantum state specified by a state-preparation-unitary $U$, such that $\ket{\psi}_n=U\ket{0}_n=\sum^{n}_{k=1}\psi_k \ket{k-1}_n$ (with $\psi_k \in \mathbb{C}$), we can prepare an exact block-encoding $U_A$ of the diagonal matrix $A = {\rm diag}(\psi_1, ...,\psi_{n})$ with $\mathcal{O}(\log(n))$ circuit complexity and a total of $\mathcal{O}(1)$ queries to a controlled-$U$ gate  with $\log(n)+3$ ancillary qubits.
\end{lemma}
In the following, we shall apply the above recipes to outline a variant of quantum gradient descent algorithm for three different classes of functions. 

\subsection{First Type of Function}
\label{sec: firstype}
Let $\xbf \equiv (x_1,x_2,...,x_n) $. The first type admits the following form $f(\xbf) = \sum_{i=1}^K a_{i,1} x_1^i + a_{i,2} x_2^i + ... + a_{i,n} x_n^i$ and $\{a_{i,1}, a_{i,2} ,..., a_{i,n} \}_{i=1}^K$ is some real coefficient. Without loss of generality, we further assume that our domain of interest is $\mathscr{D} = [-1/2,1/2]^n$ and $|f(x_1,x_2,...,x_n)| \leq 1/2$ in this domain $\mathscr{D}$. Furthermore, we assume that the maximum entry of the gradient of $f$ is bounded by some known value $P$, that is, $| \bigtriangledown f(x_1,x_2,...,x_n) |_\infty \leq P $ (where $|.|_\infty$ refers to the maximum entry of underlying vector) for all $(x_1,x_2,...,x_n) \in \mathscr{D}$. These two kinds of function is fairly common in practice, as we will see from examples in subsequent sections. The only information we need to know as input to our quantum algorithm is $K$ and all coefficients $\{a_{i,1}, a_{i,2} ,...,a_{i,n}\}_{i=1}^K$. In addition, for all $i =1,2,...,K$, we denote $s_{a_i}$ be the number of nonzero elements among $\{a_{i,j}\}_{j=1}^n$.

Most existing quantum algorithms typically rely on the encoding of classical variables $\xbf \equiv (x_1,x_2,...,x_n)$ into some quantum state $\ket{\xbf}$. Our framework exploit a different mechanism, as instead of quantum state, we encode the classical vector $ \xbf$ into a diagonal operator denoted as $\rm diag (\xbf)$, which is an $n \times n$ matrix having $(x_1,x_2,...,x_n)$ as diagonal entries and zero otherwise. In otherwords, we enact the following embedding $\xbf = \sum_{i=1}^n x_i \ket{i-1}\longrightarrow \bigoplus_{i=1}^n x_i \ket{i-1}\bra{i-1}$. By doing this way, we can leverage many arithmetic techniques from the seminal quantum singular value transformation frameworks, allowing us to manipulate these information in a subtle yet useful manner for the desired task. Recall that gradient descent algorithm is mainly composed of the following iterative operation $\xbf_{t+1} = \xbf_t - \eta \bigtriangledown f(\xbf)$ (for $t=0,1,2,...,T$). In our convention, we seek to obtain the following operation $\rm diag \big( \xbf_{t+1} \big)  = \rm diag \big( \xbf_t  - \eta \bigtriangledown f(\xbf) \big) $. First we note that the gradient of the first type is:
\begin{align}
    \bigtriangledown f(\xbf) = \begin{pmatrix}
        \sum_{i=1}^K a_{i,1} i x^{i-1}_1  \\
    \sum_{i=1}^K a_{i,2} i x_2^{i-1} \\
    \vdots \\
    \sum_{i=1}^K a_{i,n} i x_n^{i-1}
    \end{pmatrix}
\end{align}
As mentioned, we embed the classical vector $\xbf$ into a diagonal operator $\rm diag(\xbf)$, which is possible given the unitary $U$ that creates $\xbf$ (or more generally, create a larger state that includes $\xbf$ as first $n$ entries) combined with the Lemma \ref{lemma: diagonal}. Given the block encoding of $\rm diag(\xbf) \equiv \rm \big( x_1,x_2,...,x_n\big)$ and of $ \{\rm diag \big(  a_{i,1}, a_{i,2}, ..., a_{i,n} \big)\}_{i=1}^K $, which is possible due to combination of Lemma \ref{lemma: statepreparation} and Lemma \ref{lemma: diagonal}, many elegant techniques from quantum singular value transformation framework can be applied (see Appendix \ref{sec: prelim}). More concretely,  Lemma \ref{lemma: theorem56} or Lemma \ref{lemma: product}) takes $\rm diag(\xbf)$ as input and yields a block encoding of $\rm 
 diag \big( x_1^{i-1},x_2^{i-1}, ..., x_n^{i-1}  \big) $. Then Lemma \ref{lemma: product} can be used again to construct the block encoding of $\rm  diag \big( x_1^{i-1},x_2^{i-1}, ..., x_n^{i-1}  \big) \rm diag \big(  a_{i,1}, a_{i,2}, ..., a_{i,n} \big)$, which is $\rm diag \big( a_{i,1} x_1^{i-1}, a_{i,2} x_2^{i-1}, ..., a_{i,n} x_n^{i-1}  \big)$
 
Lemma \ref{lemma: scale} can insert the factor $\frac{i}{P}$ to the above operator, obtaining a block encoding of $\rm diag \big( \frac{i}{P} a_{i,1} x_1^{i-1}, \frac{i}{P} a_{i,2} x_2^{i-1}, ..., \frac{i}{P} a_{i,n} x_n^{i-1}  \big) $.
Then Lemma \ref{lemma: sumencoding} can be used to construct the block encoding of $    \frac{1}{K}\sum_{i=1}^K \rm diag \big( \frac{i}{P} a_{i,1} x_1^{i-1}, \frac{i}{P} a_{i,2} x_2^{i-1}, ..., \frac{i}{P} a_{i,n} x_n^{i-1}  \big) = \frac{1}{KP} \rm diag \Big( \bigtriangledown f(\xbf) \Big)$. Then Lemma \ref{lemma: sumencoding} can be used again to construct the block encoding of:
\begin{align}
    \frac{1}{2}\left( \rm diag(\xbf) - \frac{1}{K}\sum_{i=1}^K \rm diag \big( \frac{i}{P} a_{i,1} x_1^{i-1}, \frac{i}{P} a_{i,2} x_2^{i-1}, ..., \frac{i}{P} a_{i,n} x_n^{i-1}  \big)  \right)
    \label{eqn: 7}
\end{align}
If we define the hyperparameter $\eta = \frac{1}{KP}$ then the above operator is indeed $\frac{1}{2}\Big( \rm diag(\xbf) - \eta \rm diag \big( a_{i,1} x_1^{i-1}, a_{i,2} x_2^{i-1}, ..., a_{i,n} x_n^{i-1}  \big) \Big)= \frac{1}{2}\Big( \rm diag(\xbf) - \eta \rm diag \big( \bigtriangledown f(\xbf)  \big)   \Big) $. Therefore, if we begin at some $\xbf_0$, or more precisely $\rm diag(\xbf_0)$, then after the above procedure, we obtain a block encoding of $ \frac{1}{2}\Big( \rm diag(\xbf_0) - \eta \rm diag \big( \bigtriangledown f(\xbf_0) \big)  \Big)  =\frac{1}{2}\Big( \rm diag( \xbf_1 )   \Big) $ which contains the updated solution $\xbf_1$ as the diagonal entries. The factor $1/2$ above can be removed with a technique called amplification, see Lemma \ref{lemma: amp_amp}. Thus, we obtain a block encoding of $\rm diag(\xbf_1)$. Likewise, after the next execution, we obtain the block encoding of $\rm diag (\xbf_2)$, and continually, the block encoding of $\rm diag(\xbf_3),...,\rm diag (\xbf_T)$. A detailed analysis of complexity can be found in Appendix \ref{sec: mainalgorithm}. Here, for brevity, we summarize the result in the following theorem:
\begin{theorem}[Quantum Gradient Descent Algorithm (First Type) ]
\label{theorem: type1}
    Let $f(x_1, x_2,...,x_n):\Rbb^n \longrightarrow \Rbb$ be an $n$-dimensional real function $f(x_1,x_2, ..., x_n) = \sum_{i=1}^K \big( a_{i,1} x_1^i + a_{i,2} x_2^i + ... + a_{i,n} x_n^i   \big) $
   with known coefficients $\{ a_{i,1},a_{i,2},...,a_{i,n}\}_{i=1}^K$, on the domain $\mathscr{D} = [-1/2,1/2]^n$, and $|f(x_1,x_2,...,x_n)| \leq 1/2$ in such a domain. Let $T$ be the total iteration steps. Suppose for each $i$, there is a unitary $U_i$ of complexity $T(a_i)$ that generates some quantum state that has its first $n$ entries as $( a_{i,1},a_{i,2}, ...., a_{i,n})$. Let $T_a = \max_i \{T(a_i)\}_{i=1}^K$. Then there exists an algorithm that outputs a block encoding of a diagonal operator $\textbf{X}_T$ containing $\xbf_T$ as entries. A quantum state $\ket{\xbf_T}$ corresponding to $\xbf_T$ can also be obtained with further $\mathcal{O}(1)$ complexity. The resource cost of such algorithm are: 
   \begin{itemize}
   \item $\mathcal{O}\big(T\log (nK) \big)$ qubits
       \item   a quantum circuit of complexity  $\mathcal{O}\Big(  K^{2T} \big( T_a + \log(n)  \big) \Big)$
        \item totally $\mathcal{O}(\sum_{i=1}^K s_{a_i})$ ancilla qubits, a classical preprocessing of complexity $\mathcal{O}\left( \log n \right)$. These costs can be improved to $\mathcal{O}(1)$ if for all $i$, the state $\varpropto \sum_{j=1}^n a_{i,j} \ket{j-1}$ fall into the structured cases in Lemma \ref{lemma: statepreparation}.
   \end{itemize} 
\end{theorem}

\subsection{Second Type of Function}
\label{sec: secondtype}
Now we consider the second type of function $f(\xbf) = \sum_{i=1}^K\big( a_{i,1} x_1 + a_{i,2}x_2 + \cdots + a_{i,n}x_n  + b_i \big)^i$ with $\{b_i\}_{i=1}^K$ are known coefficients. For simplicity, we define the vector $ \textbf{a}_i \equiv (a_{i,1},a_{i,2},...,a_{i,n})^T $. Then the summation $ \big( a_{i,1} x_1 + a_{i,2}x_2 + \cdots + a_{i,n}x_n  \big) = \textbf{a}_i^T \xbf $ and simplify further $f_i(\xbf) \equiv f_i(x_1,x_2,...,x_n) = \textbf{a}_i^T \xbf + b_i$,  so the function is $f(\xbf) =  \sum_{i=1}^K \big( \textbf{a}_i^T \xbf + b_i  \big)^i$. The gradient for the function is:
\begin{align}
\bigtriangledown f(\xbf) =\begin{pmatrix}
    \sum_{i=1}^K a_{i,1} i \big( a_{i,1} x_1 + a_{i,2}x_2 + \cdots + a_{i,n}x_n + b_i \big)^{i-1} \\
    \sum_{i=1}^K a_{i,2} i \big( a_{i,1} x_1 + a_{i,2}x_2 + \cdots + a_{i,n}x_n + b_i \big)^{i-1} \\
    \vdots \\
    \sum_{i=1}^K a_{i,n} i \big( a_{i,1} x_1 + a_{i,2}x_2 + \cdots + a_{i,n}x_n + b_i \big)^{i-1} 
\end{pmatrix} = \begin{pmatrix}
        \sum_{i=1}^K a_{i,1}i \big(  \textbf{a}_i^T \xbf + b_i \big)^{i-1}  \\
        \sum_{i=1}^K a_{i,2} i\big(  \textbf{a}_i^T \xbf + b_i \big)^{i-1}  \\
        \vdots \\
        \sum_{i=1}^K a_{i,n}i  \big(  \textbf{a}_i^T \xbf + b_i\big)^{i-1}  \\
    \end{pmatrix}
\end{align}
Apparently, this gradient exhibits more complication compared to the first type's gradient, and indeed it possess considerably more technical challenge to build the operator $\rm diag \big( \bigtriangledown f(\xbf) \big)$, as the summation $ \textbf{a}_i^T \xbf$ doesn't seem to straightforward to build from the block encoding of $\rm diag(\xbf) \equiv \rm \big( x_1,x_2,...,x_n\big)$ and of $ \{\rm diag \big( \textbf{a}_i  \big) \}_{i=1}^K $. The full procedure is somewhat lengthly so we provide it in the Appendix \ref{sec: mainalgorithm}. For brevity, we summarize the key information here:
\begin{lemma}
\label{lemma: diag}
   (\textbf{First version.}) Given a block encoding of $\rm diag(\xbf)$ and of unitary $U(a_i)$ that generates a state that is, or contains $(a_{i,1},a_{i,2},...,a_{i,n})$  in its first $n$ entries for $i=1,2,...,K$, then the block encoding of $ \big(  \textbf{a}_i^T \xbf + b_i \big) \otimes \Ibb_n $ can be constructed with $\mathcal{O}(s_i)$ usage of block encoding of $\rm diag(\xbf)$ and of $ \{\rm diag \big( \textbf{a}_i  \big) \}_{i=1}^K $, and another $\mathcal{O}(1)$ two-qubit gates, where $s_i$ is the number of non-zero entries of $\textbf{a}_i$ (we need to know these indexes). We remark that for this version, the same construction can be done whenever we have block encoding of $\rm diag(\xbf)$ and of $\rm diag (a_{i,1},a_{i,2},...,a_{i,n}) $.  \\
    (\textbf{Second version.}) Given a block encoding of $\rm diag(\xbf)$ and unitary $U(\sqrt{a_i})$ that generates some state that is, or contains $ (\sqrt{a_{i,1}}, \sqrt{a_{i,2}}, ..., \sqrt{a_{i,n}})$ in its first $n$ entries. Then there exists a quantum procedure using block encoding of $\rm diag(\xbf)$ and $U(\sqrt{a_i})$ two times plus $\mathcal{O}(1)$ two qubit gates, and return the block encoding of $ \big(  \textbf{a}_i^T \xbf + b_i \big)  \otimes \Ibb_n $
\end{lemma}
We remark that the unitary$U(\sqrt{a_i})$ in the second version of the above Lemma can, in principle, be constructed using method in \cite{zhang2022quantum} -- which is arguably the most universal amplitude encoding protocol. In the following, we proceed with the first version of the above lemma. However, the same procedure and analysis holds for the second version as well. The only difference is that the second version enables certain improvement in scaling complexity (in subsequent Thm~\ref{theorem: type2}). We note that the operator $  \big(  \textbf{a}_i^T \xbf + b_i \big) \otimes \Ibb_n$ is an $n\times n$ matrix having $ \textbf{a}_i^T \xbf $ on the diagonal and zero otherwise. So we can use Lemma \ref{lemma: product} to construct the block encoding of $  \big(  (\textbf{a}_i^T \xbf)^{i-1} + b_i \big) \otimes \Ibb_n$. Remind that we have a block encoding of $\rm diag (\textbf{a}_i)$. Then we can use Lemma \ref{lemma: product} to construct the block encoding of $ \rm diag(\textbf{a}_i) \cdot  \big(  \textbf{a}_i^T \xbf + b_i \big) \otimes \Ibb_n =\rm diag \big(  a_{i,1}  \big(\textbf{a}_i^T \xbf + b_i \big)^{i-1}, a_{i,2} \big(\textbf{a}_i^T \xbf + b_i\big)^{i-1} , ..., a_{i,n }  \big(\textbf{a}_i^T \xbf  + b_i\big)^{i-1} \big) $
Then Lemma \ref{lemma: scale} can be used to insert the factor $i/P$, i.e., we obtain the block encoding of $\rm diag \big(\frac{i}{P}  a_{i,1}  \big(\textbf{a}_i^T \xbf+ b_i \big)^{i-1}, \frac{i}{P} a_{i,2} \big(\textbf{a}_i^T \xbf + b_i\big)^{i-1} , ...,  \frac{i}{P} a_{i,n }  \big(\textbf{a}_i^T \xbf + b_i \big)^{i-1} \big)  $. 
Next, Lemma \ref{lemma: sumencoding} can then be used to construct the block encoding of 
\begin{align}
     \frac{1}{K}  \sum_{i=1}^K \rm diag \big( \frac{i}{P} a_{i,1}  \big(\textbf{a}_i^T \xbf \big)^{i-1},\frac{i}{P} a_{i,2} \big(\textbf{a}_i^T \xbf\big)^{i-1} , ..., \frac{i}{P}a_{i,n }  \big(\textbf{a}_i^T \xbf \big)^{i-1} \big)  = \frac{1}{KP} \rm diag\big( \bigtriangledown f(\xbf) \big)
     \label{eqn: 10}
\end{align} 
Lemma \ref{lemma: sumencoding} allows us to construct the block encoding of $\frac{1}{2} \Big(  \rm diag(\xbf) - \frac{1}{KP} \rm diag\big( \bigtriangledown f(\xbf) \big) \Big) = \frac{1}{2} \Big(  \rm diag(\xbf) - \eta \rm diag\big( \bigtriangledown f(\xbf) \big)  \Big) $
where again we define the hyperparameter $\eta = \frac{1}{KP} $. The factor $1/2$ can be removed by amplification technique \ref{lemma: amp_amp}. Thus, similar to previous discusison, if we begin at some $\rm diag(\xbf_0)$, after the consecutive execution of the above algorithm, we obtain the block encoding of $\rm diag(\xbf_1), \rm diag(x_2), ..., \rm diag(x_T)$. The full analysis of complexity will be provided in Appendix \ref{sec: mainalgorithm}, and we state the main result here:
\begin{theorem}[Quantum Gradient Descent Algorithm (Second Type) ]
\label{theorem: type2}
    Let $f(x_1, x_2,...,x_n):\Rbb^n \longrightarrow \Rbb$ be an $n$-dimensional real function: $ f_i(x_1,x_2, ..., x_n)  =\sum_{i=1}^K \big( a_{i,1} x_1 + a_{i,2}x_2 + \cdots + a_{i,n}x_n  
 + b_i \big)^i$
   with known coefficients $\{ a_{i,1},a_{i,2},...,a_{i,n}\}_{i=1}^K$, on the domain $\mathscr{D} = [-1/2,1/2]^n$, and $|f(x_1,x_2,...,x_n)| \leq 1/2$ in such a domain. Let $T$ be the total iteration steps.  \\

\noindent
\textbf{(First version)}  Suppose that for each $i$, there is a unitary $U(a_i)$ of complexity $T(a_i)$ that generates some quantum state that has its first $n$ entries as $( a_{i,1},a_{i,2}, ...., a_{i,n})$. Let $T_a = \max_i \{T(a_i) \}_{i=1}^K$. For each $i$, define $s_i$ as the number of non-zero entries of the vector $(a_{i,1},a_{i,2}, ...., a_{i,n})$, and $S = \max_i \{ s_i \}_{i=1}^K$.  Then there exists an algorithm that outputs a block encoding of a diagonal operator $\textbf{X}_T$ containing $\xbf_T$ as entries. A quantum state $\ket{\xbf_T}$ corresponding to $\xbf_T$ can also be obtained with further $\mathcal{O}(1)$ complexity. The resource cost of such algorithm are: 
   \begin{itemize}
   \item $\mathcal{O}\big(T\log (nK) \big)$ qubits
       \item   a quantum circuit of complexity  $\mathcal{O}\Big(  (SK^2)^{T} \big( T_a +  \log(n)  \big) \Big))$
        \item totally $\mathcal{O}(\sum_{i=1}^K s_{a_i})$ ancilla qubits, a classical preprocessing of complexity $\mathcal{O}\left( \log n \right)$. These costs can be improved to $\mathcal{O}(1)$ if for all $i$, the state $\varpropto \sum_{j=1}^n a_{i,j} \ket{j-1}$ fall into the structured cases in Lemma \ref{lemma: statepreparation}.
   \end{itemize} 

\noindent
\textbf{(Second version)}  If, under mostly the same condition for function $f(x_1,x_2,...,x_n)$, except for $f_i(x_1,x_2, ..., x_n)  =\sum_{i=1}^K \big( a_{i,1} x_1 + a_{i,2}x_2 + \cdots + a_{i,n}x_n  + b_i \big)^{2i+1}$ and for each $i$, we have a unitary $U(\sqrt{a_i})$ (of complexity $T(\sqrt{a_i})$ that generates some state that contains $( \sqrt{a_{i,1}}, \sqrt{a_{j,2}}, ..., \sqrt{a_{j,n}} )$ as its first $n$ entries. Let $T_u = \max_i \{ T(\sqrt{a_i})\}_{i=1}^K $. Then there exists an algorithm that outputs a block encoding of a diagonal operator $\textbf{X}_T$ containing $\xbf_T$ as entries. A quantum state $\ket{\xbf_T}$ corresponding to $\xbf_T$ can also be obtained with further $\mathcal{O}(1)$ complexity. The resource cost of such algorithm are: 
   \begin{itemize}
   \item $\mathcal{O}\big(T\log (nK) \big)$ qubits
       \item   a quantum circuit of complexity  $\mathcal{O}\Big(  K^{2T} \big(  \log(n) + T_u)   \Big)$
        \item totally $\mathcal{O}(\sum_{i=1}^K s_{a_i})$ ancilla qubits, a classical preprocessing of complexity $\mathcal{O}\left( \log n \right)$. These costs can be improved to $\mathcal{O}(1)$ if for all $i$, the state $\varpropto \sum_{j=1}^n a_{i,j} \ket{j-1}$ fall into the structured cases in Lemma \ref{lemma: statepreparation}.
   \end{itemize}

\end{theorem}

\subsection{Third Type of Function}
\label{sec: thirdtype }
The third type of function we consider is the following $f(\xbf)  = \prod_{i=1}^K  \big( a_{i,1} x_1 + a_{i,2}x_2 + \cdots + a_{i,n}x_n  + b_i \big)^i$. Similar to previous case, we define $\textbf{a}_i \equiv (a_{i,1},a_{i,2},..., a_{i,n})^T$, so the function is equivalent to $    f( \xbf ) = \prod_{i=1}^K f_i(\xbf) = \prod_{i=1}^K \big( \textbf{a}_i^T \xbf + b_i \big)^i$
The gradient of the above function is:
\begin{align}
    \bigtriangledown f(\xbf) = \begin{pmatrix}
            \frac{\partial f_1(\xbf)}{\partial x_1} \prod_{i=2}^K f_i(\xbf) + f_1(\xbf) \frac{\partial f_2(\xbf)}{\partial x_1}\prod_{i=3}^K f_i(\xbf) + ... + \prod_{i=1}^{K-1} \frac{\partial f_K(\xbf)}{\partial x_1} \\
            \frac{\partial f_1(\xbf)}{\partial x_2} \prod_{i=2}^K f_i(\xbf) + f_1(\xbf) \frac{\partial f_2(\xbf)}{\partial x_2}\prod_{i=3}^K f_i(\xbf) + ... + \prod_{i=1}^{K-1} \frac{\partial f_K(\xbf)}{\partial x_2} \\
            \vdots \\
            \frac{\partial f_1(\xbf)}{\partial x_n} \prod_{i=2}^K f_i(\xbf) + f_1(\xbf) \frac{\partial f_2(\xbf)}{\partial x_n}\prod_{i=3}^K f_i(\xbf) + ... + \prod_{i=1}^{K-1} \frac{\partial f_K(\xbf)}{\partial x_n}
    \end{pmatrix}
\end{align}
which is apparently seemingly more complicated than the previous two cases. However, it turns out that things can be simpler. We observe that the above gradient can be decomposed as $\bigtriangledown f(\xbf) = \sum_{i=1}^K  \prod_{j=1}^{i-1} f_j(\xbf) \bigtriangledown f_i(\xbf)  \prod_{j=i+1}^K f_j(\xbf)    $. In addition, we can use a lot of recipes from the previous discussion on the second type. From prior discussion, we have a block encoding of operator $\frac{1}{P} \rm diag\Big( \bigtriangledown f_i(\xbf) \Big)$ for $i=1,2,...,K$. We also have from Lemma \ref{lemma: diag} the block encoding of operator $\big( \textbf{a}_i^T \xbf + b_i \big) \otimes \Ibb_n = \rm diag \Big( f_i(\xbf)  \Big)$. Then for any $i$, it is straightforward to use Lemma \ref{lemma: product} to construct the block encoding of the following operator $\prod_{j=1}^{i-1} \rm diag\Big( f_j(\xbf) \Big) \frac{1}{P} \rm diag\Big( \bigtriangledown f_i(\xbf) \Big) \prod_{j=i+1}^{K} \rm diag\Big( f_j(\xbf) \Big)$, which is equal to $\frac{1}{P}\rm diag\Big( \prod_{j=1}^{i-1} f_i(\xbf)  \bigtriangledown f_i(\xbf) \prod_{j=i+1}^{K} f_i(\xbf) \Big)$

Then by using Lemma \ref{lemma: sumencoding}, we can construct the block encoding of $\frac{1}{K} \sum_{i=1}^K  \frac{1}{P}\rm diag\Big( \prod_{j=1}^{i-1} f_i(\xbf)  \bigtriangledown f_i(\xbf) \prod_{j=i+1}^{K} f_i(\xbf) \Big)$, which is $  \frac{1}{KP} \rm diag \Big( \sum_{i=1}^K  \prod_{j=1}^{i-1} f_j(\xbf) \bigtriangledown f_i(\xbf)  \prod_{j=i+1}^K f_j(\xbf)     \Big) $

which is $\frac{1}{KP} \rm diag\Big( \bigtriangledown f(\xbf)\Big)$ as we point out previously. Given such an operator, by defining $\eta = 1/KP)$, we can proceed as in previous two cases: beginning with some $\rm diag(\xbf_0)$, iteratively constructing $\rm diag(\xbf_0) - \eta \rm diag (\bigtriangledown f(\xbf_0) = \rm diag (\xbf_1)$, then $\rm diag (\xbf_1) - \eta \rm diag(\bigtriangledown f(\xbf_1) = \rm diag (\xbf_2)$, ..., until $ \rm diag (\xbf_T)$. Full description and analysis will be provided in Section \ref{sec: mainalgorithm}. We state the main result in the following:
\begin{theorem}[Quantum Gradient Descent Algorithm (Third Type)]
\label{thm: type3}
    In the same context and assumption as Theorem \ref{theorem: type2}, let $f(x_1,x_2,...,x_n) = \prod_{i=1}^K \Big( a_{i,1} x_1 + a_{i,2}x_2 + \cdots + a_{i,n}x_n\Big)^i$. \\
    
\noindent
\textbf{(First version)} Suppose that for each $i$, there is a unitary $U(a_i)$ of complexity $T(a_i)$ that generates some quantum state that has its first $n$ entries as $( a_{i,1},a_{i,2}, ...., a_{i,n})$. Let $T_a = \max_i \{T(a_i) \}_{i=1}^K$. For each $i$, define $s_i$ as the number of non-zero entries of the vector $(a_{i,1},a_{i,2}, ...., a_{i,n})$, and $S = \max_i \{ s_i \}_{i=1}^K$.  Then there exists an algorithm that outputs a block encoding of a diagonal operator $\textbf{X}_T$ containing $\xbf_T$ as entries. A quantum state $\ket{\xbf_T}$ corresponding to $\xbf_T$ can also be obtained with further $\mathcal{O}(1)$ complexity. The resource cost of such algorithm are: 
   \begin{itemize}
   \item $\mathcal{O}\big(T\log (nK) \big)$ qubits
       \item   a quantum circuit of complexity  $\mathcal{O}\Big(  (SK^2)^{T} \big( T_a +  \log(n)  \big) \Big)$
        \item totally $\mathcal{O}(\sum_{i=1}^K s_{a_i})$ ancilla qubits, a classical preprocessing of complexity $\mathcal{O}\left( \log n \right)$. These costs can be improved to $\mathcal{O}(1)$ if for all $i$, the state $\varpropto \sum_{j=1}^n a_{i,j} \ket{j-1}$ fall into the structured cases in Lemma \ref{lemma: statepreparation}.
   \end{itemize} 

\noindent
\textbf{(Second version)} If, under the same condition for function $f(x_1,x_2,...,x_n)$, except for $f_i(x_1,x_2, ..., x_n)  =\sum_{i=1}^K \big( a_{i,1} x_1 + a_{i,2}x_2 + \cdots + a_{i,n}x_n  + b_i \big)^{2i+1}$ and for each $i$, we have a unitary $U(\sqrt{a_i})$ (of complexity $T(\sqrt{a_i})$ that generates some state that contains $( \sqrt{a_{i,1}}, \sqrt{a_{j,2}}, ..., \sqrt{a_{j,n}} )$ as its first $n$ entries. Let $T_u = \max_i \{ T(\sqrt{a_i})\}_{i=1}^K $. Then there exists an algorithm that outputs a block encoding of a diagonal operator $\textbf{X}_T$ containing $\xbf_T$ as entries. A quantum state $\ket{\xbf_T}$ corresponding to $\xbf_T$ can also be obtained with further $\mathcal{O}(1)$ complexity. The resource cost of such algorithm are: 
   \begin{itemize}
   \item $\mathcal{O}\big(T\log (nK) \big)$ qubits
       \item   a quantum circuit of complexity  $\mathcal{O}\Big(  K^{2T} \big(  \log(n) + T_u)   \Big)$
        \item totally $\mathcal{O}(\sum_{i=1}^K s_{a_i})$ ancilla qubits, a classical preprocessing of complexity $\mathcal{O}\left( \log n \right)$. These costs can be improved to $\mathcal{O}(1)$ if for all $i$, the state $\varpropto \sum_{j=1}^n a_{i,j} \ket{j-1}$ fall into the structured cases in Lemma \ref{lemma: statepreparation}.
   \end{itemize} 
\end{theorem}

\subsection{Discussion}
\label{sec:discussion}
From three main results, Theorem \ref{theorem: type1}, \ref{theorem: type2}, and \ref{thm: type3}, we see that a crucial factor is $T_a$ (respectively, $T_u$) which is the maximum circuit complexity (among $i=1,2,...,K$) to generate the quantum state that contains the vector $(a_{i,1}, a_{i,2}, ..., a_{i,n}) $ (or $(\sqrt{a_{i,1}}, \sqrt{a_{i,2}}, ..., \sqrt{a_{i,n}}) $) in its first entries $n$. In principle, if these coefficients are classically known, then many amplitude encoding techniques have been proposed \cite{grover2000synthesis,grover2002creating,plesch2011quantum, schuld2018supervised, nakaji2022approximate,marin2023quantum,zoufal2019quantum}. The easiest scenario is $\textbf{a}_i$ being sparse (having few non-zero entries) for all $i$. The denser vector leads to more complexity; however, we remark that an efficient procedure still exists if the coefficients obey certain conditions \cite{grover2002creating,grover2000synthesis}. 

One may ask how many iterations $T$ one needs to apply in order for the temporal solution to reach the points of the (local) minimum? In general, it is apparent that the convergence algorithm depends a lot on the behavior of function within the chosen domain. There exist a few results demonstrating convergence guarantees for gradient descent. For example, \cite{nesterov2013introductory} shows that for convex functions with Lipschitz-continuous gradients, one needs $T = \mathcal{O}\big(\frac{1}{\epsilon}\big)$ in order to achieve a vector $\xbf_T$ that is $\epsilon$-closed to the point of extrema. The works in \cite{nesterov1983method, boyd2004convex} analyzed strongly convex function, showing that $T = \mathcal{O}\big( \log \frac{1}{\epsilon}\big)$ suffices. 

There is one subtle detail of our algorithm. The core step of our method takes some diagonally block-encoded operator $\rm diag (\xbf)$ and return a block encoding of $\frac{1}{2} \Big( \rm diag (\xbf) - \eta \rm diag (\bigtriangledown f(\xbf)  \Big)$. The factor $1/2$ is then removed using Lemma \ref{lemma: amp_amp}. The condition required by such a lemma is that the operator $\Big( \rm diag (\xbf) - \eta \rm diag (\bigtriangledown f(\xbf)  \Big) $ needs to have operator norm $\leq 1/2$. More specifically, in the gradient descent algorithm with $T$ steps, if we begin with some block-encoded operator $\rm diag \big( \xbf_0 \big)$, then we need to ensure that the operator $ \rm diag \big( \xbf_0 - \eta \bigtriangledown f(\xbf_0) \big) = \rm diag (\xbf_1)$, and similarly, all subsequent operators $\rm diag(\xbf_2), ... \rm diag (\xbf_T)$, have operator norm less than $1/2$. This issue has also been discussed in related work \cite{nghiem2024simple}. It turns out that there are two conditions: the hyperparameter $\eta $ needs to be less than $\frac{1}{2PT}$, and the initial the vector $\xbf_0$ needs to be less than $\frac{1}{2} - \eta PT$, then all operators $\{ \rm diag (\xbf_t) \}_{t=0}^T $ have norm less than $1/2$. In Appendix \ref{sec: mainalgorithm}, we summary the argument and method provided in \cite{nghiem2024simple}, and point out a simple way to generate the desired initial state $\xbf_0$, or more precisely, the block-encoded operator $\rm diag(\xbf_0)$. Regarding the value of $\eta$, we remind that at a certain step (Eqn.~\ref{eqn: 7}, Eqn.~\ref{eqn: 10}), we obtained the block encoding of $\frac{1}{KP} \rm diag \Big(\bigtriangledown f(\xbf)\Big)$, then we subsequently chose $\eta = 1/KP$ as a hyperparameter for gradient descent. In order for $\eta \leq \frac{1}{2PT} $, we simply use either Lemma \ref{lemma: scale} or Lemma \ref{lemma: amp_amp} to transform the block-encoded operator $\frac{1}{KP} \bigtriangledown f(\xbf) \longrightarrow \frac{\alpha}{2PT} \bigtriangledown f(\xbf) $ for $\alpha \leq 1$, and proceed the algorithm with the hypterparameter then defined as $\eta = \frac{\alpha}{2PT}$. 

We remark that the mechanism of our framework is relying on an embedding of a vector into a diagonal operator, which is not similar to conventional quantum algorithms, such as \cite{harrow2009quantum,childs2017quantum, wiebe2012quantum}, that relies directly on quantum state. As we have seen, the output of our framework is (more precisely, the block encoding of) some diagonal operator $U$ that encodes the solution of interest $\xbf_T$ on the diagonal. As we point out in Appendix \ref{sec: mainalgorithm}, from such a diagonal operator, by performing $U\ket{\bf 0}\big( \frac{1}{\sqrt{n}} \sum_{i=0}^{n-1}\ket{i}\big)$ and performs measurement and post-select the first register to be $\ket{\bf 0}$, resulting in $\ket{\xbf_T}$, which the quantum state that corresponds to solution $\xbf_T$. 

One may wonder how the classical computer works for this problem. As straightforward as it can be, given any $\xbf = (x_1,x_2,...,x_n)$, the classical computer can compute the gradients by taking each partial derivative, resulting in the $\mathcal{O}( n K )$ time step. The update step can be performed in another $\mathcal{O}(n)$ time (that is, simply subtracting the entry by the entry). Thus, the total classical time required is $\mathcal{O}( n K T)$. So, the out method provides an exponential improvement in the number of variables $n$, while being slower in the total iteration time $T$. As also mentioned in \cite{rebentrost2019quantum}, the best domain to apply the quantum gradient descent is in a (extremely) high-dimensional setting, where only a few iterations are required. Another potential setting, as we indicated in the introduction, is when the considered function exhibits a strongly convex landscape, in which it has been rigorously proven that the total number of iterations $T$ only needs to be logarithmical in the inverse of error tolerance. 

Either with a diagonal operator $\rm diag(\xbf_T)$ or a quantum state $\ket{x_T}$ as the outcome, the motivating question is, can they be useful for practical purposes? In the next section, we point out several directions for which we can potentially leverage the quantum representation of the output for practical purposes, including solving linear systems, performing polynomial fitting and predicting unknown input, building support vector machine and classifying unseen data, assigning data to cluster, training neural network and predicting unknown data, solving ground and excited state energies. 


\section{Applications}
\label{sec: quantumlinearsolver}
The above quantum gradient descent algorithm has a logarithmic running time with respect to the number of variables or dimension. Aside from optimization context, we point out that gradient descent also find application in many directions, including solving linear system, data fitting, training neural networks, etc. Below, we shall discuss these problems in greater detail and explicitly demonstrate how our quantum framework's ability to perform gradient descent can be directly translated into these domains. 

\subsection{Solving Linear System}
\label{sec: linearsystem}
Linear system solver is apparently a vital tool in many areas of science and engineering. For example, Maxwell's equations serve as a fundamental principle to many electrical and optical engineering applications. In order to solve them numerically, one needs to discretize the equations and reduce them to a linear system of potentially large size. The quantum algorithm for solving linear systems (also called the Harrow-Hassidim-Lloyd (HHL) algorithm) was first introduced in \cite{harrow2009quantum} and immediately became one of the most influential quantum algorithms in the literature. In addition to gaining exponential speedup in the output of the solution vector stored as a quantum state, the authors also showed that matrix inversion is \textbf{BQP}-complete, which implies their optimality scaling with respect to many parameters. In a linear equation of size $n \times n$, we need to find $\xbf$ that satisfies $A \xbf = \textbf{b}$, where $A \in \mathbb{C}^{n\times n}$ and $b \in \mathbb{C}^n$. As mentioned in \cite{harrow2009quantum}, without loss of generality, assume that $A$ is Hermitian and $\textbf{b}$ has unit norm. The closed-form solution to the above system is $\xbf = A^{-1} \textbf{b}$, thus one needs to be able to invert $A$ in order to obtain the solution. More details regarding quantum linear solving shall be provided in appendix \ref{sec: detaillinearsystem}. Here we point out that alternative approach to solve linear system, based on gradient descent, exists. In fact, a few works have exploited this approach \cite{kerenidis2020quantum, gribling2021improving}, but they also work within black-box model. To find $\xbf$ that satisfies the above system, we can choose to find $\xbf$ such that it minimizes the strongly convex function $f(\xbf) = \frac{1}{2}|\xbf|^2 + || A\xbf - \textbf{b} ||^2$. where $||.||$ refers to $l_2$ Euclidean norm. A detailed expansion will be provided in Appendix \ref{sec: detaillinearsystem} showing that the above function is a combination of first type (Thm.~\ref{theorem: type1}) and second type (Thm. \ref{theorem: type2}) with $K$ in Thm.~\ref{theorem: type1} and Thm.~\ref{theorem: type2} being roughly $s$--the sparsity of $A$. Thus, our quantum gradient descent algorithm can be applied to solve the linear system. We remark that the function $f(\xbf)$ is strongly convex, so in the gradient descent algorithm, the choice of  $T = \mathcal{O}( \log \frac{1}{\epsilon})$, thus we arrive at the following
\begin{theorem}[Quantum Algorithm For Solving Linear Systems of Equations]
    \label{thm: qasolvinglinear}
    Given linear system $A\xbf = \bbf$ where $A$ is an $s$-sparse, Hermitian matrix of size $n\times n$ with conditional number $\kappa$. Suppose that entries of $A, \bbf$ are classically known. Then there is a quantum algorithm returning $\ket{\xbf} \sim A^{-1} \bbf$ in complexity $\mathcal{O}\Big(  s^2 \frac{1}{\epsilon} \log n   \Big)$
\end{theorem}
This result marks the second example demonstrating that quantum computer can solve linear system efficiently without coherent access, while the first one was provided in \cite{zhang2022quantum}. However, their work assumes that $A$ can be expressed as a linear combination of implementable unitary. Our gradient descent-based approach can work with more general structures. The only condition is that the rows of $A$ can be efficiently prepared, a problem that has been settled in \cite{zhang2022quantum}. 

\subsection{Least-square Data Fitting}
\label{sec: leastsquaredatafitting}
\begin{figure}[H]
    \centering
    \begin{subfigure}[b]{0.48\textwidth}
        \centering
        \begin{tikzpicture}
            \begin{axis}[
                width=\textwidth,
                height=6cm,
                xlabel={$x$},
                ylabel={$y$},
                legend pos=north west,
                legend cell align={left},
                axis x line=bottom,
                axis y line=left
            ]

            \addplot[domain=0:10, dashed, thick, green] {2.5 * x + 1};
            \addlegendentry{True Line}

            \addplot[domain=0:10, thick, red] {2.4 * x + 1.2};
            \addlegendentry{Fitted Line}

            \addplot[only marks, mark=*, color=blue] coordinates {
                (0, 1.5) (0.5, 2.8) (1, 4.5) (1.5, 5.1) (2, 5.8)
                (2.5, 7.3) (3, 7.8) (3.5, 9.5) (4, 10.2) (4.5, 11.5)
                (5, 13.0) (5.5, 13.9) (6, 15.5) (6.5, 17.0) (7, 18.5)
                (7.5, 19.1) (8, 21.2) (8.5, 22.9) (9, 24.1) (9.5, 25.5)
            };
            \addlegendentry{Data Points}
            \end{axis}
        \end{tikzpicture}
        \caption{ Linear function}
        \label{fig:fitlinear}
    \end{subfigure}
    \hfill
    \begin{subfigure}[b]{0.48\textwidth}
        \centering
        \begin{tikzpicture}
            \begin{axis}[
                width=\textwidth,
                height=6cm,
                xlabel={$x$},
                ylabel={$y$},
                legend pos=north west,
                legend cell align={left},
                axis x line=bottom,
                axis y line=left,
            ]

            \addplot[domain=0:5, thick, dashed, green] {3*exp(0.8*x)};
            \addlegendentry{True Function ($y = 3e^{0.8x}$)}

            \addplot[domain=0:5, thick, red] {3.1*exp(0.75*x)};
            \addlegendentry{Fitted Function ($y = 3.1e^{0.75x}$)}

            \addplot[only marks, mark=*, blue] coordinates {
                (0, 3.1) (0.5, 4.5) (1, 6.2) (1.5, 9.0) (2, 12.1)
                (2.5, 16.3) (3, 22.1) (3.5, 29.5) (4, 40.2) (4.5, 54.3)
            };
            \addlegendentry{Data Points}
            \end{axis}
        \end{tikzpicture}
        \caption{ Nonlinear function}
        \label{fig:fitnonlinear}
    \end{subfigure}
    \caption{Examples illustrating least-squares data fitting techniques applied to a linear function and a nonlinear function.}
    \label{fig: side_by_side}
\end{figure}
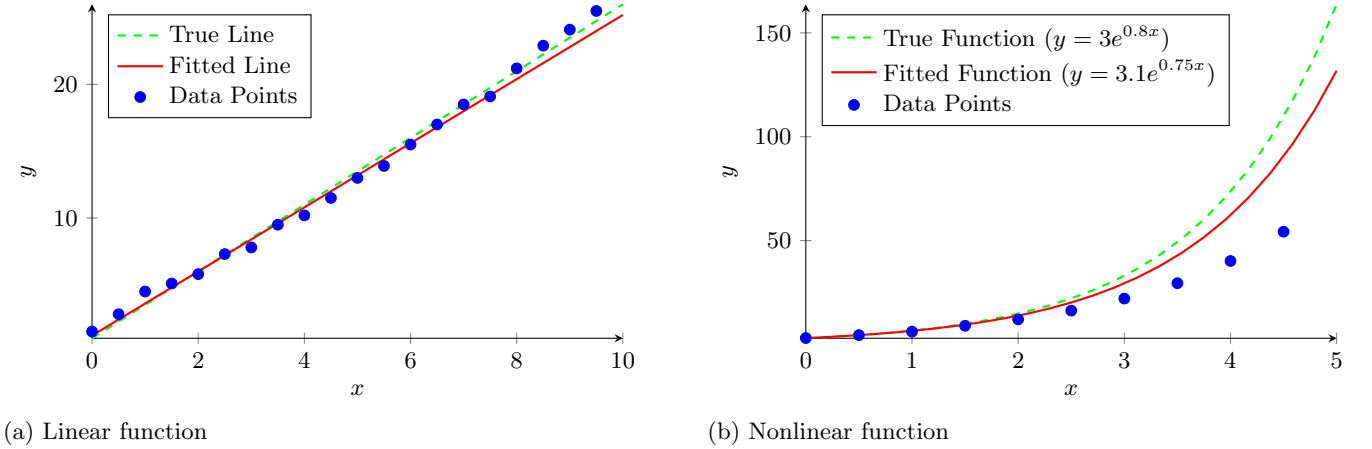

Fitting data is also another key tool in many areas of science and engineering, especially quantitative science. In experimental physics, there are countless sources of data, such as cosmological observation, particle collision, etc. In finance, stock prices, foreign exchanges, risk and credit information, insider transactions, etc., form an enormous database. Data fitting is essentially the process of finding a mathematical model that best fits the data. The ability to obtain an exact modeling yields the ability to make predictions, which is extremely useful. For instance, it allows us to testify the correctness and implications of certain theories of physics, such as relativity theory, and quantum field theory, which plays a pivotal role in our modern understanding of the physical world. We refer to Fig.~\ref{fig: side_by_side} as an illustration of data fitting in two scenarios: fitting a linear and a nonlinear function. 
The formal description of fitting technique, i.e., least-square method, is as follows. We are given a discrete set of $M$ points $\{x_i,y_i\}$ where $x_i \in \mathbb{R}^d$ for some $d$. The fit function is of the form $f(x, \lambda) = \sum_{j=1}^n f_j(x) \lambda_j$, where $\lambda = (\lambda_1, \lambda_2, ..., \lambda_n)$ and $f_j(x): \mathbb{R}^d \longrightarrow \mathbb{R}$ is some continuous function, which can be nonlinear in $x$. The optimal fit can be found by minimizing the so-called cost:
\begin{align}
    E &= \sum_{i=1}^M | f(x_i, \lambda) - y_i |^2 = | \textbf{F} \lambda - \textbf{y} |^2 
    \label{62}
\end{align}
where $\textbf{y} = (y_1, y_2, ..., y_M)$ and the matrix $\textbf{F} \in \Rbb^{M \times n}$ is defined as $\textbf{F}_{ij} = f_j(x_i)$. The above cost function shares similarity with the cost defined earlier in linear solving context. As explicitly shown in Appendix \ref{sec: detaildatafitting}, it is a combination of the first and second types of function (with $K$ roughly as $s$ -- the sparsity of \textbf{F}), so the quantum gradient descent can be applied in a straightforward manner. We remark that the above cost function can be strongly convex if the minimum eigenvalue of $\textbf{F}^T \textbf{F}$ is greater than 0, or equivalently $\textbf{F}^T \textbf{F} $ being positive definite. . We summarize the result of this section in the following
\begin{theorem}[Quantum Algorithm For Data Fitting]
    \label{thm: qadatafitting}
   Let the dataset be $\{x_i,y_i\}_{i=1}^M$. There is a quantum algorithm returning $\ket{\lambda} \sim \lambda $ that encodes the fit parameters to the function $ f(x, \lambda) = \sum_{j=1}^n f_j(x) \lambda_j$, in complexity $ \mathcal{O}\Big(  s^2 \frac{1}{\epsilon} \log n \Big)$ if $\textbf{F}^T \textbf{F} $ is positive definite. 
\end{theorem}

\noindent
\textbf{Fitting Univariate Function.} To exemplify the applicability of our method, suppose that we want to fit a single variable polynomial of degree $n$, $f(x,\lambda) = \lambda_1 x + \lambda_2 x^2 + ... + \lambda_n x^n$, and that we want to predict an unknown input $x$ (we abuse notation). Without loss of generality, assume that $n$ is some power of 2. In the appendix \ref{sec: detaildatafitting}, we show that there exists a quantum circuit of dept $\mathcal{O}(1)$ that generates the state $C \sum_{i=1}^n (\sqrt{x})^i \ket{i-1}$  where 
$$  C = \Big(  \ket{0} + z^{n/2} \ket{1}\Big) \otimes \Big(  \ket{0} + z^{n/4}\ket{1} \Big)\otimes  \Big(  \ket{0}  + z^{n/8} \ket{1} \Big) \otimes \cdots  \otimes \Big( z\ket{0} + z^2 \ket{1}  \Big)$$ 
is the normalization factor. The above result enables the second version in the context of Theorem \ref{theorem: type2},, which is more efficient.  \\

\noindent
\textbf{Predicting Unknown Input.} Suppose we have another unknown input $x$ (we abuse the notation) and we want to predict the value, then the challenge is obtaining an estimation of $f(y,\lambda)$ from the output of our quantum gradient descent algorithm, which is a block encoding of operator $\rm diag(\lambda_1,\lambda_2,...,\lambda_n)$ that contains the fit parameters. It turns out that the procedure is simple. Let $U_\lambda$ denotes the unitary block encoding of $\rm diag(\lambda_1,\lambda_2,...,\lambda_n)$. Let $U_x$ denotes the  unitary that generates the state $\ket{\Phi} = C \sum_{i=0}^{n-1} (\sqrt{x})^{i+1} \ket{i}$ that we have provided above. Then according to definition of block encoding Def.\ref{def: blockencode}, we have that $ U_{\lambda} \ket{\bf 0} \ket{\Phi} =  \ket{\bf 0} \Big( C \sum_{i=0}^{n-1} \lambda_{i+1}  (\sqrt{x})^{i+1} \ket{i}  \Big)  + \ket{\rm Garbage}$. Then we create another state $\ket{\bf 0} \ket{\Phi}$, then use either Hadamard test or SWAP test to estimate the overlaps $ \bra{\bf 0}\bra{\Phi} U_{\lambda} \ket{\bf 0} \ket{\Phi} = C^2 \sum_{i=0}^{n-1} \lambda_{i+1} x^{i+1} $, which reveals the estimation of $f(x,\lambda)$, as it is equal to $ C^2 \sum_{i=0}^{n-1} \lambda_{i+1} x^{i+1} $. Thus, the above procedure provides an end-to-end application of quantum computer, from performing the least-square fitting to predicting output.

\subsection{Support Vector Machine }
\label{sec: buildingsupportvectormachine}
\begin{figure}[H]
    \centering
    \begin{subfigure}[b]{0.45\textwidth}
    \centering
    \begin{tikzpicture}[scale = 1.5]
    \draw[thick, dashed] (0, -1.5) -- (0, 1.5) node[pos=1.0, above ] {Decision boundary};

    \draw[thick, dashed] (-0.8, -1.5) -- (-0.8, 1.5) node[pos=0.9, above left] {Margin};
    \draw[thick, dashed] (0.8, -1.5) -- (0.8, 1.5) node[pos=0.9, above right] {Margin};

    \draw[thick] (0, -1.5) -- (0, 1.5);

    \fill[red] (-1.2, 1.0) circle (2pt);
    \fill[red] (-1.1, 0.5) circle (2pt);
    \fill[red] (-1.3, -0.5) circle (2pt);
    \fill[red] (-0.8, -0.8) circle (2pt); 

    \fill[blue] (1.2, 0.7) circle (2pt);
    \fill[blue] (1.0, 0.0) circle (2pt);
    \fill[blue] (1.3, -0.5) circle (2pt);
    \fill[blue] (0.8, -0.8) circle (2pt); 

    \draw[red, thick] (-0.8, -0.8) circle (3pt);
    \draw[blue, thick] (0.8, -0.8) circle (3pt);

    \node[red] at (-1.5, 1.2) {Class 1};
    \node[blue] at (1.5, 1.2) {Class 2};

    \end{tikzpicture}
    \caption{Linear SVM}
    \label{fig: svm1}
    \end{subfigure} 
    \hfill
    \begin{subfigure}[b]{0.45\textwidth}
    \centering
        \begin{tikzpicture}[scale = 1.5]
    \draw[thick, dashed] (0, 0) circle (1.2);
    \node at (0, 1.35) {Decision boundary};

    \draw[thick, dotted] (0, 0) circle (1.0);
    \draw[thick, dotted] (0, 0) circle (1.4);
    \node at (0, 1.05) {Margin};

    \fill[blue] (0.2, 0.6) circle (2pt);
    \fill[blue] (-0.4, 0.3) circle (2pt);
    \fill[blue] (0.3, -0.4) circle (2pt);
    \fill[blue] (-0.5, -0.5) circle (2pt);

    \fill[red] (1.2, 0.0) circle (2pt);
    \fill[red] (-1.2, -0.2) circle (2pt);
    \fill[red] (0.0, -1.2) circle (2pt);
    \fill[red] (1.0, 1.0) circle (2pt);

    \draw[blue, thick] (0.3, -0.4) circle (3pt); 
    \draw[red, thick] (1.2, 0.0) circle (3pt);   

    \node[blue] at (-0.6, 0.8) {Class 1};
    \node[red] at (1.4, -0.4) {Class 2};

    \fill[black] (0, 0) circle (1pt);
    \node at (-0.2, -0.2) {Center};

\end{tikzpicture}
\caption{Nonlinear SVM}
\label{fig: smv2}
    \end{subfigure}
    \caption{Illustration of support vector machine applied to two different cases: linearly seperable and non-linearly seperable.  }
    \label{fig: svm}
\end{figure}
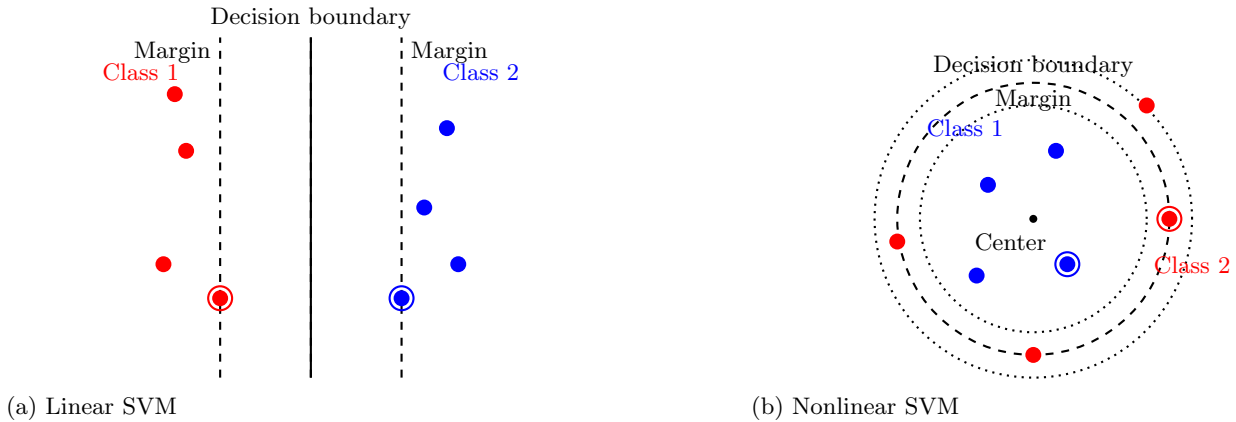

Support vector machine (SVM) is one of the most popular supervised learning algorithms, which aims to recognize hidden patterns within known data sets and predict the outcomes of unseen instances. The most popular problem arising from reality is classification; e.g., one wants to distinguish between images of different categories. More formally, given some dataset, the support vector machine works by translating features of these objects to a so-called feature vector space and finding a hyperplane that separates objects from different classes. SVM can achieve such a goal of drawing the hyperplane by maximizing the margins between two classes, while minimizing the classification errors. We refer to the examples in Figure \ref{fig: svm} for a better illustration of the principle of the support vector machine. 

We mention that the first quantum algorithm for support vector machine was introduced in \cite{rebentrost2014quantum}. Their technique is a combination of density state exponentiation \cite{lloyd2014quantum} and matrix inversion \cite{harrow2009quantum}. Thus, it requires quantum access to classical data. As we will see shortly, our work circumvent this input issue. To relate the SVM algorithm with gradient descent, we denote $\{ \xbf(s), y(s) \}_{s=1}^M$ as the set of training data, where $\xbf(s) \in \Rbb^n$ is the $n$-dimensional feature vector of the $s$-th instance, and $y(s) \in \{0,1\}$ is the corresponding label. Without loss of generalization, we consider the regime of hypercube, i.e., for all $s=1,2,...,M$, $ |\xbf(s)_i | \leq 1 $ for all $i=1,2,..,n$.  The objective of linear SVM is to find an optimal hyperplane $(\wbf, b) \equiv (w_1,w_2,...,w_n,b)$ that maximizes the margin while minimizing classification errors. The primal form of the support vector machine optimization with hinge loss is:
\begin{align}
    \min_{\wbf, b} \frac{1}{2} |\wbf|^2 + C \sum_{s=1}^M  \max \Big( 0, 1- y(s) \big(\wbf^T \xbf(s) + b \big)  \Big)
\end{align}
As shall be shown in Appendix \ref{sec: detailSMV}, the above function falls into first type \ref{theorem: type1}, and in particular, it is also strongly convex. Thus, the quantum gradient descent algorithm of Thm.~\ref{theorem: type1} can be applied, with the output being a block-encoded operator $\rm diag (\wbf ,b)$ that contains the hyperplane of interest. 
\begin{theorem}[Quantum Algorithm for Support Vector Machine]
    \label{thm: qasupportvectormachine}
    There exists a quantum algorithm that returns either an $\epsilon$-approximated block encoding of $\rm diag (\wbf, b)$ or a quantum state $\ket{\wbf} \sim (\wbf,b)^T$, with complexity $\mathcal{O}\Big( \frac{1}{\epsilon} \log n   \Big)$
\end{theorem}

\noindent
\textbf{Classifying Unseen Data:} Given the execution of the quantum gradient descent algorithm, we have obtained a block encoding of a diagonal operator $\rm diag(\wbf, b)$ (or a quantum state $\ket{\wbf,b} \sim (\wbf, b)^T$) containing weights and bias terms, which capture the hyperplane that separates between two classes. The application afterward is to test the support vector machine, by classifying some unseen data $\xbf$ (we abuse the notation, for simplicity), i.e., determining which class it belongs to. In principle, the classification can be done by computing $ \rm sign \Big(  \wbf^T \xbf + b \Big)$. We discuss three approaches for this problem. \\

$\bullet$ Assume that we have a unitary $U_x$ that generates a state $\xbf = C(x_1,x_2,...,x_n)$ where $C$ is some constant (possibly a normalization factor). Then it is simple to generate a state $\frac{1}{\sqrt{2}}\Big( \ket{0} \xbf + \ket{1} \ket{0}_n \Big) $ which contains $\frac{1}{\sqrt{2}}(\xbf,1)$ as its first $n+1$ entries. Then Lemma \ref{lemma: diag} can be used to construct the block encoding of $ \frac{1}{\sqrt{2}(n+1)} \big( \wbf^T \xbf + b\big) \otimes \Ibb$, denoted as $U_w$. From Definition \ref{def: blockencode} and Eqn.~\ref{eqn: action}, we have that $U_w \ket{\bf 0}\ket{0} = \ket{\bf 0}  \frac{1}{\sqrt{2}(n+1)} \big( \wbf^T \xbf + b\big) \ket{0} + \ket{\rm Garbage}$. Using amplitude estimation method \cite{manzano2023real, rall2023amplitude, brassard1997quantum}, one can estimate $ \frac{1}{\sqrt{2}(n+1)} \big( \wbf^T \xbf + b\big)$ and thus infer the sign.  \\

$\bullet$ If instead, we have another unitary that generates $ C(\sqrt{x_1},\sqrt{x_2},..., \sqrt{x_n}) $, then the second version of Lemma \ref{lemma: diag} can be used. The same procedure as above can be carried out to estimate $\big( \wbf^T \xbf + b\big) $ and inferring the sign.  \\

$\bullet$ Use the quantum state $\ket{\wbf,b}$ to estimate the inner product $ (\xbf,1) \ket{\wbf,b}$ by Hadamard or SWAP test, and infer the sign. As $\ket{\wbf,b} $ only differs from $(\wbf,b)^T$ by a normalization factor, the sign of $(\xbf,1) \ket{\wbf,b} $ is the same as $(\xbf,1) (\wbf,b)^T =\wbf^T \xbf + b $ . 

\subsection{Supervised Cluster Assignment}
\label{sec: supervisedlearning}
\begin{figure}[H]
    \centering
   \begin{tikzpicture}[scale = 0.6]
\node[draw, circle, fill=blue!70, minimum size=20pt] (c1) at (-3, 2) {$\mathscr{C}_1$};
\node[draw, circle, fill=red!70, minimum size=20pt] (c2) at (3, -2) {$\mathscr{C}_2$};
\node[draw, circle, fill=green!70, minimum size=20pt] (c3) at (8, 3) {$\mathscr{C}_3$};

\node[draw, circle, fill = black, minimum size = 6pt] at (-1,0) {};
\node at (-1,-0.5) [black] {Unseen data};
\foreach \x/\y in {-4/3, -3.5/2.5, -2/1.5, -3/1, -4/1} {
    \node[draw, circle, fill=blue!30, minimum size=6pt, inner sep=0pt] at (\x,\y) {};
    \draw[dashed, blue] (\x,\y) -- (c1);
}

\foreach \x/\y in {2/-1.5, 3.5/-2.5, 3/-3, 4/-1, 2.5/-3.5} {
    \node[draw, circle, fill=red!30, minimum size=6pt, inner sep=0pt] at (\x,\y) {};
    \draw[dashed, red] (\x,\y) -- (c2);
}

\foreach \x/\y in {7.5/4, 8.5/3.5, 8/2.5, 9/3, 7/3} {
    \node[draw, circle, fill=green!30, minimum size=6pt, inner sep=0pt] at (\x,\y) {};
    \draw[dashed, green] (\x,\y) -- (c3);
}

\node[anchor=west] at (-4, -4) {\textbf{Legend:}};
\draw[blue, thick] (-4, -4.5) -- ++(0.5,0) node[right, black]{Cluster 1 (blue)};
\draw[red, thick] (-4, -5) -- ++(0.5,0) node[right, black]{Cluster 2 (red)};
\draw[green, thick] (-4, -5.5) -- ++(0.5,0) node[right, black]{Cluster 3 (green)};
\draw[dashed, black] (-4, -6) -- ++(0.5,0) node[right, black]{Assigned link};

\end{tikzpicture}
    \caption{Illustration of supervised cluster assignment, which is a typical supervised learning method. Data from different classes are colored red (R), blue (B) and green (G). $M$ representative samples $\Vec{r}_i \in R,  \Vec{b}_i \in B$ , $\Vec{g}_i \in G$ (for $i=1,2,...,M$) from each class are provided. The centroid of each class is defined as $\mathscr{C}_1 = \frac{1}{|R|} \sum_i \Vec{r}_i$, $\mathscr{C}_2 = \frac{1}{|B|} \sum_i \Vec{b}_i$, and $\mathscr{C}_3 =\frac{1}{|G|} \sum_k \Vec{g}_k$.   }
    \label{fig: cluster}
\end{figure}
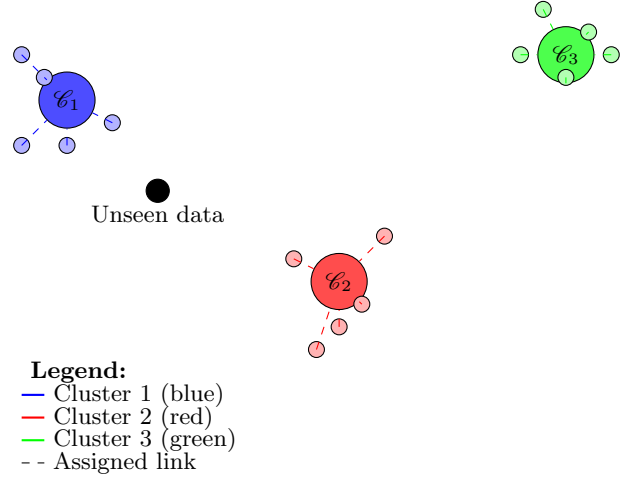

Another popular approach for supervised learning is supervised cluster assignment (SCA). See the figure \ref{fig: cluster} for an illustration. 
Instead of learning the decision boundary like SVM, SCA based on a concept called centroid. An unseen data $\xbf$ can be classified based on the distance from $\xbf$ to $\mathscr{C}_1,\mathscr{C}_2, \mathscr{C}_3$, respectively. Without loss of generalization, assume our feature space is $n$-dimensional, i.e., $\Vec{r}_i, \Vec{b}_i, \Vec{g}_i \in \Rbb^n$ for $i=1,2,..,M$. A quantum algorithm for this task was introduced in \cite{lloyd2013quantum}. Despite achieving logarithmical running time in both $n$ and $M$, their method requires oracle access to entries and norms of all $\Vec{r}_i, \Vec{b}_i, \Vec{g}_i$. It turns out that the problem can be alternatively solved with gradient descent, and hence our method is applicable. 

First, we need to find the centroid $\mathscr{C}_1$ ($\mathscr{C}_2$, $\mathscr{C}_3$ respectively) of dataset $B$ (R,G respectively), and it turns out that the solution to $\min_{\Vec{c}} \frac{1}{M} \sum_{i=1}^M | \Vec{c} - \Vec{r}_i |^2, \ \min_{\Vec{c}} \frac{1}{M}   \sum_{i=1}^M | \Vec{c} - \Vec{b}_i |^2, \ \min_{\Vec{c}} \frac{1}{M}  \sum_{i=1}^M | \Vec{c} - \Vec{g}_i |^2$, respectively, is indeed the centroid $\mathscr{C}_1$, $\mathscr{C}_2$, and $\mathscr{C}_3$. The proof is provided in Appendix \ref{sec: detailSVC}. In the same appendix, we also show that the above cost functions admit the first type \ref{theorem: type1}, hence being solvable by our quantum gradient descent method. 

Recall that the output of our algorithm (Thm.~\ref{theorem: type1}) is a block encoding of operator $\rm diag ( \mathscr{C}_1)$, $\rm diag ( \mathscr{C}_2)$, $\rm diag ( \mathscr{C}_3)$ (or quantum state $\ket{  \mathscr{C}_1} \sim  \mathscr{C}_1^T, \ket{ \mathscr{C}_2}\sim  \mathscr{C}_2, \ket{ \mathscr{C}_3}\sim  \mathscr{C}_3$). In order to classify an unseen data $\xbf$, an intuitive way is to estimate how far it is from data clusters B,R and G. A simple way is to estimate the overlaps $\mathscr{C}_1^T \xbf $, $\mathscr{C}_2^T \xbf $, $\mathscr{C}_3^T \xbf $, which can be done using the same method outlined in \textbf{Classifying Unseen Data} of previous section.

\subsection{Neural Network}
\label{sec: trainingneuralnetwork}
\begin{figure}[H]
    \centering
    \includegraphics[width = 0.5\textwidth]{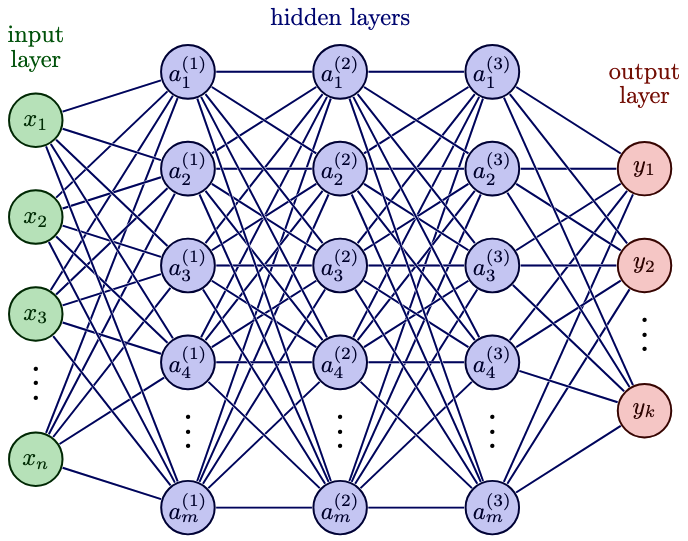}
    \caption{Description of a specific neural network. A layer consists of nodes (colored circles in the above figure). Input layer has $n$ nodes, hidden layer has $3m$ nodes and output layer has $k$ nodes. The input layer contains our dataset vectors, where each $x_i$ in the above figure is the corresponding coordinate of some feature vector $\xbf$. In the context of Algorithm \ref{algo: trainingneuralnetwork}, each $\xbf^i$ would be input layer. The output layer holds the output of such a neural network, which is the result to our problem of interest. Generally, the number of hidden layers and the value of $m$ are user-dependent. The figure above has 3 hidden layers. The number of nodes in output layer depends on purpose, for example, in binary classification problem, 1 node is sufficient to hold the label $\{0,1\}$. All nodes between different layers are connected (as indicated by the line), and there is associated weight between two connected nodes. The goal of training neural network is to obtain those weights that minimizes the loss function, via the backpropagation algorithm, for which we will describe in Algorithm \ref{algo: trainingneuralnetwork}.     }
    \label{fig: neuralnetwork}
\end{figure}

\noindent
Neural network \cite{lecun2015deep} has become an indispensable tool due to its enormous power in capturing a deep structure within a large amount of data and therefore being capable of performing pattern recognition on a complex dataset. The application of neural networks can be seen in almost every context, across various domains, such as self-driving vehicles, image classification, machine translation, fraud detection, analyzing experimental data from high-energy physics, detecting the quantum phase, and so on. More recently, inspired by the development of quantum computer, a model so-called quantum neural network has been proposed and applied widely \cite{schuld2014quest, abbas2021power, beer2020training, nguyen2024theory, farhi2018classification, cong2019quantum, altaisky2001quantum}. As neural network is already extremely popular, introduction and related tutorial for neural network can be found in many contexts, for example, \cite{lecun2015deep}. We briefly review some key concepts in the  Figure \ref{fig: neuralnetwork}, and most importantly, we elaborate how it is related to the gradient descent algorithm in Algorithm \ref{algo: trainingneuralnetwork}. 

\begin{figure}[H]
    \centering
    \includegraphics[width=0.7\textwidth]{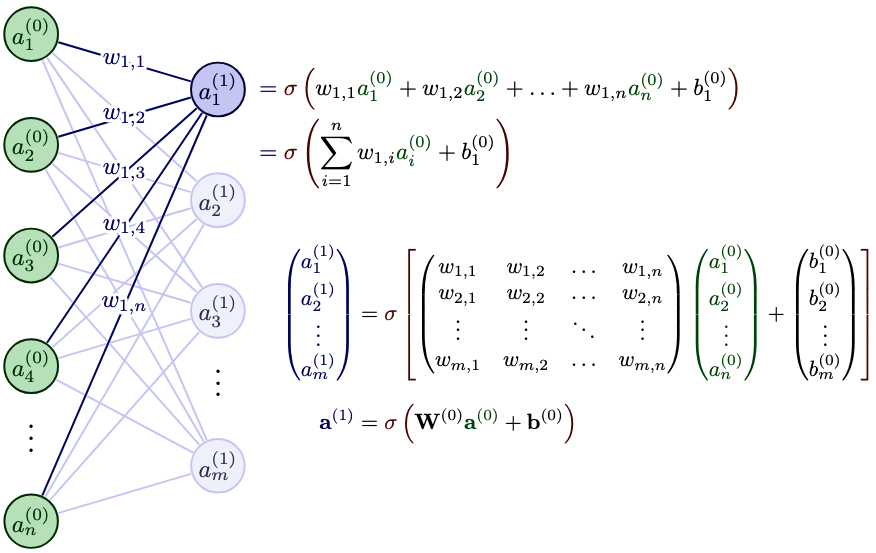}
    \caption{Illustration of feedforward step at each layer. As all the green nodes on the first layer are connected to the first node in the next layer, all information stored within these nodes are undergone a linear transformation, followed by an activation function $\sigma$ at the node $a_1^{(1)}$. The weights $w_{1,1},w_{1,2},...,w_{1,n}$ associated with connection between all green nodes and node $a_1^{(1)}$ are scalars. Thus all weights between two layers form a matrix, denoted as $\textbf{W}^{(0)}$. Essentially, the vector $\textbf{a}^{(0)}$ undergoes a linear transformation $\textbf{W}^{(0)} \textbf{a}^{(0)} + \textbf{b}^{(0)}$, and then an activation $\sigma$ is applied to all entries, resulting in the vector $\textbf{a}^{(1)} \equiv (  a^{(1)}_1, a^{(1)}_2  )^T$. Then this vector will play the same role as $\textbf{a}^{(0)}$, undergoing a linear transformation and an activation function to subsequent layers, i.e., $\textbf{a}^{(1)} \longrightarrow \sigma \big( \textbf{W}^{(1)} \textbf{a}^{(1)} + \textbf{b}^{(1)} \big)$. The feedforward processes until the final layer is reached. We remark that for the activation function, some common choices are Relu, Sigmoid, etc.   }
    \label{fig: activation}
\end{figure}

As described in Figure \ref{fig: activation}, weights between layers form a matrix. Generally speaking, for a neural network having $p$ hidden layers, there should be $p+1$ such matrices between layers. Without loss of generalization, denote them as $\textbf{W}^{(0)}$, $\textbf{W}^{(1)}$, $\textbf{W}^{(2)}$, ..., $\textbf{W}^{(p)}$. At this point, we have enough understanding of neural network and essential recipes. The standard training procedure of a neural network can be summarized as follows. 
\begin{method}[Training Neural Network]
\label{algo: trainingneuralnetwork}
    Suppose that the set of data is given as $\{\xbf(s), y(s)\}_{s=1}^M$ where $\xbf(s) \equiv ( x_1(s),x_2(s),..., x_n(s)  ) \in \mathbb{R}^n$ is the input feature vectors of the data in dimensions $n$, and $y(s)$ is the corresponding label (could be scalar or vector depending on the problem). Let the neural network having one input layer, $p$ hidden layers, each hidden layer having $m$ nodes and 1 output layer. Let $W \equiv \{\textbf{W}^{(j)} \}_{j=0}^p$ denote the weights between layers, and let $b \equiv \{ \bbf^{(j)} \}_{j=0}^p$ be the bias terms. The training algorithm proceeds as follows.
    \begin{itemize}
        \item (Feedforward) Calculate the model's output $\Tilde{y}(s)$ for an input $\xbf(s)$ for all $s=1,2,...,M$: $\Tilde{y}(s) = f(\xbf(s), W, b) $, where $f$ is a series of transformation from layer to layer (as described in Figure \ref{fig: activation}): 
        \begin{align*}
    \sigma \big( \textbf{W}^{(0)} \xbf(s) + \textbf{b}^{(0)}  \big) \longrightarrow \sigma \big(  \textbf{W}^{(1)} (  \sigma \big( \textbf{W}^{(0)} \xbf + \textbf{b}^{(0)}  \big) ) + \textbf{b}^{(1)}  \big) \longrightarrow  \cdots \longrightarrow \sigma \big(  \textbf{W}^{(p)} (  \cdots    )  + \textbf{b}^{(p)}  \big) = \Tilde{y}(s)
\end{align*}
        \item (Loss Computation) Compute the loss function $L$. Some common choices for loss function are: Mean Squared Error (MSE): $L = \frac{1}{M} \sum_{i=1}^M ( \Tilde{y}(i)- y(i))^2$, Cross-Entropy (for binary classification) $ L = -\frac{1}{M} \sum_{i=1}^M y(i) \log( \Tilde{y(i)}) $
        \item (Backpropagation) Compute the gradient of $L$ with respect to entries of weight $W$ and bias $b$, followed up by an update $W \longrightarrow  W- \eta \frac{\partial L}{\partial W}, \ 
            b \longrightarrow b - \eta \frac{\partial L}{\partial b}$
        where in the above, we make an abuse of notation: $\frac{\partial L}{\partial W}  $ actually refers to the partial derivative of loss function with respect to all entries of weight matrices $\{\textbf{W}^{(j)} \}_{j=0}^p $. Same notation abuse holds for $\frac{\partial L}{\partial b} $. 
    \end{itemize}
\end{method}
In order to apply our framework, we need the loss function to have a form similar to that mentioned in Theorem \ref{theorem: type1}, Theorem \ref{theorem: type2} or Theorem \ref{thm: type3}.  From the above description, it appears to be difficult to obtain an exact form of the loss function for a general neural network. To gain some intuition, we begin by considering two particular types of neural network that are widely used: the one with many hidden layers, each layer having few nodes (deep and narrow) and the one with few hidden layers, each layer having many nodes (wide and shallow). The complete calculation can be found in the Appendix \ref{sec: detailneuralnetwork}. In short, we have the following:
\begin{lemma}
\label{lemma: nnproperty}
    In the context of Algorithm \ref{algo: trainingneuralnetwork}, the MSE loss function with activation $\sigma(x) = \rm ReLu(x) = \max (0,x)$  is a combination of the second and third types of function described in Theorem~\ref{theorem: type2} and \ref{thm: type3}. The total number of terms is $K = \mathcal{O}\big( M m^{2^p} \big)$.
\end{lemma}
From the above result, the application of quantum gradient descent algorithm is straightforward. We deduce the following.
\begin{theorem}[Quantum Algorithm for Training Neural Network]
    \label{thm: qatrainingneuralnetwork}
   In the context of Algorithm \ref{algo: trainingneuralnetwork}, choose the activation function $\sigma = \rm ReLu$. Let $T$ be the total of iteration steps. Suppose that there exists an efficient quantum circuit that generates $\big(\sqrt{x_1(s)}, \sqrt{x_2(s)}, ..., \sqrt{x_n(s)}\big)$ for all data point $s=1,2,...,M$. There is a quantum procedure, using $\mathcal{O}\big( T \log nm^{2^p}\big)$ qubits, and a quantum circuit of complexity 
   $$\mathcal{O}\left(  M^Tm^{2^p T} \Big( \log n  +  \log (mn+m^2p) \Big) \right)$$
   that returns a block encoding of a diagonal operator $(\Wbf,\bbf)$ that contains $\Big( \Wbf^{(j)}_{i}  \Big)$
\end{theorem}
Once we finish the Algorithm \ref{thm: qatrainingneuralnetwork}, the output is a unitary block encoding of a diagonal operator that stores the weights $\big(\Wbf^{(0)},\Wbf^{(1)},...,\Wbf^{(p)}  \big)$ and bias $\big( \bbf^{0},\bbf^{1},..., \bbf^{p}\big)$. Given an unknown input $\xbf \equiv (x_1,x_2,...,x_n)$ (we abuse the notation), we are interested in the output $\Tilde{y} = f(\xbf, W,b)$ where $W,b$ are the parameters of neural network after training. The whole procedure is rather lengthy and technical, thus we refer to Appendix.~\ref{sec: detailneuralnetwork}. Here we recapitulate the result as follows.
\begin{theorem}[Quantum Algorithm for Executing Neural Network]
    Given the output of quantum algorithm \ref{thm: qatrainingneuralnetwork}, and another unknown input $\xbf \equiv (x_1,x_2,...,x_n)$, there exists a quantum circuit with complexity $\mathcal{O}\big( m^p \big)$ that output an estimation $\Tilde{y} = f(\xbf, W,b)$ .
\end{theorem}
The above result completes an end-to-end application of the quantum gradient descent algorithm. As neural network has an extremely wide range of applications, for example, classifying data from two categories, such as dog and cat, it is apparent that a quantum computer is useful for these purposes.

\subsection{Finding Ground Energy and Excited State Energy}
\label{sec: solvinggroundstateenergy}
Extracting properties of quantum system, such as ground state or excited state's energy, is one of central applications of quantum computer. Typically, a quantum system has dimension scaling exponentially in the number of its constituents, which imposes an inherent difficulty for classical devices, as storing the necessary information alone is already consuming considerable resources. There have been many proposals in this direction \cite{wang2023quantum, cortes2022quantum, wang2016quantum, lin2022heisenberg, dong2022ground, jones2019variational, nakanishi2019subspace, higgott2019variational}, leveraging various strategies, such as the variational method, the Krylov subspace method, etc, to find the targeted quantities in both the near-term and fault-tolerance regime. The essence of finding the eigenstate/eigenvalue of a quantum system described by some Hamiltonian is an optimization problem; thus, gradient descent is a natural route to the solution. In fact, the principle of variational quantum algorithms, for example \cite{mcclean2016theory}, also relies on the gradient descent method, where the gradient of the target quantity is evaluated on quantum devices and the parameters are updated by adjusting the angles of variational quantum circuit. The difference in our framework is that we do not use a variational circuit, composed of tunable gates, as a means to evaluate the gradient and update the parameters. Rather, as we have seen throughout the work, we directly quantize the gradient descent algorithm, and encode the variables of interest as a block-encoded operator. 

\noindent
\textbf{Ground State Energy.} More formally, let $H$ be the $n \times n$ Hamiltonian of interest, and we want to find $\xbf \equiv (x_1,x_2,...,x_n)$ such that the following quantity is minimized $E = \xbf^T H \xbf$. In general, it might not be possible to reduce the above quantity to any of the function types that we discussed in Thm.~\ref{theorem: type1},~\ref{theorem: type2} and ~\ref{thm: type3}. However, in reality, $H$ typically admit a certain structure, and simpler decomposition can exist. For example, if $ H = \sum_{i=1}^K a_i a_i^T$ where $a_i$ is some $n$-dimensional vector. Given such decomposition of $H$, we have $E = \sum_{i=1}^K \xbf^T a_i a_i^T \xbf = \sum_{i=1}^K (a_i^T \xbf)^2$. Apparently it shares similarity to the function considered in Theorem \ref{theorem: type2}. Now we take a closer look at the aforementioned form $H = \sum_{i=1}^K a_i a_i^T$. In what context does this particular Hamiltonian arise? The answer is many. For example, quantum state projection Hamiltonian, such as AKLT model, the Hamiltonian is of the form $H = \sum_{<i,j>} P_{ij}^{S=2} = \sum_{m=1}^K a_m a_m^T$ where $P_{ij}^{S} $ is the projector onto the spin-2 subspace of two neighboring spin-1 particles, and $a_m$ is the basis vector for the spin-2 subspace. As another example, in tensor network representation (e.g., matrix product states or projected entangled pair states, the Hamiltonian is built from bond vectors $a_i$ that encode local degrees of freedom $ H = \sum_{i=1}^K a_i a_i^T$. In quantum optics, the interaction Hamiltonian between modes in a multi-mode cavity can be expressed similarly, but $a_i$ then is a vector representing the amplitude and phase of the modes. In spin chains, a Hamiltonian projecting onto a specific spin configuration can take the form as above. 

Another particular instance where the quantity $E$ can be simplified further is a simple 1-d lattice system with nearest-neighbor interaction, such as Heisenberg model or tranverse field Ising model, where the Hamiltonian is $H = \sum_{i=1}^{N-1} H_{i,i+1}$ where $N$ is the number of sites. Each $H_{i,i+1}$ features the interacting components, and it can be $H_{i,i+1} = J_i \ \Ibb_{1,2,...,i-1} \otimes  \sigma^z_i \otimes \sigma^z_{i+1} \otimes \Ibb_{i+2,...,N}$ where $\sigma^z$ is the Pauli-z matrix, $J_i$ is some coupling constants and $\Ibb_{1,2,...,i-2}$ denotes the identity matrix acting on the first $i-2$ sites. In Appendix \ref{sec: detailgroundexcitedstate}, we show that given the aforementioned form, the energy $E = \xbf^T H \xbf $ is of the first type, provided in Thm~\ref{theorem: type1}, with $\{x_{s_1 s_2 ... s_N} \}_{s_1,s_2,...,s_N =0,1}$ being main variables. Thus, the quantum gradient descent algorithm can be applied in a straightforward manner. The output of Algorithm \ref{theorem: type1} is a block-encoded diagonal operator $\rm diag (\Phi_{\rm ground})$ that encodes the (approximated) ground state, where $\Phi_{\rm ground}$ denotes the vectorized ground state. The quantum state $\ket{\Phi_{\rm ground}}$ can be obtained with another step, as mentioned in Theorem~\ref{theorem: type1}. The value of ground state energy $E^0$ can be evaluated by measuring observables $H$ on $\ket{\Phi}_{\rm ground}$.  \\

\noindent
\textbf{Excited State Energy.} The information $(E_0, \Phi_{\rm ground}  )$ obtained from previous discussions on the ground state can be useful to find the excited state energy. Instead of finding $\xbf$ that minimizes $E = \xbf^T H \xbf$, we find $\xbf'$ that minimizes the following quantity $E_1 = \xbf'^T \big( H - E_0 \Phi_{\rm ground} \Phi_{\rm ground}^T\big) \xbf'$

It is straightforward to expand $E_1 = \xbf' H \xbf' - E_0 ( \Phi_{\rm ground}^T \xbf' )^2$. The first term $\xbf' H \xbf' $ has a form similar to Eqn.~\ref{120}, which means that it fits into Theorem~\ref{theorem: type1}. The second term, $(\bra{\Phi}_0 \xbf' )^2 $ is apparently fitting with the second type, Thm.~\ref{theorem: type2}. Thus, the function $  E_1 = \xbf' \big( H - E_0 \Phi_{\rm ground} \Phi_{\rm ground}^T\big) \xbf'$ is a combination of first and second type, so we can combine the algorithm of Theorem \ref{theorem: type1} and Theorem \ref{theorem: type2} to construct the gradient and then execute the gradient descent algorithm. The output is a block encoding of diagonalized excited state $\rm \Big( \Phi_{\rm excited} \Big)$, which can be used to produce the quantum state $\ket{\Phi_{\rm excited}}$, then the energy $E_1$ of such excited state can be found by measurement. The procedure for finding the next excited state can be proceeded similarly. We need to find $\xbf''$ such that: $E_2 =  \xbf'' \Big( H - E_0 \Phi_{\rm ground} \Phi_{\rm ground}^T - E_1 \Phi_{\rm excited}  \Phi_{\rm excited}   \Big) \xbf''$ is minimized. The above quantity is again a combination of first and second type, so it can be solved using our proposed framework. 

\subsection{Principal Component Analysis}
\label{sec: principalcomponentanalysis}
Principal component analysis (PCA) is one of the most well-known techniques to analyze large-scale data, the essential idea being to reduce dimensionality while minimizing information loss from a potential high-dimensional data set (see figure below). Quantum principal component analysis algorithm was first introduced in \cite{lloyd2013quantum} and immediately stood out as one of the most influential quantum algorithms for big data analysis. Subsequent developments include \cite{gordon2022covariance, nghiem2025new, li2021resonant, xin2021experimental}. We remark that the original QPCA \cite{lloyd2013quantum} assumes oracle/black-box access to classical data, thus it suffers from dequantization \cite{tang2021quantum}. In the following we point out that QPCA can be executed without oracle access to classical data by reformulating the PCA problem as an optimization problem, which can be solved by our quantum gradient descent algorithm. We note that the Ref.~\cite{gordon2022covariance} and \cite{nghiem2025new} also provided alternative routes to perform PCA on quantum computer without the need of coherent access. 

\begin{center}
\begin{tikzpicture}

    \foreach \x/\y in {-2.3/1.5, -2.5/1.2, -2.7/1.7, -2.2/1.1, -2.4/1.4, -2.8/1.3, -2.6/1.6} {
        \fill[red] (\x,\y) circle (2pt);
    }
    \foreach \x/\y in {2.5/1.5, 2.7/1.3, 2.6/1.7, 2.8/1.4, 2.9/1.6} {
        \fill[teal] (\x,\y) circle (2pt);
    }
    \foreach \x/\y in {-0.5/-1.8, 0.2/-1.6, -0.3/-1.4, 0.1/-1.7, -0.1/-1.5} {
        \fill[blue] (\x,\y) circle (2pt);
    }

    \draw[red, thick] (-2.5,1.4) ellipse (0.6 and 0.4);
    \draw[teal, thick] (2.7,1.5) ellipse (0.7 and 0.5);
    \draw[blue, thick] (-0.1,-1.6) ellipse (0.8 and 0.5);

    \draw[thick,->] (-3,0) -- (3,0) node[right] {PC1};
    \draw[thick,->] (0,-2) -- (0,2) node[above] {PC2};

    \draw[thick, ->, brown] (0,0) -- (1.2,0.8) node[above] {\small Feature 1};
    \draw[thick, ->, brown] (0,0) -- (-1.0,1.2) node[above] {\small Feature 2};
    \draw[thick, ->, brown] (0,0) -- (-1.5,0.5) node[left] {\small Feature 3};
    \draw[thick, ->, brown] (0,0) -- (1.7,-0.3) node[right] {\small Feature 4};
    \draw[thick, ->, brown] (0,0) -- (-1.3,-0.8) node[left] {\small Feature 5};
\end{tikzpicture}
\end{center}

Let the data points be $\mathscr{X}_1,\mathscr{X}_2,...,\mathscr{X}_M$ where each $\mathscr{X}_i \in \Rbb^n$ is a feature vector -- thereby the entries of $\xbf_i$ contain $n$ features of $i$-th data point. Define the mean of the dataset to be $\mu = \frac{ 1}{M}\sum_{i=1}^M \mathscr{X}_i $. A centered dataset is formed by shifting $\mathscr{X}_i \longrightarrow \mathscr{X}_i -\mu$ for all $i$. The covariance between $j$-th feature and $k$-th feature is defined as $Q_{jk}= \frac{1}{M}\sum_{i=1}^M (\mathscr{X}_i-\mu)_j \overline{(\mathscr{X}_i-\mu)_k} $. According to the Ref.~\cite{gordon2022covariance} (Section II.B), $Q_{jk} $ can be simplified as $Q_{jk} = \frac{1}{M} \sum_{i=1}^M (\mathscr{X}_i)_j (\mathscr{X}_i)_k - \mu_i \mu_k $. If we define a matrix of size $M \times n$ $\Sigma$ that contains $\mathscr{X}_i$ as $i$-th row. Then the so-called covariance matrix $Q = \frac{1}{M} \Sigma^T \Sigma - \mu \mu^T$ -- and this matrix is always positive semidefinite. The purpose of PCA is to find the largest eigenvalues and corresponding eigenvectors of $Q$ -- which are called \textit{principal components}. Inspired by the previous section, we aim to find a vector $\xbf$ such that the following quantity $f(\xbf) = \frac{1}{2M} \xbf^T \big( \Ibb_n - Q \big) \xbf$ is minimized. The reason why we add the terms $ \frac{1}{2M}\xbf^T \Ibb_n \xbf = \frac{1}{2M} ||\xbf||^2$ is to guarantee that the function $f(\xbf)$ is strongly convex. Direct expansion yields $f(\xbf) = \frac{1}{2M}||\xbf||^2 - \frac{1}{2M} \xbf^T \Sigma^T \Sigma \xbf  + \frac{1}{2M} \xbf^T \mu \mu^T \xbf  $. Given that $\Sigma$ is a matrix containing $\mathscr{X}_i$ as the $i$-th row, we have that $\frac{1}{2M} \xbf^T \Sigma^T \Sigma \xbf  = \frac{1}{2M} \sum_{i=1}^M \big( \mathscr{X}_i^T \xbf  \big)^2  $ and $\frac{1}{2M} \xbf^T \mu \mu^T \xbf = \frac{1}{2M}\Big( \sum_{i=1}^M \frac{1}{M} \mathscr{X}_i^T \xbf   \Big)^2  $. So we have:
\begin{align}
    f(\xbf) = \frac{1}{2M}||\xbf||^2 -  \frac{1}{2M} \sum_{i=1}^M \big( \mathscr{X}_i^T \xbf  \big)^2   +  \frac{1}{2M}\Big( \sum_{i=1}^M \frac{1}{M} \mathscr{X}_i^T \xbf   \Big)^2
\end{align}
which is the combination of type 1 (Theorem \ref{theorem: type1}) and type 2 (Theorem \ref{theorem: type2}). Thus, provided with a procedure that can efficiently generate $\{\mathscr{X}_i\}_{i=1}^M$, e.g., \cite{zhang2022quantum}, our quantum algorithm can be directly applied. In this case, the total number of terms in $f(\xbf)$ is $M$, the value of $K$ in Thm~\ref{theorem: type1} and Thm~\ref{theorem: type2} is $M$. Additionally, $f(\xbf)$ is strongly convex, so the value of $T$ only needs to be $\mathcal{O} \Big(\log \frac{1}{\epsilon}\Big)$. Hence, there is an exponential speedup in dimension $n$ of data for finding the eigenvector corresponding to the largest eigenvalue of $Q$, namely, the first principal component (PC1). To find the coordinate, say, of data $\mathscr{X}_1$ projected along the PC1, we need to evaluate $\mathscr{X}_1^T \xbf$ where we abuse the use of $\xbf$ to denote the solution to the above optimization problem. Recall that the outcome of quantum gradient descent algorithm in Thm~\ref{theorem: type1}, Thm~\ref{theorem: type2} is a block encoding of $\rm diag (\xbf)$. Then we can use Lemma \ref{lemma: diag} to construct the unitary block encoding of $\mathscr{X}_1^T \xbf \otimes \Ibb_n$, denoted as $U$ for simplicity. Then we apply $U\ket{\bf 0}\ket{0}_n = \ket{\bf 0} \big( \mathscr{X}_1^T \xbf  \big) \ket{0}_n + \ket{\rm Garbage}  $ (according to Definition \ref{def: blockencode}). By using amplitude estimation, we can estimate the value of $ \mathscr{X}_1^T \xbf $ -- which is our projected coordinate. By repeating the same procedure, we can estimate all values $\{ \mathscr{X}_i^T \xbf \}_{i=1}^M$. Hence, we can estimate $ \sum_{i=1}^M \big( \mathscr{X}_i^T \xbf  \big)^2    - \Big( \sum_{i=1}^M \frac{1}{M} \mathscr{X}_i^T \xbf   \Big)^2 $, which is also $ \xbf^T Q \xbf $ -- the largest eigenvalue. The knowledge of such eigenvalue plus $\xbf$--the first principle component, allows us to find the second principal component by proceeding similarly to the strategy provided in the discussion on \textbf{Excited State Energy} from last section.

\section{Outlook and Conclusion}
\label{sec: discussionandconclusion}
We have successfully constructed a new quantum gradient descent algorithm with end-to-end applications. Whereas most state-of-the-art quantum algorithms leverage the quantum state to represent classical data, our proposal takes an alternative route. Instead, we embed the classical information into a diagonal operator. By doing so, we can leverage many recent advancements in quantum algorithms, notably the quantum singular value transformation framework. Even though in this convention, the output of our framework is a diagonal operator that contains the solution to gradient descent, we have shown that we can recover the quantum state corresponding to solution in a simple manner from the diagonal operator representation. Aside from providing a new quantum optimization algorithm, we have affirmatively answered a highly critical question, that is, whether or not a quantum computer can be useful for practical purposes. As we have shown, many practical problems can be redefined as an optimization problem, whereby it can be solved with the gradient descent method. The ability to execute gradient descent algorithm of our proposal thus directly translates into the capability of solving linear system, performing least-square data fitting, building support vector machine, performing supervised cluster assignment, training neural network, solving ground/excited state energies. Not only yielding a quantum state corresponding to solution in these respective problems, we have shown how to leverage such a quantum output to handle extra tasks, such as predicting unseen input (in least-square fitting), classifying unseen data (in support vector machine, cluster assignment, and neural network). The probably most important advancement of our method is the elimination of coherent access to classical data, which was part of many prior quantum algorithms. Oracle, or black-box access turns out to be a big factor that, up to our belief, severely limits the reach of quantum algorithm, as the development of such operation, e.g., quantum random access memory, turns out to be quite early and thus still being incapable for large-scale application. More critically, as we mentioned earlier in the introduction, a series of work \cite{tang2018quantum,tang2019quantum,tang2021quantum, aaronson2015read} revealed the weakness of quantum computer -- it turns out that without strong input assumption, the quantum computer cannot achieve a major speed-up in many problems. 

Although the problem of gradient descent is not totally new, neither is solving a linear equation \cite{harrow2009quantum}, data fitting \cite{wiebe2012quantum}, training neural network \cite{zlokapa2021quantum}, support vector machine \cite{rebentrost2014quantum}, supervised learning \cite{lloyd2013quantum}, principal component analysis \cite{gordon2022covariance, lloyd2013quantum, tang2021quantum}, our work has indeed provided a new perspective towards all of them, bringing them closer to realistic realization and, in turn, also bringing quantum advantage closer to reality. As we indicated in the introduction, due to the so-called input and output issue, it has been suspected whether or not a quantum computer can be useful for practical purposes \cite{aaronson2015read}. The quantum gradient descent algorithm provided in this work with an end-to-end application of our quantum algorithm to concrete practical problems have affirmatively answered such an important question. In particular, as our method does not require any sort of superficial access to classical data, we believe it is not difficult to realize in the near future. Thus, it provides a great motivation for the experimental development of quantum computers. We strongly believe that many computational problems exhibit certain structures that could be translated, either directly or indirectly, into polynomial language, and that it could be handled by using our framework. What other applications could be extended from here is a great source for further exploration of quantum advantages. We conclude our discussion here and leave this open revenue for future investigation.

\section*{Acknowledgement}
The author thanks Trung V. Phan for useful feedback. The author acknowledges support from the Center for Distributed Quantum Processing, Stony Brook University.

\bibliography{ref.bib}{}
\bibliographystyle{unsrt}

\appendix
\clearpage
\newpage

\begin{center}
    \textbf{Supplementary Materials}
\end{center}

\section{Preliminaries}
\label{sec: prelim}
In this appendix we summarize the main recipes of our work, which mostly derived in the seminal work \cite{gilyen2019quantum}. We keep the statements brief and precise for simplicity, with their proofs/ constructions referred to in their original works.

\begin{definition}[Block Encoding Unitary]~\cite{low2017optimal, low2019hamiltonian, gilyen2019quantum}
\label{def: blockencode} 
Let $A$ be some Hermitian matrix of size $N \times N$ whose matrix norm $|A| < 1$. Let a unitary $U$ have the following form:
\begin{align*}
    U = \begin{pmatrix}
       A & \cdot \\
       \cdot & \cdot \\
    \end{pmatrix}.
\end{align*}
Then $U$ is said to be an exact block encoding of matrix $A$. Equivalently, we can write:
\begin{align*}
    U = \ket{ \bf{0}}\bra{ \bf{0}} \otimes A + \cdots
\end{align*}
where $\ket{\bf 0}$ refers to the ancilla system required for the block encoding purpose. In the case where the $U$ has the form 
$$ U  =  \ket{ \bf{0}}\bra{ \bf{0}} \otimes \Tilde{A} + \cdots $$
where $|| \Tilde{A} - A || \leq \epsilon$ (with $||.||$ being the matrix norm), then $U$ is said to be an $\epsilon$-approximated block encoding of $A$.
\end{definition}

The above definition has multiple natural corollaries. First, an arbitrary unitary $U$ block encodes itself. Suppose $A$ is block encoded by some matrix $U$. Next, then $A$ can be block encoded in a larger matrix by simply adding any ancilla (supposed to have dimension $m$), then note that $\Ibb_m \otimes U$ contains $A$ in the top-left corner, which is block encoding of $A$ again by definition. Further, it is almost trivial to block encode identity matrix of any dimension. For instance, we consider $\sigma_z \otimes \Ibb_m$ (for any $m$), which contains $\Ibb_m$ in the top-left corner. We further notice that from the above definition, the action of $U$ on some quantum state $\ket{\bf 0}\ket{\phi}$ is:
\begin{align}
    \label{eqn: action}
    U \ket{\bf 0}\ket{\phi} &=  \ket{\bf 0} A\ket{\phi}  + \sum_{ \textbf{j} \neq \textbf{0}} \ket{\bf j} \ket{\rm Garbage}_j \\
                            &= \ket{\bf 0} A\ket{\phi} + \ket{\rm Garbage},
\end{align}
where $\ket{\rm Garbage } \equiv \sum_{ \textbf{j} \neq \textbf{0}} \ket{\bf j} \ket{\rm Garbage}_j $ is a redundant state that is orthogonal to $\ket{\bf 0} A\ket{\phi}$.

\begin{lemma}[\cite{gilyen2019quantum} Block Encoding of a Density Matrix]
\label{lemma: improveddme}
Let $\rho = \Tr_A \ket{\Phi}\bra{\Phi}$, where $\rho \in \mathbb{H}_B$, $\ket{\Phi} \in  \mathbb{H}_A \otimes \mathbb{H}_B$. Given unitary $U$ that generates $\ket{\Phi}$ from $\ket{\bf 0}_A \otimes \ket{\bf 0}_B$, then there exists a highly efficient procedure that constructs an exact unitary block encoding of $\rho$ using $U$ and $U^\dagger$ a single time, respectively.
\end{lemma}

The proof of the above lemma is given in \cite{gilyen2019quantum} (see their Lemma 45). \\



\begin{lemma}[Block Encoding of Product of Two Matrices]
\label{lemma: product}
    Given the unitary block encoding of two matrices $A_1$ and $A_2$, then there exists an efficient procedure that constructs a unitary block encoding of $A_1 A_2$ using each block encoding of $A_1,A_2$ one time. 
\end{lemma}

\begin{lemma}[\cite{camps2020approximate} Block Encoding of a Tensor Product]
\label{lemma: tensorproduct}
    Given the unitary block encoding $\{U_i\}_{i=1}^m$ of multiple operators $\{M_i\}_{i=1}^m$ (assumed to be exact encoding), then, there is a procedure that produces the unitary block encoding operator of $\bigotimes_{i=1}^m M_i$, which requires parallel single uses of 
    $\{U_i\}_{i=1}^m$ and $\mathcal{O}(1)$ SWAP gates. 
\end{lemma}
The above lemma is a result in \cite{camps2020approximate}. 
\begin{lemma}[\cite{gilyen2019quantum} Block Encoding of a  Matrix]
\label{lemma: As}
    Given oracle access to $s$-sparse matrix $A$ of dimension $n\times n$, then an $\epsilon$-approximated unitary block encoding of $A/s$ can be prepared with gate/time complexity $\mathcal{O}\Big(\log n + \log^{2.5}(\frac{s^2}{\epsilon})\Big).$
\end{lemma}
This is presented in~\cite{gilyen2019quantum} (see their Lemma 48), and one can also find a review of the construction in~\cite{childs2017lecture}. We remark further that the scaling factor $s$ in the above lemma can be reduced by the preamplification method with further complexity $\mathcal{O}({s})$~\cite{gilyen2019quantum}.

\begin{lemma}[\cite{gilyen2019quantum} Linear combination of block-encoded matrices]
    Given unitary block encoding of multiple operators $\{M_i\}_{i=1}^m$. Then, there is a procedure that produces a unitary block encoding operator of \,$\sum_{i=1}^m \pm M_i/m $ in complexity $\mathcal{O}(m)$, e.g., using block encoding of each operator $M_i$ a single time. 
    \label{lemma: sumencoding}
\end{lemma}

\begin{lemma}[Scaling Block encoding] 
\label{lemma: scale}
    Given a block encoding of some matrix $A$ (as in~\ref{def: blockencode}), then the block encoding of $A/p$ where $p > 1$ can be prepared with an extra $\mathcal{O}(1)$ cost.  
\end{lemma}
To show this, we note that the matrix representation of RY rotational gate is
\begin{align}
   R_Y(\theta) = \begin{pmatrix}
        \cos(\theta/2) & -\sin(\theta/2) \\
        \sin(\theta/2) & \cos(\theta/2) 
    \end{pmatrix}.
\end{align}
If we choose $\theta$ such that $\cos(\theta/2) = 1/p$, then Lemma~\ref{lemma: tensorproduct} allows us to construct block encoding of $R_Y(\theta) \otimes \mathbb{I}_{{\rm dim}(A)}$  (${\rm dim}(A)$ refers to dimension of matirx $A$), which contains the diagonal matrix of size ${\rm dim}(A) \times {\rm dim}(A)$ with entries $1/p$. Then Lemma~\ref{lemma: product} can construct block encoding of $(1/p) \ \mathbb{I}_{{\rm dim}(A)} \cdot A = A/p$.  \\

The following is called amplification technique:
\begin{lemma}[\cite{gilyen2019quantum} Theorem 30; \bf Amplification]\label{lemma: amp_amp}
Let $U$, $\Pi$, $\widetilde{\Pi} \in {\rm End}(\mathcal{H}_U)$ be linear operators on $\mathcal{H}_U$ such that $U$ is a unitary, and $\Pi$, $\widetilde{\Pi}$ are orthogonal projectors. 
Let $\gamma>1$ and $\delta,\epsilon \in (0,\frac{1}{2})$. 
Suppose that $\widetilde{\Pi}U\Pi=W \Sigma V^\dagger=\sum_{i}\varsigma_i\ket{w_i}\bra{v_i}$ is a singular value decomposition. 
Then there is an $m= \mathcal{O} \Big(\frac{\gamma}{\delta}
\log \left(\frac{\gamma}{\epsilon} \right)\Big)$ and an efficiently computable $\Phi\in\mathbb{R}^m$ such that
\begin{equation}
\left(\bra{+}\otimes\widetilde{\Pi}_{\leq\frac{1-\delta}{\gamma}}\right)U_\Phi \left(\ket{+}\otimes\Pi_{\leq\frac{1-\delta}{\gamma}}\right)=\sum_{i\colon\varsigma_i\leq \frac{1-\delta}{\gamma} }\tilde{\varsigma}_i\ket{w_i}\bra{v_i} , \text{ where } \Big|\!\Big|\frac{\tilde{\varsigma}_i}{\gamma\varsigma_i}-1 \Big|\!\Big|\leq \epsilon.
\end{equation}
Moreover, $U_\Phi$ can be implemented using a single ancilla qubit with $m$ uses of $U$ and $U^\dagger$, $m$ uses of C$_\Pi$NOT and $m$ uses of C$_{\widetilde{\Pi}}$NOT gates and $m$ single qubit gates.
Here,
\begin{itemize}
\item C$_\Pi$NOT$:=X \otimes \Pi + I \otimes (I - \Pi)$ and a similar definition for C$_{\widetilde{\Pi}}$NOT; see Definition 2 in \cite{gilyen2019quantum},
\item $U_\Phi$: alternating phase modulation sequence; see Definition 15 in \cite{gilyen2019quantum},
\item $\Pi_{\leq \delta}$, $\widetilde{\Pi}_{\leq \delta}$: singular value threshold projectors; see Definition 24 in \cite{gilyen2019quantum}.
\end{itemize}
\end{lemma}
\begin{lemma}\label{lemma: qsvt}[\cite{gilyen2019quantum} Theorem 56 \textbf{Polynomial Transformation}]
\label{lemma: theorem56}  
Suppose that $U$ is an
$(\alpha, a, \epsilon)$-encoding of a Hermitian matrix $A$. (See Definition 43 of~\cite{gilyen2019quantum} for the definition.)
If $P \in \mathbb{R}[x]$ is a degree-$d$ polynomial satisfying that
\begin{itemize}
\item for all $x \in [-1,1]$: $|P(x)| \leq \frac{1}{2}$,
\end{itemize}
then, there is a quantum circuit $\tilde{U}$, which is an $(1,a+2,4d \sqrt{\frac{\epsilon}{\alpha}})$-encoding of $P(A/\alpha)$ and
consists of $d$ applications of $U$ and $U^\dagger$ gates, a single application of controlled-$U$ and $\mathcal{O}((a+1)d)$
other one- and two-qubit gates.
\end{lemma}

\section{Main Algorithm }
\label{sec: mainalgorithm}
We remind that we set the notation that for any $n$-dimensional vector $\xbf = (x_1,x_2,...,x_n)$, we use $\rm diag(\xbf)$ to refer to a diagonal matrix of size $n\times n$ having entries of $\xbf$ as diagonal entries, i.e.,
$$ \rm diag(\xbf) = \begin{pmatrix}
    x_1 & 0 & 0 & 0 \\
    0 & x_2 & 0 & 0 \\
    0 & 0 & \ddots & 0\\
    0 & 0 & 0 & x_n
\end{pmatrix}$$
\noindent
\textbf{First type of function.} The first kind of objective function $f(\xbf): \Rbb^n \rightarrow \Rbb$ of our interest is:
\begin{align}
  f(\xbf) \equiv  f(x_1,x_2,...,x_n) = \sum_{i=1}^K f_i(x_1,x_2, ..., x_n) 
  \label{5}
\end{align}
where $f_i(x_1,x_2, ..., x_n) $ is having the form:
\begin{align}
    f_i(x_1,x_2, ..., x_n)  = a_{i,1} x_1^i + a_{i,2} x_2^i + ... + a_{i,n} x_n^i
\end{align}
$\{ a_{i,j} \}$ are set of known coefficients. It is apparently that the gradient of $f_i$ is:
\begin{align}
    \bigtriangledown f_i(x_1,x_2, ..., x_n)  = \begin{pmatrix}
       a_{i,1} i x_1^{i-1} \\
       a_{i,2} i x_2^{i-1} \\
        \vdots \\
        a_{i,n} i x_n^{i-1}
    \end{pmatrix}
\end{align}
We first remind an essential tool for our subsequent construction:
\begin{lemma}[Theorem 2 in \cite{rattew2023non}; Block Encoding into a Diagonal Matrix]
     Given an $\log(n)$-qubit quantum state specified by a state-preparation-unitary $U$, such that $\ket{\psi}_n=U\ket{0}_n=\sum^{n}_{k=1}\psi_k \ket{k-1}_n$ (with $\psi_k \in \mathbb{C}$), we can prepare an exact block-encoding $U_A$ of the diagonal matrix $A = {\rm diag}(\psi_1, ...,\psi_{n})$ with $\mathcal{O}(\log(n))$ circuit complexity and a total of $\mathcal{O}(1)$ queries to a controlled-$U$ gate  with $\log(n)+3$ ancillary qubits.
\end{lemma}
Suppose that there is a unitary $U$ (of $\mathcal{O}(1)$ complexity) that generates the state:
\begin{align}
    U \ket{0}^{\otimes \log(n) + C } = \sum_{i=1}^n x_{i} \ket{i-1} + \ket{\rm Redundant}
\end{align}
where $\ket{\rm Redundant}$ refers to irrelevant state (being orthogonal to the first term). We remark that the choice of $C$ in the above is general, as long as $U$ creates a state that contains $\xbf$ as its first $n$ entries. The same assumption holds for the generation of the vector that contains $(a_{i,1},a_{i,2},....,a_{i,n})$, as we will need through the remaining. 

Then Lemma \ref{lemma: diagonal} allows us to construct the block encoding $U_X$ of a diagonal operator:
\begin{align}
    \textbf{X} = \sum_{i=1}^n x_{i} \ket{i-1} \bra{i-1} = \begin{pmatrix}
        x_1 & 0 & 0 & 0 \\
        0 & x_2 & 0 & 0 \\
        0 & 0 & \ddots & 0 \\
        0 & 0 & 0 & x_n
    \end{pmatrix} 
\end{align}
where the last notation is referring to vector $\xbf$ being embedded on diagonal line of operator $\textbf{X}$. From then, we can use Lemma \ref{lemma: theorem56} to transform the above block-encoded operator into:
\begin{align}
   \textbf{X} = \begin{pmatrix}
        x_1 & 0 & 0 & 0 \\
        0 & x_2 & 0 & 0 \\
        0 & 0 & \ddots & 0 \\
        0 & 0 & 0 & x_n
    \end{pmatrix} \longrightarrow \textbf{X}^{i-1} =  \begin{pmatrix}
        x_1^{i-1} & 0 & 0 & 0 \\
        0 & x_2^{i-1} & 0 & 0 \\
        0 & 0 & \ddots & 0 \\
        0 & 0 & 0 & x_n^{i-1}
    \end{pmatrix}
    \label{8}
\end{align}
Given that $\{ a_{i,j} \}$ are set of known coefficients, then there exist many proposals on the amplitude encoding technique \cite{grover2000synthesis,grover2002creating,plesch2011quantum, schuld2018supervised, nakaji2022approximate,marin2023quantum,zoufal2019quantum}. Here for convenience, suppose for all $i=1,2,..,K$, there is a unitary $U(a_i)$ (of complexity $T(U(a_i))$) that prepares:
\begin{align}
    U_i \ket{0}^{\otimes \log(n) +C } = \sum_{j=1}^n a_{i,j} \ket{j-1} + \ket{\rm Redundant}
\end{align}
Again using Lemma \ref{lemma: diagonal} we can prepare the block encoding of $ \sum_{j=1}^n a_{i,j} \ket{j-1}\bra{j-1} $, using a quantum circuit of complexity $\mathcal{O}\big( T(U_i) + \log(n) \big) $. Then it can be combined with Lemma \ref{lemma: scale} to construct the block encoding of $\sum_{j=1}^n \big(a_{i,j} i /P\big) \ket{j-1}\bra{j-1} $. Then we can use Lemma \ref{lemma: product} to construct the block encoding of: 
\begin{align}
   \Big( \sum_{j=1}^n \frac{a_{i,j} i}{P} \ket{j-1}\bra{j-1} \Big)  \cdot \textbf{X}^{i-1} = \frac{1}{P}\begin{pmatrix}
       a_{i,1} i  x_1^{i-1} & 0 & 0 & 0 \\
        0 & a_{i,2} i x_2^{i-1} & 0 & 0 \\
        0 & 0 & \ddots & 0 \\
        0 & 0 & 0 & a_{i,n}i x_n^{i-1}
    \end{pmatrix} 
    \label{10}
\end{align}
Repeat the construction for all $i=1,2,...,K$, then using Lemma \ref{lemma: sumencoding}, we can obtain the block encoding of:
\begin{align}
   \frac{1}{K P} \sum_{i=1}^K   \begin{pmatrix}
       a_{i,1} i x_1^{i-1} & 0 & 0 & 0 \\
        0 & a_{i,2} i  x_2^{i-1} & 0 & 0 \\
        0 & 0 & \ddots & 0 \\
        0 & 0 & 0 & a_{i,n} i x_n^{i-1}
    \end{pmatrix} 
\end{align}
Then again, using block encoding of $\textbf{X}$ and Lemma \ref{lemma: sumencoding} we can construct the block encoding of:
\begin{align}
    &= \frac{1}{2} \Big(  \textbf{X} - \frac{1}{K P} \sum_{i=1}^K   \begin{pmatrix}
       a_{i,1} i x_1^{i-1} & 0 & 0 & 0 \\
        0 & a_{i,2} i x_2^{i-1} & 0 & 0 \\
        0 & 0 & \ddots & 0 \\
        0 & 0 & 0 & a_{i,n} i x_n^{i-1}
    \end{pmatrix}   \Big) \\
    &= \frac{1}{2 }\Big( \begin{pmatrix}
        x_1 & 0 & 0 & 0 \\
        0 & x_2 & 0 & 0 \\
        0 & 0 & \ddots & 0 \\
        0 & 0 & 0 & x_n
    \end{pmatrix} - \frac{1}{K P} \sum_{i=1}^K   \begin{pmatrix}
       a_{i,1} i x_1^{i-1} & 0 & 0 & 0 \\
        0 & a_{i,2} i x_2^{i-1} & 0 & 0 \\
        0 & 0 & \ddots & 0 \\
        0 & 0 & 0 & a_{i,n} i x_n^{i-1}
    \end{pmatrix}    \Big)
    \label{13}
\end{align} 
If we choose $1/(KP) =\eta$ to be the hyperparameter for gradient descent, then we can observe that the above operator is:
\begin{align}
    \frac{1}{2} \rm diag \Big(  \xbf - \eta \bigtriangledown f (\xbf) \Big)
\end{align}
The factor 2 can be removed using amplification technique \ref{lemma: amp_amp}. Upon the removal of factor $2$, the above vector is indeed the execution of gradient descent step.  Thus, if we begin at some $\textbf{X}_0 = \rm diag (\xbf_0)$ that contains $\xbf_0$ on the diagonal, then after the first iteration, we have the block encoding of a diagonal operator $\textbf{X}_1 = \rm diag(\xbf_1)$ that contains $\xbf_1 = \xbf_0 - \eta \bigtriangledown f(\xbf_0) $ as diagonal entries. Using $\textbf{X}_1$ again as input and repeat from Equation \ref{8}, we can obtain the block encoding of $\textbf{X}_2 = \rm diag(\xbf_2)$. The process continues until $T$ iteration, we have the block encoding of operator $\textbf{X}_T = \rm diag (\xbf_T) $.\\

To analyze the complexity, we take a look at the $t$-th iteration step. Let $T(\textbf{X}_t)$ denote the circuit complexity of the unitary block encoding of $\textbf{X}_t$. It takes $i-1$ block encoding of $\textbf{X}_t$ to construct the block encoding of $\textbf{X}_t^{i-1} $ (equation \ref{8}), resulting in complexity $\mathcal{O}( i T(\textbf{X}_t)  )$. It takes another quantum circuit of complexity $\mathcal{O}\big(  T(U(a_i)) +\log(n) \big)$ to construct the block encoding of the diagonal operator $\sum_{j=1}^n (a_{i,j} i/P) \ket{j-1}\bra{j-1}$. Thus, it takes totally a quantum circuit of complexity $\mathcal{O}\big( \log(n) + T(U(a_i)) + i T(\textbf{X}_t)  \big)$ to construct the block encoding of $\Big( \sum_{j=1}^n (a_{i,j} i/P) \ket{j-1}\bra{j-1} \Big)  \cdot \textbf{X}_t^{i-1}$ (equation \ref{10}). The next step is to construct the block encoding of :
\begin{align}
   \frac{1}{K P} \sum_{i=1}^K   \Big( \sum_{j=1}^n \frac{a_{i,j}i}{P} \ket{j-1}\bra{j-1} \Big)  \cdot \textbf{X}_t^{i-1}  = \frac{1}{K P} \sum_{i=1}^K   \begin{pmatrix}
       a_{i,1} i x_1^{i-1} & 0 & 0 & 0 \\
        0 & a_{i,2} i  x_2^{i-1} & 0 & 0 \\
        0 & 0 & \ddots & 0 \\
        0 & 0 & 0 & a_{i,n} i x_n^{i-1}
    \end{pmatrix} 
\end{align}
incurring a circuit of complexity $\mathcal{O}\Big(  \sum_{i=1}^K \big( \log(n) + T(U(a_i)) + i T(\textbf{X}_t)   \big)  \Big) = \mathcal{O}\Big( K \log(n) + K T_U + K^2 T(\textbf{X}_t ) \Big)$ where we define $T_U = \max_i \{ T(U(a_i)) \}$. The final step, Eqn.\ref{13} uses one more block encoding of $\textbf{X}_t$, thus the total complexity for obtaining the block encoding of $\textbf{X}_{t+1}$ is $T(\textbf{X}_{t+1}) =  \mathcal{O}\Big( K\log(n) + K^2 T(\textbf{X}_t ) \Big) $. Using similar reasoning, we have that:
\begin{align}
    T(\textbf{X}_t ) = \mathcal{O}\Big( K\log(n) + KT_U + K^2 T(\textbf{X}_{t-1} ) \Big) 
\end{align}
By induction, we arrive at the total complexity of $\mathcal{O}\Big( ( K + K^3 + K^5 + ... + K^{2t-1} ) \big( \log(n)  + T_U\big) + K^{2t} T(\textbf{X}_0)    \Big) = \mathcal{O}\Big( K^{2t} \big( \log(n)  + T_U\big) \Big)$ where we have used the fact that $T(\textbf{X}_0) = \mathcal{O}\big( \log(n) \big) $ (which is a result of Lemma \ref{lemma: diagonal}). Thus for a total of iterations $T$, we arrive at final complexity 
$$\mathcal{O}\Big( K^{2T} \big( \log(n)  + T_U\big) \Big)$$
To comment, the value of $T_U$ is quite critical, where we remind that $T_U = \max_i \{ T(U(a_i))\}$ where each $T(U(a_i))$ is the circuit complexity of generating the state that includes the vector $\textbf{a}_i = (a_{i,1},a_{i,2},..., a_{i,n})$. If $\textbf{a}_i$ is $s$-sparse (for all $i$), then we can expect that $T(U_i) = \mathcal{O}(\log s)$ (see Ref. \cite{prakash2014quantum, schuld2018supervised} for amplitude encoding of a sparse vector).  

\noindent
\textbf{Second type of function.} Here we extend the above discussion into another kind of polynomial:
\begin{align}
    f(\xbf) = \sum_{i=1}^K f_i(x_1,x_2,...,x_n) = \sum_{i=1}^K \big( a_{i,1} x_1 + a_{i,2}x_2 + \cdots + a_{i,n}x_n  + b_i \big)^i
\end{align}
where for each $f_i(x_1,x_2,...,x_n)$, there is at most $s_i$ terms among $\{a_{i,1},a_{i,2},..., a_{i,n} \}$ being non-zero. Suppose further that we know the index of these non-zero entries and denote $S_i$ as the set of these indexes. It is easy to see that, for any $i$, $\sum_{j=1}^n a_{i,j} x_j = \sum_{j \in S_i} a_{i,j} x_j$. We define the following state:
\begin{align}
    \ket{S_i} = \frac{1}{\sqrt{s_i}} \sum_{j \in S_i } \ket{j}
\end{align}
which can be efficiently generated given that we know specifically these index $j \in S_i$. To illustrate the procedure, we exemplify a 3-quits case, which means that we have in total $2^3 = 8$ indexes. Suppose we want to regenerate a superposition:
\begin{align}
    \frac{1}{\sqrt{4}} \Big(  \ket{0} + \ket{2} + \ket{5} + \ket{7} \Big)  
\end{align}
In binary representation, it is equivalent to:
\begin{align}
     \frac{1}{\sqrt{4}} \Big( \ket{000}+ \ket{010} + \ket{101} + \ket{111} \Big)  
\end{align}
First we begin with $\ket{000}$. Apply Hadamard gate to the first qubit:
\begin{align}
    \ket{000} \longrightarrow \frac{1}{\sqrt{2}} \Big( \ket{000} + \ket{100} \Big)
\end{align}
Then apply Hadamard to the second qubit: 
\begin{align}
    \frac{1}{\sqrt{2}} \Big( \ket{000} + \ket{100} \Big) \longrightarrow \frac{1}{4} \Big( \ket{000} + \ket{010} + \ket{100} + \ket{110} \Big)
\end{align}
Then apply controlled-X gate with the first qubit as controlled qubit and the last qubit as applied qubit, then we obtain the desired superposition:
\begin{align}
    \frac{1}{4} \Big( \ket{000} + \ket{010} + \ket{101} + \ket{111} \Big)
\end{align}
In the following, we use $U_{s_i}$ to refer to the quantum circuit that generates the state $\ket{S_i}$ as defined above. 

Back to our main construction, the gradient of the function  $f$ is:
\begin{align}
    \bigtriangledown f(\xbf) = \begin{pmatrix}
        \sum_{i=1}^K i a_{i,1} \big( a_{i,1} x_1 + a_{i,2}x_2 + \cdots + a_{i,n}x_n  \big)^{i-1} \\
        \sum_{i=1}^K i a_{i,2} \big( a_{i,1} x_1 + a_{i,2}x_2 + \cdots + a_{i,n}x_n  \big)^{i-1} \\
        \vdots \\
        \sum_{i=1}^K i a_{i,n} \big( a_{i,1} x_1 + a_{i,2}x_2 + \cdots + a_{i,n}x_n  \big)^{i-1}
    \end{pmatrix}
\end{align}

\noindent
\textbf{Proof of First Version of Lemma \ref{lemma: diag}}. In the following, we provide one solution to the first version of \ref{lemma: diag}. Recall from previously that we have a unitary block encoding $U_X$ of $\Xbf = \rm diag(\xbf)$ and of the operator $\rm diag (\textbf{a}_i)$ (from Eqn. \ref{10}). So it is straightforward to use Lemma \ref{lemma: product} to construct the block encoding of:
\begin{align}
    \begin{pmatrix}
        x_1 & 0 & 0 & 0 \\
        0 & x_2 & 0 & 0 \\
        0 & 0 & \ddots & 0 \\
        0 & 0 & 0 & x_n
    \end{pmatrix} \begin{pmatrix}
       a_{i,1} & 0 & 0 & 0 \\
        0 & a_{i,2}  & 0 & 0 \\
        0 & 0 & \ddots & 0 \\
        0 & 0 & 0 & a_{i,n} 
    \end{pmatrix}  = \begin{pmatrix}
       a_{i,1} x_1 & 0 & 0 & 0 \\
        0 & a_{i,2} x_2  & 0 & 0 \\
        0 & 0 & \ddots & 0 \\
        0 & 0 & 0 & a_{i,n} x_n 
    \end{pmatrix} = \sum_{j=1}^n a_{i,j}x_j \ket{j-1}\bra{j-1}
\end{align}
Let $U$ denotes the unitary block encoding of above operator. From Definition \ref{def: blockencode} and Eqn. \ref{eqn: action}, we have:
\begin{align}
  U \big( \Ibb \otimes U_{s_i} \big) \ket{\bf 0} \ket{0}_n = U \ket{\bf 0} \ket{ S_i  }  &= \ket{\bf 0} \Big( \sum_{j=1}^n a_{i,j}x_j \ket{j-1}\bra{j-1} \Big)\frac{1}{s_i} \sum_{j \in S_i } \ket{j} + \sum_{ \textbf{j} \neq \textbf{0}} \ket{\bf j} \ket{\rm Garbage}_j \\
    &= \ket{\bf 0} \frac{1}{ \sqrt{s_i}} \sum_{j \in S_i} a_{i,j} x_j  \ket{j}  + \sum_{ \textbf{j} \neq \textbf{0}} \ket{\bf j} \ket{\rm Garbage}_j
\end{align}
where we remind that $\ket{0}_n$ refers specifically to the first computational basis state of $n$-dimensional Hilbert space, and $\ket{\bf 0}$ refers to those ancillary qubits used for block encoding purpose (see Def.\ref{def: blockencode}). We define $U_{\phi_1} = U \big( \Ibb \otimes U_{s_i} \big) $ and $U_{\phi_2}  = \Ibb \otimes U_{s_i} $, and also: 
\begin{align}
  U_{\phi_1} \ket{\bf 0} \ket{0}_n =  \ket{\phi_1} &= \ket{\bf 0} \sum_{j \in S_i} a_{i,j} x_j  \ket{j}  + \sum_{ \textbf{j} \neq \textbf{0}} \ket{\bf j} \ket{\rm Garbage}_j \\
  U_{\phi_2} \ket{\bf 0} \ket{0}_n =  \ket{\phi_2} &= \ket{\bf 0} \sum_{j \in S_i} \frac{1}{\sqrt{s_i}} \ket{j}
  \label{c2829}
\end{align}
We point out that the overlaps:
\begin{align}
    \braket{\phi_1,\phi_2} =  \frac{1}{s_i}\sum_{j \in S_i} a_{i,j}x_j = \frac{1}{s_i} \sum_{j=1}^n  a_{i,j}x_j
\end{align}
Now we consider the following Hadamard test routine: 
\begin{align}
    \frac{1}{\sqrt{2}} \Big( \ket{+}\ket{\phi_1} + \ket{-}\ket{\phi_2}  \Big)
\end{align}
which can be generated from $\ket{0}\ket{\bf 0 } \ket{0}_n$ by first applying Hadamard gate to first qubit $\ket{0}$, and then use $U_{\phi_1}, U_{\phi_2}$ controlled by $\ket{0},\ket{1}$ respectively to apply to $\ket{\bf 0}\ket{0}_n $. The above state can be equivalently written as:
\begin{align}
    \frac{1}{2} \Big( \ket{0}\big( \ket{\phi_1}+\ket{\phi_2}  \big) +\ket{1}\big(  \ket{\phi_1} - \ket{\phi_2}  \big)  \Big) 
\end{align}
Now we append another ancillary qubit $\ket{0}_a$ to obtain: 
\begin{align}
    \frac{1}{2} \Big( \ket{0}_a \ket{0}\big( \ket{\phi_1}+\ket{\phi_2}  \big) + \ket{0}_a\ket{1}\big(  \ket{\phi_1} - \ket{\phi_2}  \big)  \Big) 
\end{align}
Apply the controlled-X gate, with $\ket{1}$ as the controlled qubit, to obtain: 
\begin{align}
    \frac{1}{2} \Big( \ket{0}_a \ket{0}\big( \ket{\phi_1}+\ket{\phi_2}  \big) + \ket{1}_a \ket{1}\big(  \ket{\phi_1} - \ket{\phi_2}  \big)  \Big) 
    \label{b54}
\end{align}
We observe that the two states $\ket{0}\big( \ket{\phi_1}+\ket{\phi_2}  $ and $ \ket{1}\big(  \ket{\phi_1} - \ket{\phi_2}$ are orthogonal. So if we trace out the corresponding register, we obtain the density matrix:
\begin{align}
 \rho &=   p_0  \ket{0}_a \bra{0}_a +  p_1 \ket{1}_a \bra{1}_a \\
    &= \begin{pmatrix}
        p_0 & 0 \\
        0 & p_1
    \end{pmatrix}
    \label{rho}
\end{align}
where: 
\begin{align}
    &p_0 =  \frac{1}{4}\Big| \ket{0} \big(\ket{\phi_1} + \ket{\phi_2} \big) \Big|^2 = \frac{1 + \braket{\phi_1,\phi_2}}{2} = \frac{1+  \frac{1}{s_i}\sum_{j=1}^n x_j a_{i,j}}{2} \\
    &p_1= \frac{1 - \frac{1}{s_i}\sum_{j=1}^n x_j a_{i,j}}{2} 
\end{align}
We remark that as $\rho$ is obtained by tracing out a subsystem of a quantum state (Eqn. \ref{b54}) that we have a quantum circuit to prepare, Lemma \ref{lemma: improveddme} allows us to exactly block encode the operator $\rho$. Given that $\rho$ has the form in Eqn. \ref{rho}, we naturally obtain the block encoding of: 
\begin{align}
\begin{pmatrix}
    \frac{1+ \big( \sum_{j=1}^n x_j a_{i,j} \big)/s_i }{2}  & 0 \\
    0 & \frac{1 - \big( \sum_{j=1}^n x_j a_{i,j}\big)/ s_i}{2} 
\end{pmatrix}
\end{align}
It is simple to obtain the block encoding of $\frac{1}{2}\ket{0}\bra{0} + \frac{1}{2} \ket{1}\bra{1}$, for example, apply Hadamard gates $H^{\otimes 2} \ket{00}$ and use Lemma \ref{lemma: improveddme}. Then we can use Lemma \ref{lemma: sumencoding} to construct the block encoding of:
\begin{align}
  \frac{1}{2}\Big( \rho - \big( \frac{1}{2}\ket{0}\bra{0} + \frac{1}{2} \ket{1}\bra{1} \big) \Big) = \frac{1}{4s_i } \begin{pmatrix}
       \sum_{j=1}^n x_j a_{i,j} &  0 \\
       0  & - \sum_{j=1}^n x_j a_{i,j}
   \end{pmatrix}
\end{align}
From the above block encoding, we use Lemma \ref{lemma: tensorproduct} to construct the block encoding of:
\begin{align}
  \frac{1}{4s_i }  \begin{pmatrix}
       \sum_{j=1}^n x_j a_{i,j} &  0 \\
       0  & - \sum_{j=1}^n x_j a_{i,j}
   \end{pmatrix} \otimes \Ibb_n = \frac{1}{4s_i}  \begin{pmatrix}
       \sum_{j=1}^n x_j a_{i,j} \otimes \Ibb_n &  0 \\
       0  & - \sum_{j=1}^n x_j a_{i,j} \otimes \Ibb_n
   \end{pmatrix}
   \label{c41}
\end{align}
Since all coefficients $\{b_i\}_{i=1}^K$ are known, it is simple to construct the block encoding of $ \frac{b_i}{4s_i} \Ibb_n$ for all $i$, by using Lemma \ref{lemma: scale} plus the identity matrix, which is trivial to obtain. This step takes $\mathcal{O}(1)$ complexity. Then we use Lemma \ref{lemma: sumencoding} to construct the block encoding of: 
\begin{align}
    \frac{1}{2} \Big( \frac{b_i}{4s_i} \Ibb_n+  \frac{1}{4s_i}  \sum_{j=1}^n x_j a_{i,j} \otimes \Ibb_n \Big) = \frac{1}{8s_i} \Big( \big( \sum_{j=1}^n x_j a_{i,j} + b_i\big) \otimes \Ibb_n \Big) 
\end{align}
The above operator is again a block encoding of $\frac{1}{8s_i} \sum_{j=1}^n \big( \sum_{j=1}^n x_j a_{i,j} + b_i\big)\otimes \Ibb_n $ which is nothing but a diagonal matrix of size $ n \times n$ with entry $ \frac{1}{8s_i} \big( \sum_{j=1}^n x_j a_{i,j} + b_i\big)$. The factor $8s_i$ can be removed using amplification technique \ref{lemma: amp_amp} with further $\mathcal{O}(s_1)$ complexity. Thus, we have provided a proof of the first version of Lemma \ref{lemma: diag}. \\

\noindent
\textbf{Proof of Second Version of Lemma \ref{lemma: diag}.} We remark that the above procedure works whenever we have a unitary block encoding of $\textbf{X} \equiv \rm diag(x_1,x_2,...,x_n)$ and of $\rm diag (\textbf{a}_i)$. Another approach exists if we have a direct access to a unitary $U(\sqrt{a_i})$ that generates some vector that contains $\mathcal{C}( \sqrt{a_{i,1}}, \sqrt{a_{i,2}},..., \sqrt{a_{i,n}})$ as its first $n$ entries, i.e.:
\begin{align}
    U(a_i) \ket{0}^{\otimes \log(n) +C } = \sum_{j=1}^n \mathcal{C} \sqrt{a_{i,j}} \ket{j-1} + \ket{\rm Redundant}
\end{align}
It means that $U(\sqrt{a_i})$ is a matrix where in the first column, the first $n$ entries are $( \sqrt{a_{i,1}}, \sqrt{a_{i,2}},..., \sqrt{a_{i,n}})$:
\begin{align}
    U(a_i) = \begin{pmatrix}
        \sqrt{a_{i,1}} & \cdots \\
        \sqrt{a_{i,2}} & \cdots \\
        \vdots  & \vdots  \\
        \sqrt{a_{i,n}} & \cdots 
    \end{pmatrix}
\end{align}
where $(\cdots)$ and $(\vdots)$ refers to possibly non-zero entries. Now we use Lemma \ref{lemma: product} to construct the block encoding of:
\begin{align}
    \Xbf U(a_i) = \begin{pmatrix}
        x_1\sqrt{a_{i,1}} & \cdots \\
        x_2\sqrt{a_{i,2}} & \cdots \\
        \vdots  & \vdots  \\
        x_n\sqrt{a_{i,n}} & \cdots 
    \end{pmatrix}
\end{align}
If we take the above unitary and apply to the state $\ket{\bf 0}\ket{0}^n$, then we obtain:
\begin{align}
    \ket{\bf 0} \sum_{j=1}^n \sqrt{a_{i,j}}x_j  \ket{j-1}  +  \sum_{ \textbf{j} \neq \textbf{0}} \ket{\bf j} \ket{\rm Garbage}_j 
\end{align}
where again we remind that $\ket{\bf 0}$ is the extra qubits required for block-encode $\Xbf$ (see Definition \ref{def: blockencode}) and $\ket{\rm Garbage}$ is orthogonal to $\ket{\bf 0} \sum_{j=1}^n \sqrt{a_{i,j}}x_j  \ket{j-1}   $. By redefining:
\begin{align}
    &\ket{\phi_1}= \ket{\bf 0} \sum_{j=1}^n \sqrt{a_{i,j}}x_j  \ket{j-1} + \sum_{ \textbf{j} \neq \textbf{0}} \ket{\bf j} \ket{\rm Garbage}_j \\
    &\ket{\phi_2}= \ket{\bf 0} \sum_{j=1}^n \sqrt{a_{i,j}} \ket{j-1}  +  \ket{\bf 0}\ket{\rm Redundant}
\end{align}
Note that the state $\ket{\rm Redundant}$ is completely orthogonal to $\ket{\bf 0} \sum_{j=1}^n \sqrt{a_{i,j}}x_j  \ket{j-1} $, and $ \sum_{ \textbf{j} \neq \textbf{0}} \ket{\bf j} \ket{\rm Garbage}_j $ is orthogonal to $\ket{\bf 0}\ket{\rm Redundant}$, then the same argument beginning from Eqn. \ref{c2829} holds, allowing us to construct the block encoding of $\braket{\phi_1,\phi_2} = \sum_{j=1}^n x_j a_{i,j}$, and with Lemma \ref{lemma: tensorproduct}, we can obtain the block encoding of $\sum_{j=1}^n x_j a_{i,j} \otimes \Ibb_n $. Given that the block encoding of $b_i \Ibb_n$ is easy to obtain (using Lemma \ref{lemma: scale}), then Lemma \ref{lemma: sumencoding} can be used to construct the block encoding of $ \frac{1}{2}\Big( \sum_{j=1}^n x_j a_{i,j} + b_i \Big) \otimes \Ibb_n $. The factor $1/2$ can be removed using amplification \ref{lemma: amp_amp}. Thus, we obtain the second version of Lemma \ref{lemma: diag}. We remark that, in this case, we have the unitary $U(a_i)$ such that 
\begin{align}
    U(a_i) \ket{0}^{\otimes \log(n) +C } = \sum_{j=1}^n \sqrt{a_{i,j}} \ket{j-1} + \ket{\rm Redundant}
\end{align}
Then we can use Lemma \ref{lemma: diagonal} to construct the block encoding of operator $\rm diag( \sqrt{a_{i,1}}, \sqrt{a_{i,2}}, ...., \sqrt{a_{i,n}})$. Then we can use another such block encoding and construct the block encoding of $ \rm diag( \sqrt{a_{i,1}}, \sqrt{a_{i,2}}, ...., \sqrt{a_{i,n}}) \cdot \rm diag( \sqrt{a_{i,1}}, \sqrt{a_{i,2}}, ...., \sqrt{a_{i,n}}) = \rm diag (a_{i,1},a_{i,2},...,a_{i,n})$. It means that the access to unitary $U(\sqrt{a_i})$ can also lead to alternative approach to the first version of Lemma \ref{lemma: diag}. \\

\noindent
\textbf{Continuation of main algorithm. } Back to our main construction, given that we already obtained the block encoding of $ \big(\sum_{j=1}^n a_{i,j}x_i   \big) \otimes \Ibb_n$, we can combine it with the block encoding of $b_i\Ibb_n$, which is trivial to construct (using Lemma \ref{lemma: scale}), and use Lemma \ref{lemma: sumencoding} to construct the block encoding of $ \frac{1}{2}\big(\sum_{j=1}^n a_{i,j}x_i +b_i \big) \otimes \Ibb_n $. The factor $1/2$ can be removed using Lemma \ref{lemma: amp_amp}. As the next step, we use Lemma \ref{lemma: theorem56} to transform the block encoding of the diagonal operator $ \sum_{j=1}^n x_j a_{i,j} \otimes \Ibb_n $ into a block encoding of the following $n \times n$ matrix:
\begin{align}
    \begin{pmatrix}
       \big( \sum_{j=1}^n x_j a_{i,j}+b_i  \big)^{i-1}  & 0& 0 & 0\\
        0 & \big( \sum_{j=1}^n x_j a_{i,j}+b_i  \big)^{i-1}  & 0 & 0 \\
        0 & 0 & \ddots & 0 \\
        0 & 0  & 0 & \big( \sum_{j=1}^n x_j a_{i,j} +b_i  \big)^{i-1}
    \end{pmatrix} = \big( \sum_{j=1}^n x_j a_{i,j} +b_i \big)^{i-1} \otimes \Ibb_n
\end{align}
Recall that From Eqn. \ref{10}, we have the block encoding of 
\begin{align}
  \rm diag ( a_{i,1}, a_{i,2}, ...., a_{i,n} ) =  \begin{pmatrix}
       a_{i,1}  & 0 & 0 & 0 \\
        0 & a_{i,2}  & 0 & 0 \\
        0 & 0 & \ddots & 0 \\
        0 & 0 & 0 & a_{i,n}
    \end{pmatrix} 
\end{align}
Then Lemma \ref{lemma: product} allows us to use this block encoding with the block encoding of $\big( \sum_{j=1}^n x_j a_{i,j}+b_i  \big)^{i-1} \otimes \Ibb_n$ to construct the block encoding of:
\begin{align}
\begin{pmatrix}
     a_{i,1} \big( \sum_{j=1}^n x_j a_{i,j}+b_i  \big)^{i-1} & 0 & 0 & 0 \\
    0 &  a_{i,2} \big( \sum_{j=1}^n x_j a_{i,j}+b_i  \big)^{i-1}  & 0 & 0 \\
     0 & 0& \ddots & 0 \\
     0 & 0  & 0  &  a_{i,n} \big( \sum_{j=1}^n x_j a_{i,j} +b_i \big)^{i-1} 
\end{pmatrix}
\end{align}
Then we use Lemma \ref{lemma: scale} to multiply the above operator to a factor $i/P$, i.e., we then obtain the block encoding of: 
\begin{align}
\label{a51}
\begin{pmatrix}
     \frac{ i a_{i,1}}{ P } \big( \sum_{j=1}^n x_j a_{i,j} +b_i \big)^{i-1} & 0 & 0 & 0 \\
    0 &  \frac{ i a_{i,2}}{ P}  \big( \sum_{j=1}^n x_j a_{i,j} +b_i \big)^{i-1} & 0 & 0 \\
     0 & 0& \ddots & 0 \\
     0 & 0  & 0  & \frac{ i a_{i,n}}{ P} \big( \sum_{j=1}^n x_j a_{i,j}+b_i  \big)^{i-1}
\end{pmatrix}
\end{align}
which is essentially $ \frac{1}{P}\rm diag \Big( \frac{\partial f_i}{x_1}, \frac{\partial f_i}{x_2},...,\frac{\partial f_i}{x_n},  \Big) = \frac{1}{P}\rm diag\big( \bigtriangledown f_i(\xbf_) \big)$. Then by using Lemma \ref{lemma: sumencoding}, we can obtain the block encoding of $\frac{1}{K} \sum_{i=1}^K \frac{1}{P}\rm diag\big( \bigtriangledown f_i(\xbf_) \big)$, which is:
\begin{align}
    \sum_{i=1}^K \begin{pmatrix}
     \frac{ i a_{i,1}}{K  P } \big( \sum_{j=1}^n x_j a_{i,j} +b_i \big)^{i-1} & 0 & 0 & 0 \\
    0 &  \frac{ i a_{i,2}}{K P}  \big( \sum_{j=1}^n x_j a_{i,j}+b_i  \big)^{i-1} & 0 & 0 \\
     0 & 0& \ddots & 0 \\
     0 & 0  & 0  & \frac{ i a_{i,n}}{K P} \big( \sum_{j=1}^n x_j a_{i,j}+b_i  \big)^{i-1}
\end{pmatrix}
\label{c46}
\end{align}
Then we use Lemma \ref{lemma: sumencoding}, and another block encoding of $\textbf{X} = \rm diag( \xbf) $ to construct the block encoding of: 
\begin{align}
    & \frac{1}{2} \Big( \begin{pmatrix}
        x_1 & 0 & 0 & 0 \\
        0 & x_2 & 0 & 0 \\
        0 & 0 & \ddots & 0 \\
        0 & 0 & 0 & x_n
    \end{pmatrix} -  \sum_{i=1}^K \begin{pmatrix}
     \frac{ i a_{i,1}}{ K  P } \big( \sum_{j=1}^n x_j a_{i,j} \big)^{i-1} & 0 & 0 & 0 \\
    0 &  \frac{ i a_{i,2}}{K P}  \big( \sum_{j=1}^n x_j a_{i,j} \big)^{i-1} & 0 & 0 \\
     0 & 0& \ddots & 0 \\
     0 & 0  & 0  & \frac{ i a_{i,n}}{K P} \big( \sum_{j=1}^n x_j a_{i,j} \big)^{i-1}
\end{pmatrix}  \Big) \\ &= \frac{1}{2} \rm diag \Big( \xbf - \eta \bigtriangledown f(\xbf) \Big)
\end{align}
where we remark that we choose $\eta = 1/(KP)$ to be hyperparameter in gradient descent, and remind further that the gradient of $f$ is: 
\begin{align}
    \bigtriangledown f(\xbf) = \begin{pmatrix}
        \sum_{i=1}^K i a_{i,1} \big( a_{i,1} x_1 + a_{i,2}x_2 + \cdots + a_{i,n}x_n  \big)^{i-1} \\
        \sum_{i=1}^K i a_{i,2} \big( a_{i,1} x_1 + a_{i,2}x_2 + \cdots + a_{i,n}x_n  \big)^{i-1} \\
        \vdots \\
        \sum_{i=1}^K i a_{i,n} \big( a_{i,1} x_1 + a_{i,2}x_2 + \cdots + a_{i,n}x_n  \big)^{i-1}
    \end{pmatrix}
\end{align}
Thus we have obtained the block encoding of $(1/2) \rm diag \Big(\xbf - \eta \bigtriangledown f(x) \Big)$. The factor $1/2$ can be removed using amplification technique \ref{lemma: amp_amp}. Thus, if we begin with a block encoding of $\Xbf_0 = \rm diag (\xbf_0)$, the above algorithm yields the block encoding of $\Xbf_1 = \rm diag(\xbf_1)$, and from such operator, we repeat the algorithm again to obtain the block encoding of $\Xbf_2 =\rm diag (\xbf_2)$. Continually, for a total of $T$ iteration steps, we obtain the block encoding of $\Xbf_T = \rm diag(\xbf_T)$. 

To analyze the complexity, we consider the $t$-th iteration steps. We remind that we have set: 
\begin{align}
    \textbf{a}_i \equiv ( a_{i,1}, a_{i,2} ,..., a_{i,n}  )\\
    \sum_{j=1}^n a_{i,j} x_j = \textbf{a}_i^T \xbf
\end{align}
It takes only one block encoding of $\Xbf_t = \rm diag(\xbf_t) $ and of $ \rm diag(\textbf{a}_i)$, to construct the block encoding of $ \frac{1}{8s_i}\big(   \textbf{a}_i^T \cdot \xbf_t +b_i \big) \otimes \Ibb_n  $ (see Eqn. \ref{c41}). So the quantum circuit complexity is
\begin{align}
    \mathcal{O}\Big(  T(\Xbf_t) + T(a_i) + \log(n)  \Big)
\end{align}
where we remind that $ \mathcal{O}\Big( T(a_i)+ \log(n)\Big) $ is the quantum circuit complexity required to construct the block encoding of operator $\rm diag (\textbf{a}_i)$. It takes further $\mathcal{O}(s_1)$ usage of such block encoding to remove the factor $8s_1$, so in total it takes $\mathcal{O}(s_1)$ block encoding of $\Xbf_t$ and of $\rm diag(\textbf{a}_i)$ to construct the block encoding of  $\big(\textbf{a}_i^T \cdot \xbf_t +b_i \big) \otimes \Ibb_n $. The next step is to raise such block encoding to the power $i-1$, e.g., $\big(\textbf{a}_i^T \cdot \xbf_t +b_i \big) \otimes \Ibb_n \longrightarrow \big( \textbf{a}_i^T \cdot \xbf_t \big)^{i-1}  \otimes \Ibb_n$, hence incurring $\mathcal{O}\big( s_1 i  \big)$ total number of block encoding of $\Xbf_t$ and of $\rm diag(\textbf{a}_i)$, so the total complexity is $\mathcal{O}\Big( s_i i \big(  T(\Xbf_t) + T(a_i)  + \log(n) \big) \Big)$. The summation step (Eqn. \ref{c46}) use one block encoding of $\big( \textbf{a}_i^T \cdot \xbf_t \big)^{i-1}  \otimes \Ibb_n $ for each $i$, thus incurring a total complexity:
\begin{align}
    T(\Xbf_{t+1}) = \mathcal{O}\Big(   \sum_{i=1}^K s_i i \big(  T(\Xbf_t) + T(a_i) + \log(n) \big)  \Big) = \mathcal{O}\Big( S K^2 \big( T(\Xbf_t) + T_U + \log(n) \big) \Big)
\end{align}
where we have defined $S = \max_i \{s_i \}$ and $T_a = \max_i \{T(a_i) \}$. Similarly, we have $T(\Xbf_t) =  \mathcal{O}\Big(S K^2 T(\Xbf_{t-1}) \Big) $, and by simple induction for a total of $T$ iterations, we have:
\begin{align}
     T(\Xbf_T) = \mathcal{O}\Big( (SK^2)^{T} \big(  T(\Xbf_0) + T_U + \log(n) \big) \Big) = \mathcal{O}\Big( S^T K^{2T} \big( T_a + \log(n) \big) \Big)
\end{align}
where we have used that $T(\Xbf_0) = \mathcal{O}(\log n)$ as a result of Lemma \ref{lemma: diagonal}.\\

If we use the second version of Lemma \ref{lemma: diag}, and repeat the above procedure, then the complexity can be reduced. More concretely, at $t$-th time iteration step, it takes $\mathcal{O}(1)$ block encoding of $\Xbf_t = \rm diag(\xbf_t) $ and unitary $U(\sqrt{a_i})$ (with circuit complexity $T(\sqrt{a_i})$) to construct the block encoding of $ \big(  \textbf{a}_i^T \xbf+b_i  \big) \otimes \Ibb_n$. Thus the circuit complexity is $\mathcal{O}\Big(  T(\Xbf_t) + T(\sqrt{a_i})\Big)$. Then next step is raising $  \big(  \textbf{a}_i^T \xbf+b_i  \big) \otimes \Ibb_n $ to $i$-th power, $  \big(  \textbf{a}_i^T \xbf \big)^i \otimes \Ibb_n$, incurring a circuit complexity $\mathcal{O}\Big( i T(\Xbf_t) + iT(\sqrt{a_i}) \Big) $. Then next step is the summation, Eqn.\ref{c46} and performing the subtraction step (to obtain the block encoding of $\rm diag\Big(  \xbf_t - \eta \bigtriangledown f(\xbf_T) \Big)$, using block encoding of $  \big(  \textbf{a}_i^T \xbf+b_i  \big)^{i} \otimes \Ibb_n$ one time for each $i$, thus the total complexity is:
\begin{align}
    T(\Xbf_{t+1}) &= \mathcal{O}\Big(  \sum_{i=1}^{K} i T(\Xbf_t) + iT(\sqrt{a_i})  \Big) \\
    &=\mathcal{O}\Big(  K^2  \big( T(\Xbf_t)  + T_u \big)  \Big)
\end{align}
where we define $T_u = \max_i T(\sqrt{a_i})$. Using the same induction for a total of $T$ iterations as in previous case, we obtain the final complexity
\begin{align}
    \mathcal{O}\Big(  K^{2T} \big(  T(\Xbf_0) + T_u \big)  \Big)= \mathcal{O}\Big(  K^{2T} \big(  \log(n) + T_u  \big)  \Big)
\end{align}
where again we have used $T(\Xbf_0) = \mathcal{O}(\log n)$, as a result of Lemma \ref{lemma: diagonal}.\\

\noindent
\textbf{Third type of function.} Now we consider function of the form:
\begin{align}
    f(\xbf) &= \prod_{i=1}^K f_i(\xbf) \\
    &= \prod_{i=1}^K \big( a_{i,1}x_1+a_{i,2}x_2+...+a_{i,n} x_n + b_i \big)^i
\end{align}
Recall that the gradient of the above function is:
\begin{align}
    \bigtriangledown f(\xbf) = \begin{pmatrix}
            \frac{\partial f_1(\xbf)}{\partial x_1} \prod_{i=2}^K f_i(\xbf) + f_1(\xbf) \frac{\partial f_2(\xbf)}{\partial x_1}\prod_{i=3}^K f_i(\xbf) + ... + \prod_{i=1}^{K-1} \frac{\partial f_K(\xbf)}{\partial x_1} \\
            \frac{\partial f_1(\xbf)}{\partial x_2} \prod_{i=2}^K f_i(\xbf) + f_1(\xbf) \frac{\partial f_2(\xbf)}{\partial x_2}\prod_{i=3}^K f_i(\xbf) + ... + \prod_{i=1}^{K-1} \frac{\partial f_K(\xbf)}{\partial x_2} \\
            \vdots \\
            \frac{\partial f_1(\xbf)}{\partial x_n} \prod_{i=2}^K f_i(\xbf) + f_1(\xbf) \frac{\partial f_2(\xbf)}{\partial x_n}\prod_{i=3}^K f_i(\xbf) + ... + \prod_{i=1}^{K-1} \frac{\partial f_K(\xbf)}{\partial x_n}
    \end{pmatrix}
\end{align}
Recall from the prior discussion on the second type of function, that we have obtained the block encoding of $\frac{1}{P}\rm diag \Big( \frac{\partial f_1(\xbf)}{\partial x_1}, \frac{\partial f_1(\xbf)}{\partial x_1}, ..., \frac{\partial f_1(\xbf)}{\partial x_n}   \Big)$ (Equation \ref{a51}). Additionally, we also obtained the block encoding of $ \big(\textbf{a}_i^T\xbf + b_i\big) \otimes \Ibb_n$ for all $i$. Then it is straightforward to use Lemma \ref{lemma: product} to first construct the block encoding of $\prod_{i=2}^K \big(\textbf{a}_i^T\xbf + b_i\big) \otimes \Ibb_n =\Big(\prod_i \big(\textbf{a}_i^T\xbf + b_i\big)  \Big) \otimes \Ibb_n $. Next, using the same Lemma to construct the block encoding of $\frac{1}{P}\rm diag \Big( \frac{\partial f_1(\xbf)}{\partial x_1}, \frac{\partial f_1(\xbf)}{\partial x_1}, ..., \frac{\partial f_1(\xbf)}{\partial x_n}   \Big) \cdot \Big( \prod_{i=2}^K \textbf{a}_i^T\xbf + b_i\Big) \otimes \Ibb_n$, which is:
\begin{align}
   \frac{1}{P} \begin{pmatrix}
        \frac{\partial f_1(\xbf)}{\partial x_1} \prod_{i=2}^K f_i(\xbf) & 0 & 0 & 0 \\
        0 & \frac{\partial f_1(\xbf)}{\partial x_2} \prod_{i=2}^K f_i(\xbf) & 0 & 0 \\
        0 & 0 & \ddots & 0 \\
        0 & 0  & 0 & \frac{\partial f_1(\xbf)}{\partial x_n} \prod_{i=2}^K f_i(\xbf)
    \end{pmatrix}
    \label{67}
\end{align}
By repeating the same procedure, we can construct the block encoding of remaining operators within the gradient:
\begin{align}
    \frac{1}{P}\begin{pmatrix}
    f_1(\xbf) \frac{\partial f_2(\xbf)}{\partial x_1}\prod_{i=3}^K f_i(\xbf) & 0 & 0 & 0 \\
    0 & f_1(\xbf) \frac{\partial f_2(\xbf)}{\partial x_2}\prod_{i=3}^K f_i(\xbf)  & 0 & 0 \\
    0 & 0  & \ddots & 0 \\
    0 & 0 &  0 & f_1(\xbf) \frac{\partial f_2(\xbf)}{\partial x_n}\prod_{i=3}^K f_i(\xbf)
    \end{pmatrix} \\
   \frac{1}{P} \begin{pmatrix}
    f_1(\xbf)f_2(\xbf) \frac{\partial f_3(\xbf)}{\partial x_1}\prod_{i=4}^K f_i(\xbf) & 0 & 0 & 0 \\
    0 & f_1(\xbf)f_2(\xbf) \frac{\partial f_3(\xbf)}{\partial x_2}\prod_{i=4}^K f_i(\xbf)  & 0 & 0 \\
    0 & 0  & \ddots & 0 \\
    0 & 0 &  0 & f_1(\xbf)f_2(\xbf) \frac{\partial f_3(\xbf)}{\partial x_n}\prod_{i=4}^K f_i(\xbf)
    \end{pmatrix} \\
    \vdots \\
    \frac{1}{P}\begin{pmatrix}
    \prod_{i=1}^{K-1} f_i(\xbf) \frac{\partial f_K(\xbf)}{\partial x_1} & 0 & 0 & 0 \\
    0 & \prod_{i=1}^{K-1} f_i(\xbf) \frac{\partial f_K(\xbf)}{\partial x_2}  & 0 & 0 \\
    0 & 0  & \ddots & 0 \\
    0 & 0 &  0 & \prod_{i=1}^{K-1}f_i(\xbf) \frac{\partial f_K(\xbf)}{\partial x_n}
    \end{pmatrix} 
    \label{68}
\end{align}
Then we can use Lemma \ref{lemma: sumencoding} to construct the block encoding of their summation, which is exactly $\frac{1}{KP} \rm diag\big( \bigtriangledown f(\xbf) \big) $. Another use of Lemma \ref{lemma: sumencoding} yields the block encoding of $\frac{1}{2}\Big( \Xbf - \frac{1}{KP} \rm diag\big( \bigtriangledown f(\xbf) \big) \Big)$, and Lemma \ref{lemma: amp_amp} can be used to remove the factor $1/2$. So we can begin at some block encoding of $\Xbf_0 = \rm diag (\xbf_0)$, then construct the block encoding of $\Xbf_1 =\rm diag (\xbf_1), \Xbf_2 = \rm diag(\xbf_2),..., \Xbf_T = \rm diag( \xbf_T)$. 

To analyze the complexity, we recall that if we use the first version of Lemma \ref{lemma: diag} to construct the block encoding of $\big(\textbf{a}_i^T \xbf_t +b_i\big) \otimes \Ibb_n$ at $t$-th time step, then it takes $\mathcal{O}(s_i)$ use of block encoding of $\rm diag(\xbf_t)$, and of $\rm diag (\textbf{a}_i)$ to construct the block encoding of $( \textbf{a}_i^T \xbf +b_i ) \otimes \Ibb_n$. So the complexity is $\mathcal{O}\Big(  s_i\big( T(\Xbf_t) + T(a_i) + \log n \big)\Big)$, where we remind that $T(a_i)  + \log n$ is the complexity of obtaining the block ending of $\rm diag (\textbf{a}_i)$. The next step is to raise $( \textbf{a}_i^T \xbf +b_i ) \otimes \Ibb_n \longrightarrow ( \textbf{a}_i^T \xbf +b_i )^{i-1} \otimes \Ibb_n $, so the complexity increases as  $\mathcal{O}\Big( s_i i \big( T(\Xbf_t) +T(a_i)  + \log n \big) \Big)$. The next step is to take one block encoding of $\rm diag (\textbf{a}_i) $, and construct the block encoding of $ \frac{1}{P} \rm diag\Big( \textbf{a}_i \Big) \cdot \rm diag ( \textbf{a}_i^T \xbf +b_i )^{i-1} = \frac{1}{P} \rm diag \big(\bigtriangledown f_i(\xbf_T) \big)$. The complexity at this point is:
\begin{align}
    \mathcal{O}\Big(  s_i i \big( T(\Xbf_t) +T(a_i)  + \log n \big) \Big)
\end{align}
The next step is to construct the block encoding of (those operators in Eqn. \ref{67} and Eqn.\ref{68}: 
\begin{align}
    \rm diag  \frac{1}{P}\Big(  \prod_{j=1}^{i-1} (\textbf{a}_j^T\xbf_t +b_j)  \bigtriangledown f_i(\xbf_t) \prod_{j=i+1}^K (\textbf{a}_j^T\xbf_t +b_j)  \Big)
\end{align}
which takes $i-1$ block encodings of $ (\textbf{a}_j^T\xbf_t +b_j) \otimes \Ibb_n$ for $j=1,2,..,i-1,i+1,...,K$. We remind that the complexity for producing a block encoding of $ (\textbf{a}_j^T\xbf_t +b_j)\otimes \Ibb_n $ for any $j$ is $\mathcal{O}\Big(   s_j\big( T(\Xbf_t) + T(a_i)  + \log n \Big)$ So the total complexity is for producing the block encoding of the above operator is:
\begin{align}
    \mathcal{O}\Big(  s_i i \big( T(\Xbf_t) + T(a_i)  + \log n \big) + \sum_j s_j \big( T(\Xbf_t) + T(a_i)  + \log n \big)  \Big) 
\end{align}
The last step is to construct the block encoding of: 
\begin{align}
    \frac{1}{K} \sum_{i=1}^K  \rm diag  \frac{1}{P}\Big(  \prod_{j=1}^{i-1} (\textbf{a}_j^T\xbf_t +b_j)  \bigtriangledown f_i(\xbf_t) \prod_{j=i+1}^K (\textbf{a}_j^T\xbf_t +b_j)  \Big) = \frac{1}{KP} \rm diag \Big( \bigtriangledown f(\xbf_T)  \Big)
\end{align}
So it inflates the total complexity as:
\begin{align}
    &\mathcal{O}\Big( \sum_{i=1}^K s_i i \big( T(\Xbf_t) +T(a_i)  + \log n \big) + \sum_{j\neq i}^K s_j \big( T(\Xbf_t) +T(a_i) + \log n \big)  \Big)  \\
    &= \mathcal{O}\Big( SK^2 \big( T(\Xbf_t) + T_U + \log n \Big)
\end{align}
where again we remind that we have defined $ S = \max_i \{s_i\}$, $T_a = \max_i \{  T(a_i) \} $. Using similar induction procedure as previous case, we arrive at the final complexity:
\begin{align}
    \mathcal{O}\Big( S^T K^{2T} \Big( T_a + \log n \Big)  \Big)
\end{align}

If instead of $a_i$, we have the unitary $U(\sqrt{a_i})$ that generates the entries $(\sqrt{a_i})$, then the second of Lemma \ref{lemma: diag} is enabled. Then it takes only $\mathcal{O}(1)$ block encoding of $\Xbf_t = \rm diag(\xbf_t)$ and unitary $U(\sqrt{a_i})$ to construct the block encoding of  $( \textbf{a}_i^T \xbf +b_i ) \otimes \Ibb_n$, so the complexity is $\mathcal{O}\Big( T(\Xbf_t) + T(\sqrt{a_i}) \Big) $. Repeat the same deduction and induction as above, we reach the final complexity in this setting: 
\begin{align}
    \mathcal{O}\Big( K^{2T} \big( T_u + \log n \big) \Big)
\end{align}
where $T_u = \max_i \{ T(\sqrt{a_i}  \}$. \\

\noindent
\textbf{Initial condition.} We remark that this issue has been worked out in \cite{nghiem2024simple}, and we directly quote their procedure here (with some change of notation). At $t$-th step, suppose that we have the block encoding of $\rm diag (\xbf_t)$. We need that the operator norm of $\rm diag \big( \xbf_{t} \big) $ less than $1/2$ for all t, which means that $| \xbf_t |_\infty \leq 1/2$ where $|.|_{\infty}$ is the maximum entry (in magnitude) of $\xbf_t$. As $\xbf_t = \xbf_{t-1} - \eta \bigtriangledown f(\xbf_{t-1})$, we have that $| \xbf_t |_{\infty} \leq |\xbf_{t-1}|_\infty + \eta |\bigtriangledown f(\xbf_{t-1})|_{\infty} \leq |\xbf_{t-1}|_\infty + \eta P$. By induction, we have that for a total of $T$ iteration steps $| \xbf_T |_\infty \leq |\xbf_0|_\infty + \eta P T$. We want $|\xbf_0|_\infty \leq 1/2 $, so the first condition we need is: $\eta P T \leq 1/2$, or $\eta \leq \frac{1}{2PT}$. The second condition we need is $| \xbf_T |_\infty \leq 1/2 $ so $ \leq |\xbf_0|_\infty + \eta P T \leq 1/2$, so we need $ |\xbf_0|_\infty \leq 1/2- \eta PT$. To achieve such a condition, first we need $1/2- \eta P T >0$, which is possible by choosing $\eta$ sufficiently small, e.g., $\eta  \leq \frac{1}{2PT}$. Next, we note that we can encode the density matrix $\Ibb_M/M$ for arbitrary dimension $M$. For example, one can prepare the $2\log M$-qbit state $\sum_{i=1}^M \frac{1}{\sqrt{M}} \ket{i-1} \ket{i-1}$ and trace out the second register. The resultant density matrix can be block-encoded by Lemma \ref{lemma: improveddme}. The operator norm $|\Ibb_M/M |$ is $1/M$. By setting:
\begin{align}
    \frac{1}{M} = \frac{1}{2} - \eta PT
\end{align}
and choose $\rm diag(\xbf_0) = \frac{\Ibb}{M}$ then we condition $ |\xbf_0|_\infty \leq 1/2- \eta PT $ is met, because the maximum entry of $\xbf_0$ is also the operator norm of $\rm diag (\xbf_0)$. In fact, we even have that $\rm diag (\xbf_0) = (\frac{1}{2} - \eta PT) \Ibb $.

Now we can analyze the lower bound for $|\xbf_T|$ (note that this is $l_2$ norm, which is different from $|.|_\infty $). As $\xbf_t = \xbf_{t-1} - \eta \bigtriangledown f(\xbf_{t-1})$, we have $|\xbf_t| \geq \Big| |\xbf_{t-1}|  - \eta |\bigtriangledown f(\xbf_{t-1})| \Big| \geq \Big| |\xbf_{t-2} | - \eta |\bigtriangledown f(\xbf_{t-2})| - \eta |\bigtriangledown f(\xbf_{t-1})| \Big|$. For a total of $T$ iterations, we have $|\xbf_T| \geq \Big| |\xbf_0| -  \eta  \sum_{t=0}^{T-1} |\bigtriangledown f(\xbf_t)| \Big|$. We have the following relation: for a $n$-dimensional vector $\xbf$, $|\xbf| \leq \sqrt{n } |\xbf|_{\infty}$. So for all $t$, we have $|\bigtriangledown f(\xbf_t)| \leq \sqrt{n} |\bigtriangledown f(\xbf_t)|_{\infty} = \sqrt{n} P$. We also choose $\rm diag(\xbf_0) =  (\frac{1}{2} - \eta PT) \Ibb$, which implies $|\xbf_0| = \sqrt{n } ( \frac{1}{2}-\eta PT   )$.  Thus we have:
\begin{align}
    |\xbf_T| &\geq \Big| \sqrt{n} (\frac{1}{2}-\eta PT) - \eta \sqrt{n} PT   \Big| \\
    &= \Big|  \frac{1}{2} -  2 \eta PT   \Big| \sqrt{n} \\
    &\geq \Big| -\eta PT  \Big| \sqrt{n} \\
    &= \sqrt{n} \eta PT
\end{align}

\noindent
\textbf{Obtaining Quantum State Corresponding to Solution. } Our gradient descent/quantum linear solving algorithm returns a block encoding of a diagonal operator $\Xbf = \rm diag(\xbf)$ (we abuse notation $\xbf$ to denote the solution $\xbf_T$ after $T$ iterations) that contains the solution:
\begin{align}
     \textbf{X} &= \begin{pmatrix}
           x_1  & 0 & \cdots  & 0 \\
        0 & x_2 & \cdots & 0\\
        0 & 0 & \ddots & 0 \\
        0 & 0 & \cdots & x_n
    \end{pmatrix}
\end{align}
Denote $U(\textbf{X})$ as a unitary block encoding of $\textbf{X}$. The question now is how to obtain $\ket{\xbf}$ from the above unitary $U(\textbf{X})$ ? The answer is simple and comes from the definition \ref{def: blockencode}. More concretely, from equation \ref{eqn: action} and we choose $\ket{\phi} = \frac{1}{\sqrt{n}} \sum_{i=1}^n \ket{i-1}$ then we have that:
\begin{align}
    U(\textbf{X}) \ket{\bf 0} \ket{\phi} &= \ket{\bf 0} \textbf{X} \ket{\phi} + \ket{\rm Garbage} \\
    &= \ket{\bf 0} \big(  \frac{1}{\sqrt{n}} \sum_{i=1}^n x_i \ket{i-1}   \big) + \ket{\rm Garbage}
\end{align}
If we measure the ancilla register and post-select on $\ket{\bf 0}$ then the probability of measuring $\ket{\bf 0}$ is:
\begin{align}
    p_{\bf 0} = \frac{1}{n} \sum_{i=1}^n |x_i|^2
\end{align}
which means that it depends a lot on the value of $|x_i|^2$. We have shown before that $|\xbf| \geq \sqrt{n }\eta PT$, thus $p_{\bf 0} \geq (\eta PT)^2$. If we choose, for example, $\eta = \frac{1}{4PT}$, which is a valid choice because $\eta < \frac{1}{2PT}$, then $p_{\bf 0} \in \mathcal{O}(1)$. Once we obtain the state $\ket{\xbf}$ that corresponds to the solution, we can perform further measurements to obtain any expectation value $\bra{\xbf } M \ket{\xbf}$ (for some Hermitian matrix $M$) associated with $\xbf$.

\section{Review of Prior Quantum Gradient Descent Algorithms}
\label{sec: reviewpriorgradient}
\noindent
\textbf{The work in Ref.~\cite{rebentrost2019quantum}.} The work in \cite{rebentrost2019quantum} considers $f: \Rbb^n \longrightarrow \Rbb$ to be a homonogeous polynomial of even degree $2p$, which admits the following tensor algebraic form:
\begin{align}
    f(\xbf) = \frac{1}{2} \bra{\xbf}^{\otimes p} A \ket{\xbf}^{\otimes p}
    \label{eqn: A}
\end{align}
where $A$ is a matrix of size $n^p \times n^p $ of bounded matrix norm $|A| = \Gamma$, with the following decomposition:
\begin{align}
    A = \sum_{\alpha =1}^K A^{\alpha}_1 \otimes A^{\alpha}_2 \cdots \otimes A^{\alpha}_p,
\end{align}
where $K$ is some constant (not the same as $K$ appeared in our main theorems \ref{theorem: type1},\ref{theorem: type2}, \ref{thm: type3}). Furthermore, they also assume that $A$ is sparse (with sparsity $s$) and that its entries can be accessed through an oracle. Furthermore, they imposed a spherical constraint, where, at any iteration step, the solution is normalized: $\xbf \longrightarrow \ket{\xbf} \equiv \xbf/|\xbf|$. The above tensor structure of the objective function $f$ allows the gradient to be be conveniently written as:
\begin{align}
    \bigtriangledown f(\xbf) &= D(\ket{\xbf}) \ket{\xbf}
\end{align}
where $D(\xbf)$ is an $n \times n$ matrix:
\begin{align}
     D(\xbf) &= \sum_{\alpha=1}^K\sum_{m=1}^p \Big( \prod_{n=1, n\neq m}^p  \bra{\xbf} A_n^{\alpha} \ket{\xbf}\Big) A_m^{\alpha} \\
     &=  \Tr_{1,2,...,p-1} ( \ket{\xbf}\bra{\xbf} ^ {\otimes (p-1)} \otimes \mathbb{I}) M_D
\end{align}
and 
\begin{align}
    M_D = \sum_{\alpha=1}^K\sum_{m=1}^p \Big( \bigotimes_{n=1,n\neq  m}^p A_{n}^{\alpha} \Big) \bigotimes A_m^{\alpha}. 
\end{align}
While the algorithm of \cite{rebentrost2019quantum} is far more detailed and complicated, we capitulate their main idea and flow here. The operator $M_D$ has the following property:
\begin{align}
    M_D = \sum_{j=1}^p Q_j A Q_j
\end{align}
where $Q_j$ is consisted mostly of SWAP operators that swap between two registers. Oracle access to $A$ allows them to use the tool from quantum simulation \cite{berry2007efficient, berry2012black, berry2015hamiltonian} to simulate $\exp(-i A t)$ for some $t>0$, then they use the Lie-Suzuki-Trotter formula to approximate:
\begin{align}
    exp(- i M_D t) \approx \prod_{j=1}^p Q_j \exp(-i A t) Q_j
\end{align}
Then they use such exponentiation $\exp(-i M_D T)$ combined with the method introduced \cite{lloyd2013quantum} to achieve the approximation:
\begin{align}
    \Tr_{1,2,...,p-1} \exp(-i M_D t) \ket{\xbf}\bra{\xbf}^{\otimes p} \exp(-i M_D t) \approx \exp\big(-i D(\ket{\xbf}) t \big) \ket{\xbf}\bra{\xbf} \exp \big(-iD(\xbf)t \big)
\end{align}
Finally, they use method of \cite{wiebe2012quantum} to uses the above operation to approximately construct the state that is roughly as:
\begin{align}
   \ket{\bf 0} \xbf +  \ket{\bf 1}\Big(  \eta D(\ket{\xbf})\xbf\Big) + \ket{\rm Garbage}
\end{align}
Then they transform the state $\ket{\bf 1}$ into $-\ket{\bf 0} + \ket{\rm Remaining}$, so the above state is transformed to:
\begin{align}
   &\ket{\bf 0} \ket{\xbf} -  \ket{\bf 0}\Big(  \eta D(\ket{\xbf}) \xbf\Big) - \ket{\rm Remaining} + \ket{\rm Garbage} \\
   &=\ket{\bf 0} \Big( \ket{\xbf}- \eta D(\ket{\xbf}) \ket{\xbf}   \Big)  + \ket{\rm Garbage}
\end{align}
If $\ket{\xbf } = \ket{\xbf_t}$ which is the solution at $t$-th step, then by measuring the ancilla and post-select the outcome being $\ket{\bf 0}$, the result in the main register is a quantum state corresponding to $\ket{\xbf_t} - \eta D(\ket{\xbf_t})\ket{\xbf_t} = \xbf_{t+1}$, e.g., $\ket{\xbf_{t+1}}$. The spherical constraint is naturally placed under the effect of measurement. For a gradient descent with $T$ steps, the complexity complexity for producing $\ket{x_T}$ is 
$$ \mathcal{O}\Big( \frac{p^{5T} \Gamma^{3T} s^{T}}{\epsilon^{4T} } \log n   \Big)$$

\noindent
\textbf{The work in Ref.~\cite{nghiem2024simple}}. This work considers a function $f(\xbf)$ composed by sum of monomials:
\begin{align}
    f(\xbf)  = \sum_{i=1}^K f_i(\xbf) =  \sum_{i=1}^K a_i x_1^{i_1} x_2^{i_2} ... x_n^{i_n}
\end{align}
where $i_1,i_2,...,i_n \in \mathbb{Z}$. Extra assumptions include $|f(\xbf)| \leq 1$, $| \bigtriangledown f(\xbf) |_{\infty} \leq P$ for $\xbf \in [-1/2,1/2]^n$. The gradient of the above function is: 
\begin{align}
    \bigtriangledown f(\xbf) = \sum_{i=1}^K \bigtriangledown f_i(\xbf) =  \begin{pmatrix}
        \sum_{i=1}^K a_i i_1 x_1^{i_1-1} x_2^{i_2} ... x_n^{i_n} \\
        \sum_{i=1}^K a_i  i_2 x_1^{i_1} x_2^{i_2-1} ... x_n^{i_n} \\
        \vdots \\
        \sum_{i=1}^K a_i i_n x_1^{i_1} x_2^{i_2} ... x_n^{i_n-1}
    \end{pmatrix}
\end{align}
The technique used in Ref~\cite{nghiem2024simple} is somewhat similar to ours, as they also encode the classical variables $\xbf \equiv (x_1,x_2,...,x_n)$ as a diagonal operator $\rm diag (\xbf)$. They also aimed to produce the diagonal operator proportional to  $\rm diag \Big( \bigtriangledown f(\xbf) \Big)$. They do so by constantly applying Lemma \ref{lemma: singleentry} to construct $x_j \ket{k-1}\bra{k-1}$ for $j,k=1,2,...,n$, and manually construct the gradient based on the above equation. For instance, we consider the first entry, which is also the partial derivative of $f_i$ with respect to $x_1$:
\begin{align}
    \frac{\partial f_i(\xbf)}{\partial x_1} =  a_i i_1 x_1^{i_1-1} x_2^{i_2} ... x_n^{i_n}
\end{align}
From the block encoding of $\rm diag (\xbf)$, Lemma \ref{lemma: singleentry} allows us to construct the block encoding of $x_1\ket{0}\bra{0}, x_2\ket{0}\bra{0}, ..., x_n \ket{0}\bra{0}$. Then Lemma \ref{lemma: product} can be used to construct the block encoding of $x_1^{i_1-1} \ket{0}\bra{0},x_2^{i_2}\ket{0}\bra{0}, ..., x_n^{i_n}\ket{0}\bra{0}$. Then Lemma \ref{lemma: scale} can be used to construct the block encoding of $ \frac{a_i}{P} i_1 x_1^{i_1-1} x_2^{i_2} ... x_n^{i_n}\ket{0}\bra{0} = \frac{1}{P} \frac{\partial f_1(\xbf)}{\partial x_1} \ket{0}\bra{0}$. The same procedure can be used to construct the block encoding of $\frac{1}{P}\frac{\partial f_1(\xbf)}{\partial x_j} \ket{j-1}\bra{j-1}$ for remaining $j=2,3,...,n$. Then by using Lemma \ref{lemma: sumencoding}, it is possible to construct the block encoding of an operator that is proportional to 
$$ \sum_{j=1}^n \frac{\partial f_i(\xbf)}{\partial x_j}\ket{j-1}\bra{j-1}  = \rm diag \Big( \bigtriangledown f_1(\xbf)  \Big)$$
The same procedure yields the block encoding of operators $\sim \rm diag \Big( \bigtriangledown f_2(\xbf)\Big), \rm diag \Big( \bigtriangledown f_3(\xbf) \Big), ..., \rm diag \Big(  \bigtriangledown f_K(\xbf) \Big)$, where $\sim$ to refer to being proportional. Then Lemma \ref{lemma: sumencoding} allows us to construct the block encoding of $\sim  \sum_{i=1}^K \rm diag \Big(  \bigtriangledown f_i(\xbf) \Big) = \rm diag \Big( f(\xbf)  \Big) $. Then, Lemma \ref{lemma: sumencoding} is used again to construct the block encoding of $\sim \rm diag (\xbf) - \rm diag \Big( f(\xbf)   \Big)$. Thus, if one begins at some block-encoded operator $\rm diag \big( \xbf_0 \big)$, then one can construct the block encoding of $\sim \Big(  \rm diag (\xbf_0) - \rm diag \Big( f(\xbf_0)   \Big) = \rm diag \big( \xbf_1 \big)$, and then $\rm diag \Big(  \rm diag (\xbf_1) - \rm diag \Big( f(\xbf_1)   \Big) = \rm diag \big(\xbf_2 \big) $, and so on, until $T$-th iteration step, then one reach $\sim \rm diag \big( \xbf_T \big)$. 

Despite the fact the objective function $f(\xbf)  =  \sum_{i=1}^K f_i(\xbf) = \sum_{i=1}^K a_i x_1^{i_1} x_2^{i_2} ... x_n^{i_n} $ considered in Ref.~\cite{nghiem2024simple} is very general, the algorithm described above is also efficient when for each $f_i(\xbf)=a_i x_1^{i_1} x_2^{i_2} ... x_n^{i_n} $, many of the values among $\{i_1,i_2,...,i_n\}$ are zero, and $K$ needs to be small. More specifically, for each $f_i(\xbf)$, let $v_i$ denotes the number of variables of $f_i(\xbf)$ (having power $> 0$), $d_i$ denotes the total degree of $f_i(\xbf)$. Let $v = \max \{v_i\}_{i=1}^K$ and $d = \max \{d_i\}_{i=1}^K $. Then as worked out in \cite{nghiem2024simple}, the total complexity is 
$$ \mathcal{O}\Big(  \log(n) (K^2 d v^2 )^T \Big)$$
To demonstrate the limitation of the above algorithm, we revisit the problem of building a support vector machine from Section.~\ref{sec: buildingsupportvectormachine}. The loss function that we want to minimize is:
\begin{align}
    L = \frac{1}{2} \sum_{i=1}^n w_i^2 + C \sum_{s=1}^M \max \Big( 0, 1- y(s) \big( \sum_{i=1}^n w_i \xbf(s)_i + b \big) \Big)
\end{align}
which contains $K = n$ sum of monomial $w_i^2$. Thus, the algorithm has polynomial scaling in dimension $n$. The same issue exists for performing least-square data fitting and training neural network. So the method proposed in Ref.~\cite{nghiem2024simple} is not useful for these practical problems. At the same, as we have shown throughout the work, our proposal can handle these scenarios, thus proving to be better than \cite{nghiem2024simple}, even though there is certain overlap in terms of technique.

\section{More Details for Solving Linear System}
\label{sec: detaillinearsystem}
In quantum setting, instead of exactly acquiring the solution, the goal is to output a quantum state $\ket{\xbf}$ that corresponds to the solution $\xbf$ of the above linear system. Within the oracle/black box model, as mentioned in \cite{harrow2009quantum}, one has an efficient procedure that computes the entries of $A$, as well as an efficient means to prepare the quantum state $\ket{\textbf{b}} \equiv \textbf{b}$. Given these conditions, roughly speaking, the authors of \cite{harrow2009quantum} leveraged the Hamiltonian simulation techniques provided in \cite{berry2007efficient,berry2012black,berry2014high} to perform the exponentiation $\exp(-iAt)$, then combined with quantum phase estimation \cite{kitaev1995quantum, abrams1999quantum} and controlled phase rotations to invert the phase register that stores the eigenvalues of $A$, thus achieving matrix inversion. The result of \cite{harrow2009quantum} is a quantum state $\ket{\xbf}$ corresponding to the solution of the linear equation above. Thus, as mentioned in Ref. \cite{harrow2009quantum}, the quantum linear solver will be most useful if instead of knowing exactly the solution, one needs to estimate certain quantities associated with the solution, for example $\bra{x}M\ket{x}$ where $ M$ is one observable. The HHL algorithm described above has motivated many subsequent constructions that achieve improvement in certain aspects. For example, Ref. \cite{childs2017quantum} introduced an exponentially improved quantum linear solver with respect to error tolerance, Ref. \cite{wossnig2018quantum} introduced a different model to solve dense linear systems with quadratic speed-up compared to Ref. \cite{harrow2009quantum} (and also the best known classical solution), Ref. \cite{clader2013preconditioned} provided an enhanced quantum linear solver incorporating preconditioning technique that expands the reach of original quantum linear solver \cite{harrow2009quantum}, that could deal systems with high conditional number. Furthermore, a series of works \cite{bravo2023variational,an2022quantum, costa2022optimal, subacsi2019quantum} provided an alternatively physically inspired route to solve linear systems, for example, adiabatic process, variational approach, etc. More on the application side, the authors in Ref. \cite{rebentrost2014quantum} showed how to extend the technique from the HHL algorithm to build a quantum support vector machine, which is a very famous machine learning algorithm. Nevertheless, all these aforementioned works are either heuristic (variational quantum linear solver) or assuming the oracle/black-box procedure, which severely limits their practical capability. 

Recall that two functions are:
\begin{align}
     f(\xbf) = \frac{1}{2}|\xbf|^2 + |A\xbf -\bbf|^2
\end{align}
Remark that the above function can be equivalently written as:
\begin{align}
    f( \xbf) &= \frac{1}{2}|\xbf|^2 +  (  A\xbf - \textbf{b} )^T ( A\xbf - \textbf{b} ) \\
    &= \frac{1}{2}|\xbf|^2+ |A\xbf|^2 + |\textbf{b}|^2 - 2 \textbf{b}^T A \xbf 
\end{align}
Let
\begin{align}
    A = \begin{pmatrix}
        A_{11} &  A_{12} & \cdots & A_{1n} \\
        A_{21} & A_{22}  & \cdots & A_{2n} \\
        \vdots & \vdots &  \ddots & \vdots \\
        A_{n1} & A_{n2} & \cdots & A_{nn} 
    \end{pmatrix}, \xbf = \begin{pmatrix}
        x_1 \\
        x_2 \\
        \vdots \\
        x_n
    \end{pmatrix}, 
    \textbf{b} = \begin{pmatrix}
        b_1 \\
        b_2\\
        \vdots \\
        b_n
    \end{pmatrix}
\end{align}
So we have that:
\begin{align}
    A \xbf = \begin{pmatrix}
        \sum_{i=1}^n A_{1 i} x_i \\
        \sum_{i=1}^n A_{ 2 i} x_i \\
        \vdots \\
        \sum_{i=1}^n A_{n i} x_i
    \end{pmatrix} \longrightarrow 
    | A \xbf|^2 = (  \sum_{i=1}^n A_{1 i} x_i )^2 + (\sum_{i=1}^n A_{ 2 i} x_i)^2 +  \cdots + (\sum_{i=1}^n A_{n i} x_i)^2 
\end{align}
Similarly: 
\begin{align}
    & \textbf{b}^T A\xbf =   \sum_{i=1}^n b_i \sum_{j=1}^n A_{ij} x_j =\big(\sum_{i=1}^n b_i A_{i1} \big) x_1 + \big(\sum_{i=1}^n b_i A_{i2} \big) x_2 + ... + \big(\sum_{i=1}^n b_i A_{in} \big) x_n   \\
    & |\textbf{b}|^2 = \sum_{i=1}^n b_i^2
\end{align}
Therefore, the function $f(\xbf)$ is:
\begin{align}
    f(\xbf) = \frac{1}{2}\sum_{i=1}^n x_i^2 + (  \sum_{i=1}^n A_{1 i} x_i )^2 + (\sum_{i=1}^n A_{ 2 i} x_i)^2 +  \cdots + (\sum_{i=1}^n A_{n i} x_i)^2  \\ + \big(\sum_{i=1}^n b_i A_{i1} \big) x_1 + \big(\sum_{i=1}^n b_i A_{i2} \big) x_2 + ... + \big(\sum_{i=1}^n b_i A_{in} \big) x_n +  \sum_{i=1}^n b_i^2
    \label{37}
\end{align}
If all the coefficients $\{a_{ij} \}_{ij=1}^n$ and $\{b_i\}_{i=1}^n$ are classically known then the above equation fits perfectly to the context of Theorem \ref{theorem: type1} and \ref{theorem: type2}. 
Thus, we can apply our quantum gradient descent algorithm to solve the linear equation. An important aspect worth discussing is, in the context of Theorem \ref{theorem: type2}, what should be the value of $S$ and $K$ for the above equation ? If $A$ is dense, the apparently $S= n$ and $K=n$, and the quantum gradient descent will have running time being polynomial in $n$, which is not efficient. However, if $A$ is sparse, for example, each row of $A$ has at most $s$ non-zero entries, then for $i=1,2,...,n$ the row vector $ (  A_{i1}, A_{i2}, ..., A_{in} )$ has sparsity $s$, which means that $S$ in Theorem \ref{theorem: type2} is equal to $s$. However, as there is the summation 
$$(  \sum_{i=1}^n A_{1 i} x_i )^2 + (\sum_{i=1}^n A_{ 2 i} x_i)^2 +  \cdots + (\sum_{i=1}^n A_{n i} x_i)^2  $$
The value of $K$ is equal to $n$, which still leads to polynomial scaling. In order to achieve more efficient scaling, we need many of composing terms in the above summation to be zero, which means that the matrix $A$ needs to have many of its rows to contain zero entries only, and all the remaining rows contain at most $s$ non-zero entries each. This is similar to many existing contexts, including \cite{harrow2009quantum, childs2017quantum, clader2013preconditioned, wiebe2012quantum} where the efficient quantum algorithm is only possible for highly sparse matrices. The quite intriguing aspect of our quantum gradient descent applied to linear system solving is that, the complexity does not depend on conditional number of the matrix, which implies the ability to deal with more type of linear systems (see \cite{clader2013preconditioned} for detailed discussion on linear systems with large conditional number). At the same time, our gradient descent-based linear solver has an exponential scaling on the iteration times $T$. We attribute it as a trade-off for the independence of running time with respect to conditional number. Furthermore, it could be guaranteed that the solution to the linear system is corrected if the above function $f(\xbf)$ defined in Eqn. \ref{37} is (strongly) convex, which means that the local minima are also global minima and $T = \mathcal{O}\big( \log \frac{1}{\epsilon}\big)$ is sufficient to reach the desired minima.

\section{Details for Least-Square Data Fitting}
\label{sec: detaildatafitting}
Recall that the objective is minimizing:
\begin{align}
    E = | \textbf{F} \lambda - \textbf{y}|^2
\end{align}
As worked out in \cite{wiebe2012quantum}, the closed form solution to the above problem is based on Moore-Penrose pseudoinverse, i.e.
\begin{align}
    \lambda = ( \textbf{F}^\dagger \textbf{F})^{-1} \textbf{F}^\dagger \textbf{y}
\end{align}
Thus, in order to obtain the optimal fit parameters, one needs to be able to perform matrix inversion and multiplication. Quantum data fitting algorithm \cite{wiebe2012quantum} was proposed as an adaptation of the quantum linear solving algorithm. Similar to linear system setting \cite{harrow2009quantum}, a very crucial assumption made in \cite{wiebe2012quantum} is the oracle/black-box access to entries of $\textbf{F}$. Given such a tool, instead of inverting a matrix, the algorithm in \cite{wiebe2012quantum} adjusted a technical step from the original HHL algorithm, e.g., from phase inversion to phase multiplication, to perform multiplication of the given matrix to desired vector $\textbf{y}$. Thereby, the complexity of data fitting algorithm \cite{wiebe2012quantum} is essentially similar to matrix inversion one. 

Given the above discussion, it is reasonable to expect that our method can also be applied to the data fitting problem in \cite{wiebe2012quantum}. Indeed, the formalism of the data-fitting problem can naturally lead to an optimization problem, which could be solved by the gradient descent method. The equation \ref{62} is exactly what we got in the linear solving context (Equation \ref{eqn: linearcost}), thus, the quantum gradient descent developed in this work could be extended simply to deal with data fitting problem as well, extending the scope of its capability, meanwhile not involving the oracle/black-box access requirement. 

Recall that in the main text, we indicated that we want to fit the following function:
\begin{align}
    f(x,\lambda) = \lambda_1 x + \lambda_2 x^2 + ... + \lambda_n x^n
\end{align}
Without loss of generality, assume that $n$ is some power of 2. In the context of Theorem \ref{theorem: type2}, if we desire a more efficient running time, then we need to be able to use the second version of Lemma \ref{lemma: diag}, which means that we need an efficient means to prepare some state that contains $( \sqrt{x},\sqrt{x^2}, ..., \sqrt{x^n})$  in its first $n$ entries. For brevity, we define $z = \sqrt{x}$. We observe the following examples:
\begin{align}
    &\ket{00} + z\ket{01} + z^2 \ket{10} + z^3 \ket{11} = \Big(\ket{0}+z^2\ket{1} \Big) \otimes \Big(\ket{0} + z\ket{1} \Big) \\
    &\ket{000} + z\ket{001} + z^2 \ket{010} + z^3\ket{011} + z^4 \ket{100} + z^5 \ket{101} + z^6 \ket{110} + z^7 \ket{111} =  \Big( \ket{0} + z^4\ket{1} \Big) \otimes  \Big( \ket{0}+ z^2 \ket{1}  \Big) \otimes \Big(\ket{0} + z\ket{1} \Big)
\end{align}
More generally, we have (in digital convention):
\begin{align}
    \sum_{i=0}^{n-1} z^i \ket{i}  = \Big(  \ket{0} + z^{n/2} \ket{1}\Big) \otimes \Big(  \ket{0} + z^{n/4}\ket{1} \Big) \otimes \Big(  \ket{0}  + z^{n/8} \ket{1} \Big) \otimes  \cdots \otimes \Big( \ket{0} + z \ket{1}  \Big) \\
    \longrightarrow \sum_{i=0}^{n-1} z^{i+1} \ket{i}  = \Big(  \ket{0} + z^{n/2} \ket{1}\Big) \otimes \Big(  \ket{0} + z^{n/4}\ket{1} \Big)\otimes  \Big(  \ket{0}  + z^{n/8} \ket{1} \Big) \otimes \cdots  \otimes \Big( z\ket{0} + z^2 \ket{1}  \Big)
\end{align}
We begin with $\log(M)$ qubits initialized in $\ket{0}^{\otimes\log(n)}$. Then for the $k$-th qubit, we use single qubit rotation gate and rotate $k$-th qubit according to the angle:
\begin{align}
    \ket{0} \longrightarrow  \frac{1}{\sqrt{1+ z^{2n/2^k}}} \Big(  \ket{0} + z^{M/2^k} \ket{1}\Big)
\end{align}
and for the last qubit, we rotate as:
\begin{align}
    \ket{0} \longrightarrow \frac{1}{\sqrt{z^2+z^4}} \Big(  z\ket{0}+z^2\ket{1} \Big)
\end{align}
Then with a circuit of complexity $\mathcal{O}(1)$, we obtain the desired state:
\begin{align}
    C \Big(  \ket{0} + z^{n/2} \ket{1}\Big) \otimes \Big(  \ket{0} + z^{n/4}\ket{1} \Big)\otimes  \Big(  \ket{0}  + z^{n/8} \ket{1} \Big) \otimes \cdots  \otimes \Big( z\ket{0} + z^2 \ket{1}  \Big)
\end{align}
where the normalization factor:
\begin{align}
    C = \prod_{k=1}^{\log(n)-1} \frac{1}{\sqrt{z^2+z^4}} \frac{1}{\sqrt{z^2+z^4}} 
\end{align}
So the algorithm of Theorem \ref{theorem: type2} can be applied in a straightforward manner. \\

\section{Details for Support Vector Machine}
\label{sec: detailSMV} 
Recall that the objective of linear SVM is to find an optimal hyperplane that maximizes the margin while minimizing classification errors. The primal form of the support vector machine optimization with hinge loss is:
\begin{align}
    L = \frac{1}{2} |\wbf|^2 + C \sum_{s=1}^M  \max \Big( 0, 1- y(s) \big(\wbf^T \xbf(s) + b \big)  \Big)
\end{align}
where $\wbf \equiv (w_1,w_2,...,w_n)  \in \Rbb^n $ is the weight vector, $b$ is the bias term, $C$ is regularization parameter. The objective is to find $\wbf, b$ such that the above quantity $L$ is minimized. It is straightforward to show that:
\begin{align}
    L = \frac{1}{2} \sum_{i=1}^n w_i^2 + C \sum_{s=1}^M \max \Big( 0, 1- y(s) \big( \sum_{i=1}^n w_i \xbf(s)_i + b \big) \Big)
\end{align}
We remark that the contribution of the term $C \sum_{s=1}^M \max \Big( 0, 1- y(s) \big( \sum_{i=1}^n w_i \xbf(s)_i + b \big) \Big) $ is only non-zero if: 
\begin{align}
    y(s) \big( \sum_{i=1}^n w_i \xbf(s)_i + b \big)  < 1
\end{align}
corresponding to those points that lie within the margin, or misclassified. Therefore, the loss function $L$ is essentially equal to:
\begin{align}
    L &= \frac{1}{2} \sum_{i=1}^n w_i^2 + C \sum_{s=1}^N \Big( 1-y(s) \big( \sum_{i=1}^n w_i \xbf(s)_i + b \big)  \Big) \\
    &= \frac{1}{2} \sum_{i=1}^n w_i^2  + \sum_{i=1}^n \Big( \sum_{s=1}^N -y(s) x(s)_i  \Big) w_i + CbN + CN 
\end{align}
where $N$ is the number of points lying within the margin. This form of function fits exactly to the context of Theorem \ref{theorem: type1}, therefore, the quantum gradient descent algorithm for the first type of polynomial can work. \\

\section{Details for Supervised Cluster Assignment   }
\label{sec: detailSVC}
Recall that we are trying to find:
\begin{align}
    \min_{\Vec{c}}  \frac{1}{M} \sum_{i=1}^M | \Vec{c} - \Vec{r}_i |^2
\end{align}
To prove this that the vector $\Vec{c}$ that minimizes the above function is indeed the centroid, let $\Vec{c} \equiv (c_1,c_2,..,c_n)^T, \  \Vec{r}_i \equiv (r_{i,1},r_{i,2},...,r_{i,n})^T$ be the coordinates. We have that:
\begin{align}
| \Vec{c} - \Vec{r}_i |^2 &= \sum_{j=1}^n | c_j - r_{i,j}|^2 \\
\longrightarrow  \sum_{i=1}^M | \Vec{c} - \Vec{r}_i |^2 &= \sum_{i=1}^M \sum_{j=1}^n | c_j - r_{i,j}|^2  = \sum_{j=1}^n  \sum_{i=1}^M  | c_j - r_{i,j}|^2 
\end{align}
We have the following basic inequality:
\begin{align}
    M \sum_{i=1}^M  | c_j - r_{i,j}|^2  \geq | \sum_{i=1}^M ( c_j - r_{i,j} ) |^2 \geq 0
\end{align}
By dividing both sides of the above equation to $M^2$, we obtain $\frac{1}{M}\sum_{i=1}^M  | c_j - r_{i,j}|^2 \geq 0 $, thus implying $ \frac{1}{M} \sum_{j=1}^n  \sum_{i=1}^M  | c_j - r_{i,j}|^2  \geq 0$. The equality occurs if and only if, for all $j=1,2,..,n$:
\begin{align}
   \sum_{i=1}^M ( c_j - r_{i,j} ) = 0 \\
   \longrightarrow M c_j - \sum_{i=1}^M r_{i,j} = 0 \\
   \longrightarrow c_j = \frac{\sum_{i=1}^M r_{i,j}}{M}
\end{align}
Recall that $\Vec{c} \equiv  (c_1,c_2,...,c_n)^T $ and $\Vec{r}_i \equiv (r_{i,1},r_{i,2},...,r_{i,n})^T$, so we have:
\begin{align}
    \Vec{c} = \frac{1}{M} \sum_{i=1}^M \Vec{r}_i
\end{align}
For the cost function, we have the following:
\begin{align}
 \frac{1}{M}  \sum_{i=1}^M | \Vec{c} - \Vec{r}_i |^2 &= \frac{1}{M} \sum_{i=1}^M \sum_{j=1}^n ( c_j-r_{i,j} )^2 \\
                            &= \frac{1}{M} \sum_{i=1}^M  \sum_{j=1}^n \big( c_j^2  - 2 c_j r_{i,j} + r_{i,j}^2 \big) \\
                            &= \frac{1}{M }\sum_{i=1}^M \Big( \sum_{j=1}^n c_j^2 - \sum_{j=1}^n2 c_j r_{i,j} + \sum_{j=1}^n r_{i,j}^2 \Big) \\
                            &=  \sum_{j=1}^n c_j^2 - \sum_{j=1}^n  2 c_j \big( \frac{\sum_{i=1}^M r_{i,j}}{M}  \big)  + \frac{1}{M}\sum_{i=1}^M \sum_{j=1}^n r_{i,j}^2
\end{align}
This cost function apparently belongs to the first type, \ref{theorem: type1}, thus can be solved by our quantum gradient descent method.

\section{Details For Neural Networks}
\label{sec: detailneuralnetwork}
\begin{figure}[H]
\centering
\begin{tikzpicture}
    \begin{axis}[
        axis lines = middle,
        xlabel = $x$,
        ylabel = $y$,
        xmin = -1.2, xmax = 1.2,
        ymin = -0.2, ymax = 1.2,
        legend style = {font=\small},
        samples = 100
    ]
        \addplot[
            domain = -1:1,
            samples = 100,
            thick,
            blue
        ] {max(0, x)};
        
        \addplot[
            domain = -1:1,
            samples = 100,
            thick,
            red,
        ] {(5/8)*x^2 + (1/2)*x};

        \addplot[
            domain = -1:1,
            samples = 100,
            thick,
            green,
        ] { (-21/32)*x^4 + (35/32)*x^2 + (1/2)*x};
    \end{axis}
\end{tikzpicture}

\caption{Illustration of approximating ReLu function $\max(0,x)$ (blue line) with polynomial function $\frac{5}{8}x^2 + \frac{1}{2}x$ (red line) and $ \frac{-21}{32}x^4 + \frac{35}{32}x^2 + \frac{1}{2}x $ (green line). }
\label{fig: reluapproximation}
\end{figure}
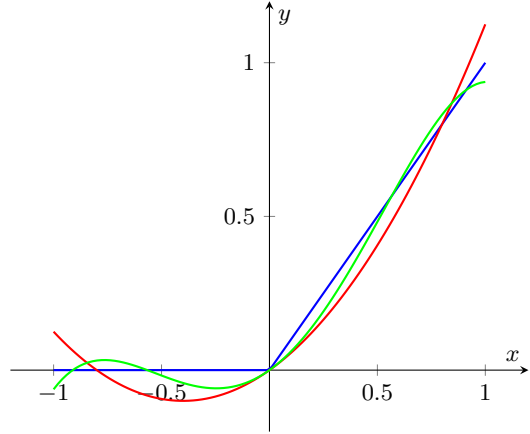

This appendix is devoted to show that the loss function for the training neural network is similar to the form we consider in the main text, Thm. \ref{theorem: type1} and Thm. \ref{theorem: type2}. We shall examine two different types of neural network. The first type is wide and shallow, and the second type is deep but narrow.  \\
\noindent
\textbf{Wide and Shallow Network:}\\
For a neural network which is wide and shallow, it means that the number of hidden layers is small and the number of nodes in each layer is very high. For concreteness, we consider only 1 hidden layer that has $m$ nodes. From the input layer $\xbf = (x_1,x_2,...,x_n)$, we have the following transformation at the first layer: 
\begin{align}
    \begin{pmatrix}
        x_1 \\
        x_2 \\
        \vdots \\
        x_n 
    \end{pmatrix} \longrightarrow \begin{pmatrix}
        \sigma \Big( (\wbf^{(0)}_1)^T \xbf + b_1^{(0)}  \Big) \\
        \sigma \Big( (\wbf^{(0)}_2)^T \xbf + b_2^{(0)}  \Big) \\
        \vdots \\
        \sigma \Big( (\wbf^{(0)}_m)^T \xbf+ b_m^{(0)} \Big)
    \end{pmatrix} \equiv \begin{pmatrix}
        a^{(1)}_1 \\
        a^{(1)}_2 \\
        \vdots \\
        a^{(1)}_m 
    \end{pmatrix}
\end{align}
After that first layer, there is another round of matrix application, resulting in the output layer as:
\begin{align}
    \Tilde{y} &= \sigma\Big( w^{(1)}_1 a^{(1)}_1  + w^{(1)}_2 a^{(1)}_2 + ... + w^{(1)}_m a^{(1)}_m \Big) \\
      &= \sigma \left(  w^{(1)}_1  \sigma \Big( (\wbf^{(0)}_1)^T \xbf + b_1^{(0)}  \Big) + w_2^{(1)} \sigma \Big( (\wbf^{(0)}_2)^T \xbf + b_2^{(0)}  \Big) + ... + w_m^{(1)} \sigma \Big( (\wbf^{(0)}_m)^T \xbf+ b_m^{(0)} \Big) \right)
\end{align}
The loss function is:
\begin{align}
    L &= \frac{1}{M} \sum_{s=1}^M \big(\Tilde{y}(s) - y(s) \big)^2 \\
    &= \frac{1}{M} \sum_{s=1}^M \Tilde{y}^2(s) + \frac{1}{M} \sum_{i=1}^M y^2(s) - \frac{2}{M}\sum_{s=1}^M \Tilde{y}(s) y(s)
\end{align}
The activation function $\sigma$ we choose is Relu($x$) = $\max(0,x)$. Since this function is not continuous on the interval $[-1,1]$, we approximate it with a polynomial (see Figure \ref{fig: reluapproximation}):
\begin{align}
    f(x) = \frac{5}{8}x^2 + \frac{1}{2}x 
\end{align}

So for a single instance $s$, we have:
\begin{align}
    \Tilde{y}(s) &= \frac{5}{8} \left(  w^{(1)}_1  \sigma \Big( (\wbf^{(0)}_1)^T \xbf + b_1^{(0)}  \Big) + w_2^{(1)} \sigma \Big( (\wbf^{(0)}_2)^T \xbf + b_2^{(0)}  \Big) + ... + w_m^{(1)} \sigma \Big( (\wbf^{(0)}_m)^T \xbf+ b_m^{(0)} \Big) \right)^2 + \\ 
    & \frac{1}{2} \left(  w^{(1)}_1  \sigma \Big( (\wbf^{(0)}_1)^T \xbf + b_1^{(0)}  \Big) + w_2^{(1)} \sigma \Big( (\wbf^{(0)}_2)^T \xbf + b_2^{(0)}  \Big) + ... + w_m^{(1)} \sigma \Big( (\wbf^{(0)}_m)^T \xbf+ b_m^{(0)} \Big) \right)  \\
    &= \frac{5}{8} \left( \sum_{i,j=1}^m  w^{(1)}_i w^{(1)}_j  \sigma \Big( (\wbf^{(0)}_i)^T \xbf + b_i^{(0)}  \Big)  \sigma \Big( (\wbf^{(0)}_j)^T \xbf + b_j^{(0)}  \Big)   \right) + \\
    & \frac{1}{2} \left( \sum_{i=1}^m w^{(1)}_i  \sigma \Big( (\wbf^{(0)}_i)^T \xbf + b_i^{(0)}  \Big)  \right)
\end{align}
which leds to:
\begin{align}
    \Tilde{y}(s)^2 &= \frac{25}{64} \left( \sum_{i,j=1}^m  w^{(1)}_i w^{(1)}_j  \sigma \Big( (\wbf^{(0)}_i)^T \xbf + b_i^{(0)}  \Big)  \sigma \Big( (\wbf^{(0)}_j)^T \xbf + b_j^{(0)}  \Big)   \right)^2 + \\ & \frac{5}{8} \left( \sum_{i,j=1}^m  w^{(1)}_i w^{(1)}_j  \sigma \Big( (\wbf^{(0)}_i)^T \xbf + b_i^{(0)}  \Big)  \sigma \Big( (\wbf^{(0)}_j)^T \xbf + b_j^{(0)}  \Big)   \right)\left( \sum_{i=1}^m w^{(1)}_i  \sigma \Big( (\wbf^{(0)}_i)^T \xbf + b_i^{(0)}  \Big)  \right)  + \\
    & \frac{1}{4} \left( \sum_{i=1}^m w^{(1)}_i  \sigma \Big( (\wbf^{(0)}_i)^T \xbf + b_i^{(0)}  \Big)  \right)^2
\end{align}
By expanding further:
\begin{align}
     \sigma \Big( (\wbf^{(0)}_i)^T \xbf + b_i^{(0)}  \Big) = \frac{5}{8}\Big(  (\wbf^{(0)}_i)^T \xbf + b_i^{(0)} \Big)^2 + \frac{1}{2} \Big( (\wbf^{(0)}_i)^T \xbf + b_i^{(0)} \Big)
\end{align}
we have:
\begin{align}
    \Tilde{y}(s) &= \frac{5}{8} \left( \sum_{i,j=1}^m  w^{(1)}_i w^{(1)}_j  \left( \frac{5}{8}\Big(  (\wbf^{(0)}_i)^T \xbf + b_i^{(0)} \Big)^2 + \frac{1}{2} \Big( (\wbf^{(0)}_i)^T \xbf + b_i^{(0)} \Big) \right) \left( \frac{5}{8}\Big(  (\wbf^{(0)}_j)^T \xbf + b_j^{(0)} \Big)^2 + \frac{1}{2} \Big( (\wbf^{(0)}_j)^T \xbf + b_j^{(0)} \Big) \right) \right) + \\
    & \frac{1}{2} \left( \sum_{i=1}^m w^{(1)}_i \frac{5}{8}\Big(  (\wbf^{(0)}_i)^T \xbf + b_i^{(0)} \Big)^2 +  w^{(1)}_i\frac{1}{2} \Big( (\wbf^{(0)}_i)^T \xbf + b_i^{(0)} \Big)  \right) 
\end{align}
While it seems cumbersome to expand the analytical expression further, we observe that the terms within the expansion above shares similar form to what we have considered in the main text, e.g., Thm.~\ref{theorem: type2}, Thm.~\ref{thm: type3}:
\begin{align}
    (\wbf^{(0)}_i)^T \xbf + b_i^{(0)}  \xleftrightarrow{\text{correspondence}} \textbf{a}_i \xbf + b_i \\
   w^{(1)}_i w^{(1)}_j \Big( (\wbf^{(0)}_i)^T \xbf + b_i^{(0)} \Big)  \Big(  (\wbf^{(0)}_j)^T \xbf + b_j^{(0)} \Big)^2\xleftrightarrow{\text{correspondence}} \prod_k \Big( \textbf{a}_k^T \xbf + b_k \Big)^k \\
    \frac{5}{8}\Big(  (\wbf^{(0)}_i)^T \xbf + b_i^{(0)} \Big)^2 + \frac{1}{2} \Big( (\wbf^{(0)}_i)^T \xbf + b_i^{(0)} \Big) \xleftrightarrow{\text{correspondence}} \sum_k \Big(  \textbf{a}_i^T \xbf + b_i \Big)^k
\end{align}
Thus, in principle, our quantum gradient descent, e.g., theorem \ref{theorem: type2}, \ref{thm: type3} can be applied. 

\noindent
\textbf{Deep and Narrow Network:}\\
Here we examine a deep but narrow network, which is more or less a direct expansion from the above calculation. Suppose that we have $p$ hidden layers and that each layer only has 2 nodes ($p >> 2$). 
Following the description in Algorithm \ref{algo: trainingneuralnetwork} and Figure \ref{fig: activation}, from the input layer $\xbf \equiv (x_1,x_2,...,x_n)$, we have the transformation:
\begin{align}
    \begin{pmatrix}
        x_1 \\
        x_2 \\
        \vdots \\
        x_n 
    \end{pmatrix} \longrightarrow
     \begin{pmatrix}
         \sigma ( w^{(0)}_{1,1} x_1 + w^{(0)}_{1,2} x_2 + ... + w^{(0)}_{1,n} x_n  +  b_1^{(0)}   )\\
         \sigma ( w^{(0)}_{2,1} x_1 + w^{(0)}_{2,2} x_2 + ... + w^{(0)}_{2,n} x_n  +  b_2^{(0)}  )
     \end{pmatrix} \equiv  
     \begin{pmatrix}
         a^{(1)}_1 \\
         a^{(1)}_2 
     \end{pmatrix}
\end{align}
More concretely, we have the first entry:
\begin{align}
   a^{(1)}_1 = \sigma ( w^{(0)}_{1,1} x_1 + w^{(0)}_{1,2} x_2 + ... + w^{(0)}_{1,n} x_n  +  b_1^{(0)}   ) &= \sigma\Big(  (\wbf^{(0)}_1)^T \xbf  + b^{(0)}_1 \Big) \\
    &= \frac{5}{8} \Big( (\wbf^{(0)}_1)^T \xbf  + b^{(0)}_1  \Big)^2 + \frac{1}{2} \Big(  \sum_{i=1}^n  (\wbf^{(0)}_1)^T \xbf  + b^{(0)}_1  \Big) 
    \label{b5}
\end{align}
Likewise, the second entry is:
\begin{align}
  a^{(2)}_2= \sigma ( w^{(0)}_{2,1} x_1 + w^{(0)}_{2,2} x_2 + ... + w^{(0)}_{2,n} x_n  +  b_2^{(0)}   ) &=  \frac{5}{8} \Big( (\wbf^{(0)}_2)^T \xbf  + b^{(0)}_2  \Big)^2 + \frac{1}{2} \Big(  \sum_{i=1}^n  (\wbf^{(0)}_2)^T \xbf  + b^{(0)}_2  \Big) 
\end{align}
Then we have the transformation at the next, second layer:
\begin{align}
    \begin{pmatrix}
        a^{(1)}_1\\
         a^{(1)}_2
     \end{pmatrix} \longrightarrow 
    \begin{pmatrix}
   \sigma \Big(  w^{(1)}_{1,1}  a^{(1)}_1 + w^{(1)}_{1,2} a^{(1)}_2 \Big)\\
    \sigma\Big( w^{(1)}_{2,1}  a^{(1)}_1  + w^{(1)}_{2,2} a^{(1)}_2  \Big)
    \end{pmatrix} \equiv \begin{pmatrix}
        a^{(2)}_1 \\
        a^{(2)}_2 
    \end{pmatrix}
\end{align}
Again we expand:
\begin{align}
   a^{(2)}_1 =  \sigma \Big(  w^{(1)}_{1,1}  a^{(1)}_1 + w^{(1)}_{1,2} a^{(1)}_2 \Big) &= \frac{5}{8} \Big(w^{(1)}_{1,1}  a^{(1)}_1 + w^{(1)}_{1,2} a^{(1)}_2  \Big)^2 + \frac{1}{2} \Big(  w^{(1)}_{1,1}  a^{(1)}_1 + w^{(1)}_{1,2} a^{(1)}_2\Big)   \\
     &= \frac{5}{8}\left( \big( w^{(1)}_{1,1}\big)^2 \big( a^{(1)}_1 \big)^2 +  \big( w^{(1)}_{1,2}\big)^2 \big( a^{(1)}_2 \big)^2 + 2w^{(1)}_{1,1} w^{(1)}_{1,2} a^{(1)}_1 a^{(1)}_2 \right) + \frac{1}{2} \left(  w^{(1)}_{1,1}  a^{(1)}_1 + w^{(1)}_{1,2} a^{(1)}_2  \right)
\end{align}
Similarly:
\begin{align}
    a^{(2)}_2 =  \frac{5}{8}\left( \big( w^{(1)}_{2,1}\big)^2 \big( a^{(1)}_1 \big)^2 +  \big( w^{(1)}_{2,2}\big)^2 \big( a^{(1)}_2 \big)^2 + 2w^{(1)}_{2,1} w^{(1)}_{2,2} a^{(1)}_1 a^{(1)}_2 \right) + \frac{1}{2} \left(  w^{(1)}_{2,1}  a^{(1)}_1 + w^{(1)}_{2,2} a^{(1)}_2  \right)
\end{align}
The transformation at next, third layer is:
\begin{align}
    \begin{pmatrix}
        a^{(2)}_1 \\
        a^{(2)}_2
    \end{pmatrix}\longrightarrow \begin{pmatrix}
        \sigma \left( w^{(2)}_{1,1} a^{(2)}_1 + w^{(2)}_{1,2} a^{(2)}_2   \right) \\
        \sigma \left(  w^{(2)}_{2,1} a^{(2)}_1 + w^{(2)}_{2,2} a^{(2)}_2 \right)
    \end{pmatrix}
\end{align}
and so on until the $p$-th layer. \\

Both examples above, including a wide/shallow network and a deep/narrow network, exhibit a lengthly and complicated analytical form for the loss function. However, the above derivation intuitively suggests that the loss function is the combination of the second (Theorem \ref{theorem: type2}) and third type of function (Theorem \ref{thm: type3}) that we considered. To estimate the scaling, we need to gauge how many terms are there in the expansion having form like $(\wbf^{(0)}_i)^T \xbf + b^{(0)}_i$, or some power of it. We refer to those terms as basic terms, since we have noted before that $(\wbf^{(0)}_i)^T \xbf + b_i^{(0)}  \xleftrightarrow{\text{correspondence}} \textbf{a}_i \xbf + b_i $. The number of basic terms correspond to $K$ in the context of Theorem.~\ref{theorem: type2} and Theorem.~\ref{thm: type3}. For the purpose of gauging, it is quite simple, as we point out the following pattern. Consider a neural network having $p$ hidden layers and there are $m$ nodes in each layer. Suppose that the input layer has $n$ nodes (featuring $n$-dimensional data), and output layer has $1$ node. Then we have the transformation:
\begin{align}
    \begin{pmatrix}
        x_1\\
        x_2\\
        \vdots \\
        x_n
    \end{pmatrix} \longrightarrow \begin{pmatrix}
        a^{(1)}_1 \\
        a^{(1)}_2 \\
        \vdots \\
        a^{(1)}_m 
    \end{pmatrix} \longrightarrow \begin{pmatrix}
        a^{(2)}_1 \\
        a^{(2)}_2 \\
        \vdots \\
        a^{(2)}_m 
    \end{pmatrix}  \longrightarrow \cdots \longrightarrow \begin{pmatrix}
        a^{(p)}_1 \\
        a^{(p)}_2 \\
        \vdots \\
        a^{(p)}_m 
    \end{pmatrix} 
\end{align}
For each $i=1,2,...,m$, the entry $a^{(1)}_i = \sigma\Big( \wbf^{(0)}_i\xbf + b^{(0)}_i  \Big) = \frac{5}{8} \Big(  \wbf^{(0)}_i \xbf + b^{(0)}_i \Big)^2  + \frac{1}{2}\Big( \wbf^{(0)}_i\xbf + b^{(0)}_i \Big) $ consists of two summands. The combination:
\begin{align}
    w^{(1)}_{1,1} a^{(1)}_1 + w^{(1)}_{1,2} a^{(1)}_2 + ... + w^{(1)}_{1,m} a^{(1)}_m + b^{(1)}_1 = (\wbf^{(1)}_1)^T \textbf{a}^{(1)} + b^{(1)}_1
\end{align}
then has $2m$ basic terms.  The application of the activation function 
\begin{align}
     a^{(2)}_1 &=\sigma \left( (\wbf^{(1)}_1)^T \xbf + b^{(1)}_1\right)  \\
            &= \frac{5}{8} \left( w^{(1)}_{1,1} a^{(1)}_1 + w^{(1)}_{1,2} a^{(1)}_2 + ... + w^{(1)}_{1,m} a^{(1)}_m + b^{(1)}_1\right)^2 + \frac{1}{2} \left ( w^{(1)}_{1,1} a^{(1)}_1 + w^{(1)}_{1,2} a^{(1)}_2 + ... + w^{(1)}_{1,m} a^{(1)}_m + b^{(1)}_1\right)
\end{align}
inflates the total number of terms from $2m$ to $\mathcal{O}(m^2)$. Thus each entry $a^{(2)}_j$ for $j=1,2,...,m$ has totally $\mathcal{O}(m^2)$ basic terms. Next, we have the transformation:
\begin{align}
    w^{2}_{1,1} a^{(2)}_1 + w^{(2)}_{1,2} a^{(2)}_2 + ... + w^{(2)}_{1,m} a^{(2)}_m + b^{(2)}_1 = (\wbf^{(2)}_1)^T \xbf + b^{(2)}_1
\end{align}
having $m$ terms under the summation, and each $a^{(2)}_j$ has $\mathcal{O}(m^2)$ basic terms, so there are $\mathcal{O}(m^3)$ basic terms. The application of activation function $ \sigma \left( (\wbf^{(2)}_1)^T \xbf + b^{(2)}_1\right) =a^{(3)}_1$ increases the total of basic terms to $\mathcal{O}(m^6)$. By continuing in a similar way, the first entry at 4-th layer $a^{(4)}_1$ has $\mathcal{O}( m\cdot m^6)^2 = \mathcal{O}(m^{14})$ basic terms. By using induction, we arrive at at the output layer, there are totally $\mathcal{O}(m^{2^p})$ basic terms. \\

\noindent
\textbf{Classical Algorithm for Training Neural Network.} The best classical algorithm to train neural networks works exactly as Algorithm \ref{algo: trainingneuralnetwork}, however, instead of explicitly working out the loss function and compute at the end, it can proceed layer by layer for a faster running time. More concretely, we consider a general network network (Fig. \ref{fig: neuralnetwork}) with an input layer that has $n$ nodes, $p$ hidden layers each layer having $m$ nodes, and an output layer having 1 node. At the first layer, it computes:
\begin{align}
    \textbf{a}^{(1)} = \sigma\big( \textbf{W}^{(0)} \xbf + \textbf{b}^{(0)}  \big) 
\end{align}
which takes time complexity $\mathcal{O}\big( mn   \big)$. Then, at the second layer, it computes: 
\begin{align}
    \textbf{a}^{(2)} = \sigma \big( \textbf{W}^{(1)} \textbf{a}^{(1)} + \textbf{b}^{(1)}    \big)
\end{align}
which takes further $\mathcal{O}\big( m^2 \big)$ time. Continually, at the third layer:
\begin{align}
    \textbf{a}^{(3)} = \sigma \big(   \textbf{W}^{(2)} \textbf{a}^{(2)} + \textbf{b}^{(2)} \big)
\end{align}
incurring further $\mathcal{O}(\big( m^2 \big)$ time. The same computation proceeds with remaining $p-2$ layers and output layer, so the total time complexity is $\mathcal{O}\big( mn + p m^2  \big)$. The loss function is computed straightforwardly according to Algorithm \ref{algo: trainingneuralnetwork}, and its gradient is evaluated numerically as follows:
\begin{align}
    \frac{\partial L}{\partial W} = \frac{L(W+\delta) - L(W) }{\partial W}
\end{align}
Thus, classical algorithm computes the loss function using two different weights set $W+\delta$ and $W$. The last step is backpropagation, which updates the parameters, taking further $\mathcal{O}( mn+ m^2 p  )$ time complexity as there are $p$ hidden layers, so in total there are $\mathcal{O}( mn+ m^2 p  )$ number of weights. The same procedure applies for $M$ data points, resulting in a total of time $\mathcal{O}\Big(  M(mn+m^2p) \Big)$  The memory usage for the classical algorithm is apparently $\mathcal{O}\big(  T( mn+ m^2 p  ) \big)$. 

The above discussion implies that the quantum gradient descent algorithm yields an exponential speed-up with respect to $n$ --the dimension of data point, meanwhile being polynomially slower in $M$ --the total number of data point and $m$ --the width of neural network. Thus, quantum algorithm achieves the best performance compared to classical counterpart in the regime where $n$ is very large, and $m = \mathcal{O}(1)$, and $M$ is considerably smaller than $n$. \\

\noindent
\textbf{Quantum Algorithm for Executing Neural Network.} Here we describe a very important routine, which provides an end-to-end application of the above quantum algorithm. Recall that once we finish the Algorithm \ref{thm: qatrainingneuralnetwork}, the output is a unitary block encoding of a diagonal operator that stores the weights $\big(\Wbf^{(0)},\Wbf^{(1)},...,\Wbf^{(p)}  \big)$ and bias $\big( \bbf^{0},\bbf^{1},..., \bbf^{p}\big)$. Given another input $\xbf \equiv (x_1,x_2,...,x_n)$ (we abuse the notation), how can we obtain the output $\Tilde{y}$ ? The answer is revealed in the following discussion.

Suppose that the outcome of algorithm underlying Thm.~\ref{thm: qatrainingneuralnetwork} be the block encoding of:
\begin{align}
\label{outcomeA}
    A = & \bigoplus_{i=1}^m  \begin{pmatrix}
       w^{(0)}_{i,1}  & 0 & 0 & 0 \\
       0 & w^{(0)}_{i,2} & 0 & 0 \\
       0 & 0 & \ddots & 0 \\
       0 & 0 & 0 & w^{(0)}_{i,n} 
   \end{pmatrix} \bigoplus_{i=1}^m \begin{pmatrix}
       w^{(1)}_{i,1}  & 0 & 0 & 0 \\
       0 & w^{(1)}_{i,2} & 0 & 0 \\
       0 & 0 & \ddots & 0 \\
       0 & 0 & 0 & w^{(1)}_{i,m} 
   \end{pmatrix} \bigoplus \cdots \bigoplus \begin{pmatrix}
       w^{(p)}_{1}  & 0 & 0 & 0 \\
       0 & w^{(0)}_{2} & 0 & 0 \\
       0 & 0 & \ddots & 0 \\
       0 & 0 & 0 & w^{(p)}_{m} 
   \end{pmatrix} \\ & \bigoplus  \begin{pmatrix}
        b^{(0)}_{1} & 0 & 0 & 0 \\
         0 & b^{(0)}_{2} & 0 &  0 \\
         0 & 0 & \ddots & 0 \\
        0 &  0 &  0 & b^{(0)}_{m} 
    \end{pmatrix} \bigoplus \cdots \bigoplus \begin{pmatrix}
        b^{(p)}_{1} 
    \end{pmatrix}
\end{align}
For simplicity, for $j=1,2,...,p$ and $i=1,2,...,m$, we denote:
\begin{align}
    W_i^0 = \begin{pmatrix}
       w^{(0)}_{i,1}  & 0 & 0 & 0 \\
       0 & w^{(0)}_{i,2} & 0 & 0 \\
       0 & 0 & \ddots & 0 \\
       0 & 0 & 0 & w^{(0)}_{i,n} 
   \end{pmatrix}, W_i^j =\begin{pmatrix}
       w^{(j)}_{i,1}  & 0 & 0 & 0 \\
       0 & w^{(j)}_{i,2} & 0 & 0 \\
       0 & 0 & \ddots & 0 \\
       0 & 0 & 0 & w^{(j)}_{i,m} 
   \end{pmatrix},   \textbf{B}^j = \begin{pmatrix}
        b^{(j)}_{1} & 0 & 0 & 0 \\
         0 & b^{(j)}_{2} & 0 &  0 \\
         0 & 0 & \ddots & 0 \\
        0 &  0 &  0 & b^{(j)}_{m}
    \end{pmatrix}
\end{align}
In Appendix \ref{sec: proofWB}, we prove the following lemmas:
\begin{lemma}
\label{lemma: WB}
    Given block encoding of $A$ as above, there exists a quantum circuit composed of one block encoding of $A$, and another complexity $\mathcal{O}(1)$ circuit, that is a block encoding of $W_i^j$, $\Bbf^j$ for any $i,j$. 
\end{lemma}
and we recall a result from the relevant work \cite{nghiem2024simple}:
\begin{lemma}[Lemma 2 in \cite{nghiem2024simple}]
    Given a unitary block encoding of a diagonal matrix $n \times n$ $\textbf{X} = \sum_{j=1}^{n} x_j \ket{j-1}\bra{j-1}$ of complexity $\mathcal{O}(\log(n))$, then given two integers $1\leq j,k \leq n$ there exists a quantum circuit of complexity $\mathcal{O}(\log(n))$ (using block encoding of $\textbf{X}$ two times), and $\log(n) +3$ ancilla qubits that produces the block encoding of a matrix of 1 entry $x_j \ket{k}\bra{k}$.
    \label{lemma: singleentry}
\end{lemma}
To proceed, we apply the above Lemma to $\textbf{B}^j$ to obtain the block encoding of $b^{(0)}_i \ket{0}\bra{0}$  for all $i,j$. If we have an efficient means to prepare the input $\xbf = (x_1,x_2,...,x_n)$, then the first version of Lemma \ref{lemma: diag} allows us to construct the block encoding of $\frac{1}{n} \Big( (\wbf^{0}_i)^\dagger \xbf + b^{(0)} \Big) \ket{0}\bra{0}$  for any $j=1,2,...,p$. The factor $n$ can be removed using amplification \ref{lemma: amp_amp} with further $\mathcal{O}(n)$ complexity. Otherwise, if we have an efficient means to prepare  $ (\sqrt{x_1},\sqrt{x_2},..., \sqrt{x_n})$, then the second version of Lemma \ref{lemma: diag} can be used instead, allowing us to prepare the block encoding of $ \Big( (\wbf^{(0)}_i)^\dagger \xbf + b^{(0)}\Big) \ket{0}\bra{0} $ more efficiently. The block encoding of $ \Big( (\wbf^{(0)}_i)^\dagger \xbf + b^{(0)}_i \Big) \ket{0}\bra{0}  $ can be transformed to block encoding of $  \Big( (\wbf^{(0)}_i)^\dagger \xbf + b^{(0)}_i \Big) \ket{i}\bra{i}$ by noting that there is a complexity 1 circuit composed of $X$ gates that transform $\ket{0}\longrightarrow \ket{i}$. Then we use Lemma \ref{lemma: sumencoding} to construct the block encoding of: 
\begin{align}
    \frac{1}{m} \sum_{i=1}^m \Big( (\wbf^{0}_i)^\dagger \xbf + b^{(0)} \Big) \ket{i-1}\bra{i-1}= \frac{1}{m}\begin{pmatrix}
      \sum_{i=1}^n w^{(0)}_{1,i} x_i + b^{(0)}_1  & 0 & 0 & 0 \\
       0 & \sum_{i=1}^n w^{(0)}_{2,i}  x_i +b^{(0)}_2   & 0 & 0 \\
       0 & 0 & \ddots & 0 \\
       0 & 0 & 0 &  \sum_{i=1}^n w^{(0)}_{m,i }  x_i + b^{(0)}_m 
   \end{pmatrix}
   \label{77}
\end{align}
The factor $m$ can be removed by using amplification method \ref{lemma: amp_amp}, with further complexity $\mathcal{O}\big( m \big)$. In order to apply the activation function $\sigma$, we leverage the quantum singular value transformation \cite{gilyen2019quantum}, as recapitulated in Lemma \ref{lemma: theorem56} of the Appendix \ref{sec: prelim}. Then, we obtain the block encoding of: 
\begin{align}
    \begin{pmatrix}
        \sigma\big( \sum_{i=1}^n w^{(0)}_{1,i} x_i + b^{(0)}_1 \big)  & 0  & 0 & 0 \\
        0 &   \sigma \big( \sum_{i=1}^n w^{(0)}_{2,i} x_i + b^{(0)}_2 \big) & 0 & 0  \\
        0 & 0& \ddots & 0 \\
        0 & 0 & 0 & \sigma \big( \sum_{i=1}^n w^{(0)}_{m,i} x_i + b^{(0)}_m \big) 
    \end{pmatrix} 
\end{align}
We remind that the diagonal entries above are indeed the entries of $a^{(1)} $ (see Figure \ref{fig: activation}, \ref{fig: neuralnetwork}), which means that we have successfully ``implemented'' the first layer. The proceeding of the next layer is straightforward from Equation, as we just need to replace the input $\textbf{X} = \rm diag(\xbf) $ by the above operator, and replace $\{\wbf^{(0)}_i\}_{i=1}^m$ in the above by $\{ \wbf^{(1)}_i\}_{i=1}^m$. The feedforward process till the $p$-th layer is carried out in a straightforward way as discussed. The final piece that we need to handle is the output layer. Suppose that after the $p$-th layer, we obtain a block encoding of the operator $diag( a^{(p)}_1, a^{(p)}_2, ..., )$
\begin{align}
     \begin{pmatrix}
       a^{(p)}_1  & 0 & 0 & 0 \\
       0 &  a^{(p)}_2 & 0 & 0 \\
       0 & 0 & \ddots & 0 \\
       0 & 0 & 0 &  a^{(p)}_m
   \end{pmatrix}
\end{align}
Lemma \ref{lemma: sumencoding} allows us to construct the block encoding of:
\begin{align}
    \begin{pmatrix}
      w^{(p)}_1 a^{(p)}_1  & 0 & 0 & 0 \\
       0 &  w^{(p)}_2 a^{(p)}_2 & 0 & 0 \\
       0 & 0 & \ddots & 0 \\
       0 & 0 & 0 &  w^{(p)}_m a^{(p)}_m
   \end{pmatrix} = \sum_{k=1}^m w^{(p)}_k a^{(p)}_k \ket{k}\bra{k}
\end{align}
We note that under the action of Hadamard gates $H^{\otimes \log m}$, we obtain the following:
\begin{align}
    H^{\otimes \log m} \ket{k}\bra{k} H^{\otimes \log m} = \frac{1}{m}\ket{0}\bra{0}_m + \rho^1_{\rm redundant}
\end{align}
where $\rho^1_{\rm redundant}$ refers to redundant terms being orthogonal to the first term, i.e, $\Tr \big( \ket{0}\bra{0}_m  \cdot \rho^1_{\rm redundant} \big)  =0$; additionally, $\ket{0}_m$ denotes specifically the computational basis state of $m$-dimensional Hilbert space. It is trivial that $H^{\otimes \log m}$ is unitary, thus is naturally block-encode itself. So we can use Lemma \ref{lemma: product} to construct the block encoding of:
\begin{align}
    H^{\otimes \log m} \Big( \sum_{k=1}^m w^{(p)}_k a^{(p)}_k \ket{k}\bra{k} \Big)  H^{\otimes \log m}  = \frac{\sum_{k=1}^m w^{(p)}_k a^{(p)}_k}{m} \ket{0}\bra{0} + (\cdots)
\end{align}
From Lemma \ref{lemma: WB}, we have obtained the block encoding of $b^{(p)}_1$, which is a scalar. Given that $\Ibb_m$ is trivial to block-encoded (see discussion below Definition \ref{def: blockencode}), it is simple to construct the block encoding of $b^{(p)}_1  \otimes \Ibb_m$ and of $(b^{(p)}_1 \otimes \Ibb_m)/m$(using Lemma \ref{lemma: tensorproduct}). We further note that:
\begin{align}
    (b^{(p)}_1  \otimes \Ibb_m)/m = \frac{b^{(p)}_1 }{m}  \ket{0}\bra{0}_m + \rho^2_{\rm redundant}
\end{align}
where $\rho^2_{\rm redundant}$ refers to another redundant term and $\Tr \big( \ket{0}\bra{0}_m  \cdot \rho^2_{\rm redundant} \big)  =0$. Then we can use Lemma \ref{lemma: sumencoding} to construct the block encoding of:
\begin{align}
   \frac{1}{2} \Big( \frac{\sum_{k=1}^m w^{(p)}_k a^{(p)}_k}{m} \ket{0}\bra{0}_m + \rho^1_{\rm redundant} + \frac{b^{(p)}_1}{m}  \otimes \Ibb_m  \Big)= \frac{\sum_{k=1}^m w^{(p)}_k a^{(p)}_k + b^{(p)}_1}{2m}  \ket{0}\bra{0}_m +  \frac{1}{2} \big( \rho^1_{\rm redundant}+ \rho^2_{\rm redundant} )
\end{align}
Let $U$ denotes the unitary block encoding of the above operator. According to Definition \ref{def: blockencode} and its below discussion (Eqn. \ref{eqn: action}), we have:
\begin{align}
    U \ket{\bf 0}\ket{0}_m &= \ket{\bf 0}\Big( \frac{\sum_{k=1}^m w^{(p)}_k a^{(p)}_k + b^{(p)}_1}{2m}  \ket{0}\bra{0}_m + \frac{1}{2} \big( \rho^1_{\rm redundant}+ \rho^2_{\rm redundant} )  \Big)  \ket{0}_m + \ket{\rm Garbage} \\
    &= \ket{\bf 0}  \Big( \frac{\sum_{k=1}^m w^{(p)}_k a^{(p)}_k + b^{(p)}_1}{2m}  \ket{0}_m + \frac{1}{2} \big( \rho^1_{\rm redundant}+ \rho^2_{\rm redundant} )\ket{0}_m \Big)  + \ket{\rm Garbage}
\end{align}
where we remind that $\ket{\bf 0}$ denotes the ancillary qubits required for block encoding purpose,$ \ket{\rm Garbage}$ denotes irrelevant states that are completely orthogonal to $\ket{\bf 0}  $, e.g., $ \Tr \big( \ket{\rm Garbage}\bra{\rm Garbage} \cdot \ket{\bf 0}\bra{\bf 0} \big) = 0$.  Then if we perform measurement on the ancillary system, and the $\log m$ qubit system, the probability of measuring $\ket{\bf 0}\ket{0}_m$ is:
\begin{align}
    | \bra{\bf 0}\bra{0}_m U \ket{\bf 0}\ket{0}_m  |^2 = \Big( \frac{\sum_{k=1}^m w^{(p)}_k a^{(p)}_k + b^{(p)}_1}{2m}  \Big)^2
\end{align}
Thus, one can repeat the measurement $\mathcal{O}(1/\epsilon^2)$ times to estimate such a probability up to additive error $\epsilon$. We remark that the complexity can be improved to $\mathcal{O}(1/\epsilon)$ by using amplitude estimation \cite{brassard1997quantum}. The value of $\frac{\sum_{k=1}^m w^{(p)}_k a^{(p)}_k + b^{(p)}_1}{2m}  $ is the output the neural network, up to a scaling factor $m$.

\section{Details for Solving Ground State and Excited State Energy}
\label{sec: detailgroundexcitedstate}
\noindent
\textbf{Explicit decomposition of energy $\xbf^T H \xbf$}. Here we show that the energy $E =\xbf^T H \xbf$, for $H = \sum_{i=1}^{N-1} H_{i,i+1}$ is of the first type (Thm~\ref{theorem: type1}). We have the following decomposition:
\begin{align}
    \sigma^z = \ket{0}\bra{0} - \ket{1}\bra{1}, \ \Ibb_{1,2,...,i-2} = \sum_{j=1}^{2^{i-2}} \ket{j}\bra{j}, \ \Ibb_{i+2,...,N} = \sum_{k=1}^{2^{N-i-1}} \ket{k}\bra{k}
\end{align}
So we have:
\begin{align}
    &\Ibb_{1,2,...,i-2} \otimes  \sigma^z_i \otimes \sigma^z_{i+1} \otimes \Ibb_{i+2,...,N} \\
    &= \sum_{j=0}^{2^{i-1}-1} \ket{j}\bra{j} \Big(  \ket{0}\bra{0}_i - \ket{1}\bra{1}_i   \Big) \Big(  \ket{0}\bra{0}_{i+1} - \ket{1}\bra{1}_{i+1}\Big) \sum_{k=0}^{2^{N-i-1}-1} \ket{k}\bra{k} \\
    &= \sum_{j=0}^{2^{i-1}-1} \ket{j}\bra{j} \Big( \ket{00}\bra{00} - \ket{01}\bra{01} - \ket{10}\bra{10} + \ket{11}\bra{11}  \Big)_{i,i+1} \sum_{k=0}^{2^{N-i-1}-1} \ket{k}\bra{k} \\
    &= \sum_{j=0}^{2^{i-1}-1}\sum_{k=0}^{2^{N-i-1}-1}  \ket{j}\ket{00}_{i,i+1}\ket{k} \bra{j}\bra{00}_{i,i+1} \bra{k} -\sum_{j=0}^{2^{i-1}-1}\sum_{k=0}^{2^{N-i-1}-1}  \ket{j}\ket{01}_{i,i+1}\ket{k} \bra{j}\bra{01}_{i,i+1}\bra{k}  \\ & - \sum_{j=0}^{2^{i-1}-1}\sum_{k=0}^{2^{N-i-1}-1}  \ket{j}\ket{10}_{i,i+1}\ket{k} \bra{j}\bra{10}_{i,i+1}\bra{k} + \sum_{j=0}^{2^{i-1}-1}\sum_{k=0 }^{2^{N-i-1}-1}  \ket{j}\ket{11}_{i,i+1}\ket{k} \bra{j}\bra{11}_{i,i+1}\bra{k}
\end{align}
So the energy is $\xbf^T H \xbf = \xbf^T \sum_{i=1}^{N-1} H_{i,i+1} \xbf = \xbf^T \sum_{i=1}^{N-1} \big( J_i \Ibb_{1,2,...,i-2} \otimes  \sigma^z_i \otimes \sigma^z_{i+1} \otimes \Ibb_{i+2,...,N} \big) \xbf $. Now we consider a specific term:
\begin{align}
    \xbf^T \Big(  \sum_{j=0}^{2^{i-1}-1}\sum_{k=0}^{2^{N-i-1}-1}  \ket{j}\ket{00}_{i,i+1}\ket{k} \bra{j}\bra{00}_{i,i+1}\bra{k} \Big) \xbf  =  \sum_{j=0}^{2^{i-1}-1}\sum_{k=1}^{2^{N-i-1}-1}  \Big( \bra{j}\bra{00}_{i,i+1}\bra{k} \xbf \Big)^2
\end{align}
For convenience, we adopt the representation of $\xbf \equiv (x_1,x_2,...,x_n) = \sum_{m=1}^n x_m \ket{m-1}$ in the binary string format as $\sum_{m=1}^n x_m \ket{m-1} \equiv  \sum_{s_1,s_2,...,s_N =0,1} x_{s_1 s_2 ... s_N} \ket{s_1s_2...s_N}$ where $N = \log n$. Thus, the dot product:
\begin{align}
    \sum_{j=0}^{2^{i-1}-1}\sum_{k=0}^{2^{N-i-1}-1}  \Big( \bra{j}\bra{00}_{i,i+1}\bra{k} \xbf \Big)^2 = \sum_{s_1,s_2,...,s_{i-1},s_{i+2},...,s_N = 0,1}  \big( x_{s_1s_2...s_{i-2} 00 s_k...s_N} \big)^2 
\end{align}
Similarly, we have:
\begin{align}
     \sum_{j=0}^{2^{i-1}-1}\sum_{k=0}^{2^{N-i-1}-1}  \Big( \bra{j}\bra{01}_{i,i+1}\bra{k} \xbf \Big)^2 = \sum_{s_1,s_2,...,s_{i-1},s_{i+2},...,s_N = 0,1}  \big( x_{s_1s_2...s_{i-1} 01 s_k...s_N} \big)^2 \\
     \sum_{j=0}^{2^{i-1}-1}\sum_{k=0}^{2^{N-i-1}-1}  \Big( \bra{j}\bra{10}_{i,i+1}\bra{k} \xbf \Big)^2 = \sum_{s_1,s_2,...,s_{i-1},s_{i+2},...,s_N = 0,1}  \big( x_{s_1s_2...s_{i-1} 10 s_k...s_N} \big)^2 \\
     \sum_{j=0}^{2^{i-1}-1}\sum_{k=0}^{2^{N-i-1}-1}  \Big( \bra{j}\bra{11}_{i,i+1}\bra{k} \xbf \Big)^2 = \sum_{s_1,s_2,...,s_{i-1},s_{i+2},...,s_N = 0,1}  \big( x_{s_1s_2...s_{i-1} 11 s_k...s_N} \big)^2
\end{align}
So the energy is:
\begin{align}
    E &= \xbf^T \sum_{i=1}^N \big( J_i \ \Ibb_{1,2,...,i-2} \otimes  \sigma^z_i \otimes \sigma^z_{i+1} \otimes \Ibb_{i+2,...,N} \big) \xbf \\
    &= \sum_{i=1}^N  \Big(   \sum_{s_1,s_2,...,s_{i-1},s_{i+2},...,s_N = 0,1}  J_i\big( x_{s_1s_2...s_{i-2} 00 s_k...s_N} \big)^2  +J_i\big( x_{s_1s_2...s_{i-1} 01 s_k...s_N} \big)^2  \\ &+  J_i\big( x_{s_1s_2...s_{i-1} 10 s_k...s_N} \big)^2 +J_i \big( x_{s_1s_2...s_{i-1} 11 s_k...s_N} \big)^2  \Big)
    \label{120}
\end{align}
which is exactly the first type of function considered in Thm~\ref{theorem: type1}. \\

\noindent
\textbf{Accelerating excited state energy finding.} We remind that once we finish with the ground state, then we find the first excited state energy by minimizing the function:
\begin{align}
    E_1 = \xbf'^T \big( H - E_0 \Phi_{\rm ground} \Phi_{\rm ground}^T\big) \xbf'
\end{align}
As mentioned, the above function fits into context of Thm~\ref{theorem: type2}. To gain more efficiency, which means that we need to be able to use the second version of Lemma \ref{lemma: diag}, then we need a means to generate some state that contains the squared of entries of $\Phi_{\rm ground}$. In order to achieve such a goal, we need the following:
\begin{lemma}[Positive Power Exponent \cite{gilyen2019quantum},\cite{chakraborty2018power}]
\label{lemma: positive}
    Given an $\delta$-approximated block encoding of a positive matrix $A$ such that 
    $$ \frac{\Ibb}{\kappa} \leq A \leq \Ibb. $$
   Let $\delta = \epsilon/ (\kappa \log^3(\frac{\kappa}{\epsilon})) $ and $c \in (0,1)$. Then we can implement an $\epsilon$-approximated block encoding of $A^c/2$ in time complexity $\mathcal{O}( \kappa T_A \log^2 (\frac{\kappa}{\epsilon})  )$, where $T_A$ is the complexity of block encoding of $A$. 
\end{lemma}
Recall from previous discussion that we have obtained the block encoding of 
$$\rm diag (\Phi_{\rm ground}) \equiv \rm diag \Big( (\Phi_{\rm ground})_1, (\Phi_{\rm ground})_2, ...,  (\Phi_{\rm ground})_n   \Big) $$
where $( (\Phi_{\rm ground})_i)$ refers to $i$-th entry of $\Phi_{\rm ground}$. Then we can use Lemma \ref{lemma: positive} with $c=1/2$ to transform:
\begin{align}
    \rm diag \Big( (\Phi_{\rm ground})_1, (\Phi_{\rm ground})_2, ...,  (\Phi_{\rm ground})_n   \Big)  \longrightarrow \rm diag \Big( \sqrt{(\Phi_{\rm ground})_1}, \sqrt{(\Phi_{\rm ground})_2}, ...,  \sqrt{(\Phi_{\rm ground})_n  } \Big) 
\end{align}
Denote $U_\Phi$ as the unitary block encoding of the above operator. Then according to Definition \ref{def: blockencode} (and property in Eqn. \ref{eqn: action}), we have that:
\begin{align}
    U_\Phi \ket{\bf 0} \sum_{i=1}^n \frac{1}{\sqrt{n}} \ket{i-1} = \ket{\bf 0} \sum_{i=1}^n \frac{ \sqrt{(\Phi_{\rm ground})_i}}{\sqrt{n}} \ket{i-1} + \ket{\rm Garbage}
\end{align}
Measuring the first ancillary system and post-select $\ket{\bf 0}$ we obtain the quantum state that is $\sim \sum_{i=1}^n  \sqrt{(\Phi_{\rm ground})_i} \ket{i-1}$, which contains $\Big( \sqrt{(\Phi_{\rm ground})_1}, \sqrt{(\Phi_{\rm ground})_2}, ...,  \sqrt{(\Phi_{\rm ground})_n  } \Big)$ (up to some normalization factor) in its first $n$ entries.

\section{Proof of Lemma \ref{lemma: WB} }
\label{sec: proofWB}
\noindent
\textbf{Proof of Lemma \ref{lemma: WB}.} Essentially we need to prove that given a block encoding of $\bigoplus_{i=1}^n W_i$ where $W_i$ is some $n \times n$ operator, then the block encoding of $W_i$ for any $i$ is obtainable. To show this, we note that $ \bigoplus_{i=1}^n W_i = \sum_{i=1}^n \ket{i-1}\bra{i-1} \otimes W_i$. From $\ket{0}$, it is simple to construct a circuit $U$ of complexity 1, composed soly of $X$ gates, that transform $\ket{0}\longrightarrow \ket{i-1}$ for any $i$. By virtue of Lemma \ref{lemma: improveddme}, we can use such circuit to construct the block encoding of $\ket{i-1}\bra{i-1}$. Then Lemma \ref{lemma: product} allows us to construct the block encoding of:
\begin{align}
    \ket{i-1}\bra{i-1} \cdot \Big(\sum_{i=1}^n \ket{i-1}\bra{i-1} \otimes W_i  \Big) = \ket{i-1}\bra{i-1}\otimes W_i
\end{align}
Then we use Lemma \ref{lemma: product} again to construct the block encoding of:
\begin{align}
    U\Big( \ket{i-1}\bra{i-1}\otimes W_i \Big) U^\dagger = \ket{0}\bra{0}\otimes W_i
\end{align}
According to Definition \ref{def: blockencode}, it is also a block encoding of $W_i$.

\end{document}